\documentclass[11pt,a4paper]{article}

\usepackage{jheppub}
\usepackage{amsfonts}
\usepackage{amssymb}
\usepackage{comment}
\usepackage{epsfig}
\usepackage{graphicx}
\usepackage{amsmath}
\usepackage{tabu}
\usepackage{subfig}
\usepackage{float}
\usepackage{epstopdf}
\usepackage{multirow}
\usepackage{verbatim}
\usepackage{color}

\newcommand{\bea}{\begin{eqnarray}}
\newcommand{\eea}{\end{eqnarray}}
\newcommand{\bi}{\begin{itemize}}
\newcommand{\ei}{\end{itemize}}
\newcommand{\ben}{\begin{enumerate}}
\newcommand{\een}{\end{enumerate}}
\newcommand{\be}{\begin{equation}}
\newcommand{\ee}{\end{equation}}
\newcommand{\ba}{\begin{align}}
\newcommand{\ea}{\end{align}}
\newcommand{\T}{\mathbb{T}}
\newcommand{\mP}{\mathbb{P}}
\newcommand{\F}{\mathcal{F}}
\newcommand{\comments}[1]{}
\def\nn{\nonumber}

\def\LVS{{\scriptscriptstyle \rm LVS}}

\def\b{{\scriptscriptstyle \rm bulk}}
\def\KK{{\scriptscriptstyle \rm KK}}
\def\W{{\scriptscriptstyle \rm W}}

\def\dS{{\scriptscriptstyle \rm dS}}

\newcommand\vo{{\mathcal{V}}}
\newcommand{\mbb}{\mathbb}

\newcommand{\mc}{\mathcal}

\newcommand{\beqa}{\begin{eqnarray}}
\newcommand{\eeqa}{\end{eqnarray}}

\title{Chiral Global Embedding of Fibre Inflation Models}

\author[a,b,c]{Michele Cicoli,}
\author[a,b]{David Ciupke,}
\author[a,b]{Victor A. Diaz,}
\author[a]{Veronica Guidetti,}
\author[d]{Francesco Muia,}
\author[c]{Pramod Shukla}

\affiliation[a]{\small Dipartimento di Fisica e Astronomia, Universit\`a di Bologna, \\ via Irnerio 46, 40126 Bologna, Italy}
\affiliation[b]{\small INFN, Sezione di Bologna, viale Berti Pichat 6/2, 40127 Bologna, Italy}
\affiliation[c]{\small Abdus Salam ICTP, Strada Costiera 11, Trieste 34151, Italy}
\affiliation[d]{\small Rudolf Peierls Centre for Theoretical Physics, University of Oxford, \\ 1 Keble Rd., Oxford OX1 3NP, UK}

\emailAdd{mcicoli@ictp.it}
\emailAdd{ciupke@bo.infn.it}
\emailAdd{diaz@bo.infn.it}
\emailAdd{veronica.guidetti@studio.unibo.it}
\emailAdd{francesco.muia@physics.ox.ac.uk}
\emailAdd{shukla.pramod@ictp.it}

\abstract{We construct explicit examples of fibre inflation models which are globally embedded in type IIB orientifolds with chiral matter on D7-branes and full closed string moduli stabilisation. The minimal setup involves a Calabi-Yau threefold with $h^{1,1}=4$ K\"ahler moduli which features multiple K3 fibrations and a del Pezzo divisor supporting non-perturbative effects. We perform a consistent choice of orientifold involution, brane setup and gauge fluxes which leads to chiral matter and a moduli-dependent Fayet-Iliopoulos term. After D-term stabilisation, the number of K\"ahler moduli is effectively reduced to $3$ and the internal volume reduces to the one of fibre inflation models. The inflationary potential is generated by suitable string loop corrections in combination with higher derivative effects. We analyse the inflationary dynamics both in the single-field approximation and by numerically deriving the full multi-field evolution in detail. Interestingly, we find that the K\"ahler cone conditions set strong constraints on the allowed inflaton field range.}


\begin{document}

\maketitle

\section{Introduction}
\label{intro}

Cosmic inflation is an early period of accelerated expansion of our universe which can provide a solution to the flatness and horizon problems of standard Big Bang cosmology. Moreover, quantum fluctuations during inflation can source primordial perturbations that caused the formation of large scale structures and the temperatures anisotropies observed in the cosmic microwave background.

From a microscopic point of view, inflation is expected to be driven by the dynamics of a scalar field undergoing a slow-roll motion along a very shallow potential that mimics a positive cosmological constant. An important feature of inflationary models is the distance travelled by the inflaton in field space during inflation since it is proportional to the amount of primordial gravitational waves which get produced \cite{Lyth:1996im}. From an effective field theory point of view, in small field models with a sub-Planckian inflaton excursion, dimension six operators can easily spoil the flatness of the inflationary potential. On the other hand, quantum corrections to large field models with a trans-Planckian field range lead to an infinite series of unsuppressed higher-dimensional operators which seem to bring the effective field theory approach out of control.

These dangerous operators can be argued to be absent or very suppressed only in the presence of a symmetry whose origin can only be postulated from an effective field theory perspective but can instead be derived from an underlying UV complete theory. For this reason inflationary model building in string theory has received a lot of attention \cite{McAllister:2007bg, Baumann:2009ni, Cicoli:2011zz, Burgess:2013sla}. Besides the presence of additional symmetries, string compactifications naturally provide many 4D scalars which can play the r\^ole of the inflaton. Promising inflaton candidates are type IIB K\"ahler moduli which parametrise the size of the extra dimensions and enjoy non-compact rescaling symmetries inherited from the underlying no-scale structure \cite{Burgess:2014tja}. 

Identifying a natural inflaton candidate with an appropriate symmetry that protects the flatness of its potential against quantum corrections is however not sufficient to trust inflationary model building in string compactifications. In fact, three additional requirements to have a successful string inflationary model are ($i$) full moduli stabilisation, ($ii$) a global embedding into consistent Calabi-Yau orientifolds with D-branes and fluxes and ($iii$) the realisation of a chiral visible sector. 

The first condition is crucial to determine all the energy scales in the model and to check the stability of the inflationary dynamics by controlling the behaviour of the scalar directions orthogonal to the inflaton one. The second condition is instead fundamental to guarantee the consistency of the inflation model from the microscopic point of view by checking the cancellation of all D-brane tadpoles and Freed-Witten anomalies and the actual generation of all the effects needed to stabilise the moduli and to develop the inflationary potential. Finally the requirement of having a model which can give rise to inflation and reproduce at the same time a chiral visible sector is crucial for two main reasons: to ensure the absence of any dangerous interplay between chirality and moduli stabilisation which can forbid the generation of D-terms or non-perturbative effects needed to fix the moduli \cite{Blumenhagen:2007sm}, and to determine the post-inflationary evolution of our universe starting from the reheating process where the inflaton energy density gets converted into the production of visible sector degrees of freedom \cite{Green:2007gs, Brandenberger:2008kn, Barnaby:2009wr, Cicoli:2010ha}. Other important post-inflationary issues which can affect the predictions of important inflationary observables like the number of efoldings $N_e$, the scalar spectral index $n_s$ and the tensor-to-scalar ratio $r$ are periods of matter domination due to light moduli \cite{Dutta:2014tya, Cicoli:2016olq, Bhattacharya:2017ysa}, the production of axionic dark radiation from moduli decays \cite{Cicoli:2012aq, Higaki:2012ar, Hebecker:2014gka, Cicoli:2015bpq}, non-thermal dark matter \cite{Acharya:2008bk, Allahverdi:2013noa, Aparicio:2015sda}, moduli-induced baryogenesis \cite{Kane:2011ih, Allahverdi:2016yws} and the interplay between the inflationary and the supersymmetry breaking scale \cite{Conlon:2008cj, He:2010uk, Antusch:2011wu, Buchmuller:2014pla}.

A comprehensive global chiral model which satisfies all these conditions for models where the inflaton is a local blow-up mode \cite{Conlon:2005jm} has been recently constructed in \cite{Cicoli:2017shd}. The chiral visible sector lives on D3-branes at an orientifolded singularity and full closed string moduli stabilisation in a dS vacuum is achieved by following the LVS procedure \cite{Balasubramanian:2005zx, Cicoli:2008va}. The main limitation of this model is the emergence of an $\eta$-problem associated with the presence of large $g_s$ corrections to the effective action which tend to spoil the flatness of the inflationary potential if their flux-dependent coefficients are not tuned small. 

In this regard, fibre inflation models \cite{Cicoli:2008gp} look more promising. In these constructions, the inflaton is a fibration modulus which remains exactly massless when only the leading order no-scale breaking effects are included. The inflationary potential is then generated only at subleading order by a combination of string loop corrections \cite{Berg:2005ja, Berg:2007wt, Berg:2014ama, Haack:2015pbv} and higher derivative terms \cite{Ciupke:2015msa, Grimm:2017okk}. This hierarchy of scales is guaranteed by the extended no-scale cancellation and provides a natural solution to the $\eta$-problem \cite{Cicoli:2007xp}. This solution can also be understood from the point of view of an effective non-compact rescaling symmetry for the K\"ahler moduli \cite{Burgess:2014tja}. 

Different versions of fibre inflation models have been constructed so far depending on the microscopic nature of the effects which drive the inflationary dynamics: Kaluza-Klein and winding string loops \cite{Cicoli:2008gp}, Kaluza-Klein loops and $\mc{O}(\alpha'^3)$ $F^4$ terms \cite{Broy:2015zba}, and winding $g_s$ loops combined with higher derivative terms \cite{Cicoli:2016chb}. In all cases the inflationary potential is plateau-like and takes a simple form with a constant term and negative exponentials. Additional positive exponentials show up with coefficients which are naturally very small and give rise to a rising behaviour at large field values. Ref. \cite{Burgess:2016owb} provided a generalised description of fibre inflation models showing how they can reproduce the correct spectral index observed by Planck \cite{Ade:2015lrj, Ade:2015xua} while the predicted value of the tensor-to-scalar ratio is in the range $0.001 \lesssim r\lesssim 0.01$. Such a large value of $r$ is compatible with the fact that these are large field models where the inflaton range is around $5$ Planck units. An effective supergravity description of fibre inflation models as $\alpha$-attractors has also been recently given in \cite{Kallosh:2017wku}. 

Despite all these successes, fibre inflation models are still lacking a complete global embedding into chiral string compactifications. However a first step forward has already been made in \cite{Cicoli:2016xae} where these inflationary models have been successfully embedded in consistent type IIB orientifolds with moduli stabilisation but without a chiral visible sector. In order to have a viable inflationary and moduli stabilisation mechanism, the internal Calabi-Yau manifold has to have at least $h^{1,1}=3$ K\"ahler moduli and its volume form has to feature a K3 or $T^4$ fibration over a $\mbb{P}^1$ base and a rigid  shrinkable blow-up mode \cite{Cicoli:2008va, Cicoli:2011it}. Starting from concrete Calabi-Yau threefolds with these topological properties, ref. \cite{Cicoli:2016xae} provided several different examples with an explicit choice of orientifold involution and D3/D7 brane setups which are globally consistent and can generate corrections to the 4D effective action that can fix all closed string moduli inside the K\"ahler cone and reproduce the form of the inflationary potential of fibre inflation models. However the case with $h^{1,1}=3$ is too simple to allow for non-trivial D7 worldvolume fluxes which give rise to chiral matter. In fact, non-zero gauge fluxes induce moduli dependent Fayet-Iliopoulos terms which, in combination with soft term contributions for $U(1)$-charged matter fields, would lift the leading order flat direction, making the inflaton too heavy to drive inflation. 

In this paper we shall extend the results of \cite{Cicoli:2016xae} by considering more complicated Calabi-Yau threefolds with $h^{1,1}=4$ in order to build global fibre inflation models with a chiral visible sector. After analysing the topological conditions on the underlying compactification manifold to allow a successful chiral global embedding of fibre inflation models, we find that the simplest examples involve Calabi-Yau threefolds with $3$ K3 divisors and a toroidal-like volume with a diagonal del Pezzo divisor suitable to support non-perturbative effects to freeze the moduli. The internal volume is therefore controlled by $3$ K\"ahler moduli and can equivalently be seen as different K3 fibrations over $3$ different $\mbb{P}^1$ bases. After searching through the Kreuzer-Skarke list of Calabi-Yau manifolds embedded in toric varieties \cite{Kreuzer:2000xy}, we find several concrete examples which admit these topological features. 

We then focus on one of them and describe several possible choices of orientifold involution, D-brane setup and gauge fluxes which satisfy global consistency conditions and generate perturbative $g_s$ and $\alpha'$ corrections to the 4D K\"ahler potential and non-perturbative effects in the superpotential that are suitable to both stabilise the moduli and reproduce the typical potential of fibre inflation models. In particular, non-zero gauge fluxes induce chiral matter on D7-branes wrapped around smooth combinations of the four-cycles which control the overall volume.\footnote{We do not consider K3 fibred cases where the visible sector lives on D3 branes at singularities since they would lead to dark radiation overproduction \cite{Angus:2014bia}.} Moreover, a moduli-dependent Fayet-Iliopoulos term lifts one of the K\"ahler moduli, so that after D-term stabilisation the effective number of K\"ahler moduli is reduced to $3$ and the internal volume simplifies to the standard expression of fibre inflation models used in the examples of \cite{Cicoli:2016xae}.

After computing all relevant loop and higher derivative effects in full detail, we analyse the resulting inflationary dynamics finding an interesting result: the K\"ahler cone bounds set severe constraints on the allowed inflaton field range when they are combined with other phenomenological requirements, like the generation of the correct amplitude of the power spectrum by the inflaton quantum fluctuations, and consistency conditions like the stability of the inflaton evolution against possible orthogonal runaway directions, the fact that the gravitino mass remains always smaller than any Kaluza-Klein scale in the model and finally that dangerous higher derivative effects do not spoil the flatness of the inflationary potential before achieving enough efoldings of inflation.\footnote{These last two consistency conditions are qualitatively similar since the superspace derivative expansion is under control if $m_{3/2}\ll M_\KK$ \cite{Cicoli:2013swa}.} Because of this tension, we also perform a full multi-field numerical analysis of the inflationary evolution showing how an early period of accelerated expansion occurs generically. On the other hand, the inflaton quantum fluctuations can generate the right amplitude of the density perturbations only if the microscopic parameters take appropriate values. 

We believe that our results make fibre inflation models more robust since they represent the first concrete models which are globally consistent and chiral. Nonetheless several issues still need to be investigated further. The most important ones are the inclusion of an explicit uplifting mechanism to realise a dS vacuum, a thorough derivation of the perturbative corrections to the 4D effective action and a better determination of the Calabi-Yau K\"ahler cone, going beyond its approximated expression inherited from the toric ambient space. We leave the study of these issues for the future. 

This paper is organised as follows. In Sec. \ref{MinReq}, after presenting a basic review of fibre inflation models, we summarise the minimal requirements that are needed for the construction of a fully consistent global embedding with a chiral visible sector. In Sec. \ref{ExA} we provide a concrete Calabi-Yau example, describing the orientifold involution, the D-brane setup, the choice of gauge fluxes and the resulting chiral spectrum, Fayet-Iliopoulos term and inflationary potential generated by $g_s$ and $\alpha'$ effects. The inflationary evolution is analysed in full detail in Sec. \ref{InfAn} by focusing first on the single-field approximation and by studying then the multi-field dynamics. In Sec. \ref{Concl} we draw our conclusions and we discuss a few open issues. App. \ref{ExB} contains additional explicit chiral global examples.

\section{Chiral global inflationary models}
\label{MinReq}

Let us begin by briefly reviewing the setup of fibre inflation and proceed afterward by displaying the minimal requirements for a successful chiral global embedding of fibre inflation models.

\subsection{Fibre inflation in a nutshell}

Fibre inflation models are based on a class of type IIB orientifold flux compactifications with D3/D7-branes and O3/O7-planes where the Calabi-Yau (CY) threefold takes a so-called `weak Swiss-cheese' form:
\be
\vo = f_{3/2}(\tau_j)  - \sum_{i=1}^{N_{\rm small}} \lambda_i \tau_i^{3/2} \quad\text{with}\quad j=1,...,N_{\rm large}\,,
\label{voGen}
\ee
where $h^{1,1} = N_{\rm large} + N_{\rm small}$ and $f_{3/2}$ is a homogeneous function of degree $3/2$. In these models, the stabilisation of the K\"ahler moduli is performed in two steps. Firstly, the total volume $\vo$ as well as the volumes of the $N_{\rm small}$ rigid blow-up divisors $\tau_i$ are fixed following the LVS procedure \cite{Balasubramanian:2005zx, Cicoli:2008va} where the leading order $\alpha'^3$ corrections to the K\"ahler potential \cite{Becker:2002nn, Minasian:2015bxa, Bonetti:2016dqh} are balanced against non-perturbative contributions to the superpotential \cite{Kachru:2003aw}. This leaves $N_{\rm flat} = N_{\rm large} -1 = h^{1,1} - N_{\rm small} - 1$ flat directions which are natural inflaton candidates. These directions can receive a potential at subleading order by $g_s$ corrections due to the exchange of Kaluza-Klein (KK) and winding modes \cite{Berg:2005ja, Berg:2007wt, Berg:2014ama, Haack:2015pbv, Cicoli:2007xp} as well as by $(\alpha')^3$ $F^4$-terms \cite{Ciupke:2015msa, Grimm:2017okk}. In the simplest fibre inflation models $h^{1,1}=3$ and $N_{\rm small}=1$, so that $N_{\rm flat}=1$. This leading order flat direction corresponds to a K\"ahler modulus $\tau_f$ which parametrises the volume of a K3 surface and the total scalar potential schematically looks like \cite{Cicoli:2008gp, Broy:2015zba, Cicoli:2016chb, Burgess:2016owb}:
\be
V = V_\LVS (\vo ,\tau_s ) + V_\dS (\vo) + V_{\rm inf} (\vo ,\tau_s, \tau_f)\,,
\ee
where $V_{\rm inf} (\vo ,\tau_s, \tau_f) = V_{g_s}^\KK + V_{g_s}^\W + V_{F^4} \ll V_\LVS (\vo ,\tau_i )$ is the inflationary potential. $V_\LVS$ is the leading order LVS potential which fixes $\vo$ and $\tau_s$, $V_\dS$ is an uplifting contribution to get a dS vacuum which can originate from anti D3-branes \cite{Kachru:2003aw, antiDdS}, hidden sector T-branes \cite{Cicoli:2015ylx} or non-perturbative effects at singularities \cite{Cicoli:2012fh}, while $V_{g_s}^\KK$, $V_{g_s}^\W$ and $V_{F^4}$ are respectively KK and winding string loops and $F^4$ terms. 

In fibre inflation models, the underlying CY threefold is a K3 fibration over a $\mbb{P}^1$ base which has two decompactification limits, corresponding to either the K3 fibre or the base growing large. Thus, kinematically it is expected that the fibre volume can traverse several Planck units. These LVS inflationary models present a variety of distinct features that make them very promising candidates to realise large field inflation and to discuss explicit global embeddings:
\ben
\item The de Sitter uplift is independent of the inflaton. This is contrary to a hypothetical KKLT embedding \cite{Kachru:2003aw}, where the uplift would be inflaton-dependent and, thus, large field inflation would typically destroy the KKLT minimum.

\item The back-reaction of heavy moduli is incorporated and under control, in particular, due to the fact that moduli stabilisation is done in two steps and the leading order potential is independent of the inflaton because of the extended no-scale cancellation \cite{Cicoli:2007xp}. This is in contrast with the majority of large field models of inflation \cite{Westphal:2015eva}.

\item The possibility to achieve tensor-to-scalar ratios between $r \sim 0.01$ and $r \sim 0.001$ which can be tested by future CMB observations \cite{Remazeilles:2015hpa, Remazeilles:2017zlb}.
\een
An explicit realisation of fibre inflation not only places several constraints on the underlying CY geometry, but also on the setup of D-branes and O-planes. In the following section we list the sufficient requirements to build a viable global model which also allows for a chiral visible sector.

\subsection{Requirements for chiral global embedding}

The simplest global embedding of fibre inflation models requires at least three K\"ahler moduli \cite{Cicoli:2016xae}. However, in order to incorporate also a chiral visible sector we need at least $h^{1,1}=4$ K\"ahler moduli. Here we will focus on obtaining chiral matter on D7-branes wrapped around a suitable divisor with world-volume gauge fluxes turned on. In this case D7 gauge fluxes induce a D-term potential for the K\"ahler moduli that fixes a particular combination thereof. Thus, D-term fixing and the leading order LVS stabilisation mechanism leave just a single flat direction, in our case a K3 fibre, which will play the r\^ole of the inflaton. In order to obtain a viable chiral global model we require the following ingredients and consistency conditions:
\ben
\item A Calabi-Yau with $h^{1,1}=4$ featuring three large cycles and a shrinkable rigid divisor, so that the internal volume takes the form (\ref{voGen}) with $N_{\rm small}=1$. In the explicit example described in Sec. \ref{ExA} the volume simplifies further to: 
\be
\vo=c_a\,\sqrt{\tau_1\,\tau_2\,\tau_3}-c_b\,\tau_s^{3/2}\,,
\label{VoLu}
\ee
with $c_a>0$ and $c_b>0$. Each of the $3$ moduli $\tau_1$, $\tau_2$ and $\tau_3$ controls the volume of a K3 surface while $\tau_s$ parametrises the size of a `diagonal' del Pezzo divisor \cite{Cicoli:2011it}. D-term stabilisation will fix $\tau_3\propto \tau_2$ while the standard LVS procedure will freeze the overall volume $\vo\simeq c_a\,\sqrt{\tau_1\,\tau_2\,\tau_3}$ and the blow-up mode $\tau_s$. The leading order flat direction can be parametrised by $\tau_1$ which will drive inflation. 

\item An orientifold involution and a D3/D7-brane setup with gauge fluxes on the visible D7-brane stacks such that tadpole cancellation is satisfied with enough room for bulk three-form fluxes to be turned on for complex structure and dilaton stabilisation. The D-brane and O-plane setup must also allow for the generation of KK- and/or winding string loop corrections which have the correct form to generate a suitable inflationary potential.

\item A choice of world-volume fluxes which cancels all Freed-Witten anomalies \cite{Minasian:1997mm, Freed:1999vc} but leads, at the same time, to just a single moduli-dependent Fayet-Iliopoulos (FI) term \cite{Dine:1987xk, Dine:1987gj} in order to leave a leading order inflationary flat direction by lifting just one of the two flat directions leftover by the LVS stabilisation mechanism. 

\item There should be no chiral intersection between the visible sector and the del Pezzo divisor supporting non-perturbative effects required for LVS moduli fixing as otherwise the prefactor of the non-perturbative superpotential would be vanishing \cite{Blumenhagen:2007sm}. The absence of these dangerous chiral intersections should be guaranteed by an appropriate choice of gauge fluxes. 

\item Moduli stabilisation and inflation have to take place inside the CY K\"ahler cone and the effective field theory should be well under control with $\langle \vo \rangle \gg 1$ and $g_s \ll 1$. 

\item In order to trust inflationary model building within an effective field theory, the following hierarchy of scales should be satisfied from horizon exit to the end of inflation:
\be
m_{\rm inf} < H < m_{3/2} < M_\KK^{(i)} < M_s < M_p \,,
\ee 
where $m_{\rm inf}$ is the inflaton mass, $H$ is the Hubble constant, $m_{3/2}$ is the gravitino mass which sets the mass scale of all the heavy moduli during inflation, $M_\KK^{(i)}$ denote various KK scales associated with bulk modes and open string excitations on D7-branes wrapped around four-cycles, $M_s$ is the string scale and $M_p$ is the reduced Planck mass $M_p = 2.4\cdot 10^{18}$ GeV. Notice that, apart from $M_p$, all these energy scales are moduli dependent and so evolve during inflation. After stabilising $\vo$ and $\tau_s$ $\grave{a}$ la LVS and fixing one large modulus in terms of another large direction via setting the FI-term to zero, we find that the `reduced' moduli space of the inflationary direction is in fact a compact interval. Therefore the field space available for inflation is kinematically finite (albeit in general trans-Planckian), a feature of the model which has so far been overlooked. We will state the precise phenomenological and consistency conditions for successful inflation in Sec. \ref{InfAn}.
\een

\section{A chiral global example}
\label{ExA}

In this section, we shall present all the topological and model-building details of the global embedding of fibre inflation models into explicit chiral CY orientifolds with $h^{1,1} = 4$.

\subsection{Toric data}

Let us consider the following toric data for a CY threefold whose volume takes the form $\vo=c_a\,\sqrt{\tau_1\,\tau_2\,\tau_3}-c_b\,\tau_s^{3/2}$ discussed above: 
\begin{table}[H]
  \centering
 \begin{tabular}{|c||cccccccc|}
\hline
     & $x_1$  & $x_2$  & $x_3$  & $x_4$  & $x_5$ & $x_6$  & $x_7$ & $x_8$   \\
    \hline
 4 & 0 & 0  & 0 & 1 & 1 & 0 & 0  & 2   \\
 4 & 0 & 0  & 1 & 0 & 0 & 1 & 0  & 2   \\
 4 & 0 & 1  & 0 & 0 & 0 & 0 & 1  & 2   \\   
 8 & 1 & 0  & 0 & 1 & 0 & 1 & 1  & 4   \\   
 \hline
 & dP$_7$ & NdP$_{11}$  &  NdP$_{11}$ & K3 & NdP$_{11}$ & K3 & K3 & SD \\
 \hline
 \end{tabular}
 \end{table}
The Hodge numbers are $(h^{2,1}, h^{1,1}) = (98, 4)$, the Euler number is $\chi=-188$, while the Stanley-Reisner ideal is:
\be
{\rm SR1} =  \{ x_1 x_4, \, x_1 x_6,\, x_1 x_7, \, x_2 x_7, \, x_3 x_6, \, x_4 x_5 x_8, \, x_2 x_3 x_5 x_8 \}\,. \nn
\ee
This corresponds to the polytope ID $\#1206$ in the CY database of Ref. \cite{Altman:2014bfa}. A detailed divisor analysis using \texttt{cohomCalg} \cite{Blumenhagen:2010pv, Blumenhagen:2011xn} shows that the divisor $D_1$ is a del Pezzo dP$_7$ while each of the divisors $\{D_4, \, D_6,\, D_7 \}$ is a K3 surface. Moreover, each of the divisors $\{D_2, \, D_3,\, D_5 \}$ is a `rigid but not del Pezzo' surface with $h^{1,1} =12$ which we denote as NdP$_{11}$ while $D_8$ is a `special deformation' divisors with Hodge diamond:
\bea
{\rm SD} \equiv
\begin{tabular}{ccccc}
    & & 1 & & \\
   & 0 & & 0 & \\
  23 & & 160 & & 23 \\
   & 0 & & 0 & \\
    & & 1 & & \\
  \end{tabular} \nn
	\eea
The intersection form in the basis of smooth divisors $\{D_1, D_4, D_6, D_7\}$ can be written as:
\be
I_3= 2 \, D_4 \, D_6 \, D_7 + 2 \, D_1^3 \,.
\label{I3A}
\ee
Writing the K\"ahler form in the above basis of divisors as $J=t_1\, D_1 + t_4\, D_4 + t_6 \, D_6+ t_7\, D_7$  and using the intersection
polynomial (\ref{I3A}), the CY overall volume becomes:
\be
\vo = 2\, t_4\, t_6\, t_7 + \frac{t_1^3}{3}\,.
\label{VolA}
\ee
The K\"ahler cone conditions can be derived from the following generators of the K\"ahler cone: 
\be
K_1 = - \, D_1 + D_4 + D_6 + D_7\,, \qquad K_2 =  D_7\,, \qquad K_3 =  D_4\,, \qquad K_4 =  D_6\,.
\ee
Expanding the K\"ahler form as $J= \sum_{i=1}^4 r_i K_i$, the K\"ahler cone is defined via the following conditions on the two-cycle moduli:
\be
r_1 = -\, t_1 > 0\,, \qquad r_2 = t_1 + t_7 > 0\,, \qquad r_3 = t_1 + t_4 >0\,, \qquad r_4 = t_1 + t_6 > 0\,.
\label{KCCA}
\ee
Notice that this expression of the CY K\"ahler cone is only approximate since it is inherited from the K\"ahler cone of the ambient toric variety.\footnote{If the same CY threefold can be realised as a hypersurface embedded in different ambient spaces, the CY K\"ahler cone is approximated as the intersection of the K\"ahler cones of the different toric varieties \cite{Altman:2014bfa}.} However this procedure can either overcount some curves of the CY threefold, for example if they do not intersect with the CY hypersurface, or miss some of them, if they cannot be obtained as the intersection between two divisors of the ambient space and the CY hypersurface. Hence the actual CY K\"ahler cone can turn out to be either larger or smaller. This analysis would require a deeper investigation which is however beyond the scope of this paper.\footnote{We however expect that the CY K\"ahler cone cannot get smaller. In fact, if this were the case, there should exist an extra constraint from requiring the positivity of a curve of the CY which is trivial in the ambient space. But this does not seem to be possible since each CY divisor is inherited from a single toric divisor (i.e. we do not have a toric divisor which splits into two CY divisors, and so where $h^{1,1}$ of the CY is larger than $h^{1,1}$ of the ambient space). In fact, if this trivial curve existed, it should have a dual divisors, and so $h^{1,1}$ of the CY should be larger than $h^{1,1}$ of the ambient case, which is however not the case.} Here we just mention that this analysis has been performed in detail in \cite{Cicoli:2012vw} where the CY K\"ahler cone turned out to be larger than the approximated version. 

The four-cycle moduli, which can be computed as $\tau_i = \partial_{t_i}\vo$, look like:
\be
\tau_1 = t_1^2\,, \qquad \tau_4 = 2\, t_6\, t_7\,, \qquad \tau_6 = 2\, t_4\, t_7\,, \qquad \tau_7 = 2\, t_4\, t_6\,,
\ee
and so, using the K\"ahler cone conditions (\ref{KCCA}), the overall volume reduces to:
\be
\vo = t_4\tau_4-\frac13\, \tau_1^{3/2} = t_6\tau_6-\frac13\, \tau_1^{3/2}
= t_7\tau_7-\frac13\, \tau_1^{3/2} = \frac{1}{\sqrt{2}}\,\sqrt{\tau_4\, \tau_6\, \tau_7} -\frac13\, \tau_1^{3/2}\,,
\ee
which shows clearly that the CY threefold $X$ features three K3 fibrations over different $\mP^1$ bases. 
The second Chern class of $X$ is given by:
\be
c_2 (X) = D_4 \, D_5 + 4\, D_5^2 + 12\, D_5 \, D_6 + 12\, D_5\, D_7 + 12\, D_6\, D_7\,,
\ee
which results in the following values of the topological quantities $\Pi_i=\int_X c_2 \wedge \hat{D}_i$:
\be
\Pi_1 = 8\,, \quad \Pi_2 = \Pi_3 =16\,, \quad \Pi_4 =24\,,
\quad \Pi_5 =16\,, \quad \Pi_6 = \Pi_7 =24\,, \quad\Pi_8 =128\,.
\label{Pi'sA}
\ee
The intersection curves between two coordinate divisors are given in Tab. \ref{Tab2} while their volumes are listed in Tab. \ref{Tab3}. 
\begin{table}[H]
  \centering
 \begin{tabular}{|c||c|c|c|c|c|c|c|c|}
\hline
 & $D_1$ & $D_2$ & $D_3$ & $D_4$ & $D_5$ & $D_6$ & $D_7$ & $D_8$   \\
    \hline
 $D_1$ & $\mc{C}_3$  & $\T^2$              & $\T^2$              & $\emptyset$ &  $\T^2$             & $\emptyset$ & $\emptyset$ & $\mc{C}_3$ \\
 $D_2$ & $\T^2$      & $\mP^1\sqcup \mP^1$ & $\mP^1\sqcup \mP^1$ & $\T^2$      & $\mP^1\sqcup \mP^1$ & $\T^2$      & $\emptyset$ & $\mc{C}_3$ \\
 $D_3$ & $\T^2$      & $\mP^1\sqcup \mP^1$ & $\mP^1\sqcup \mP^1$ & $\T^2$      & $\mP^1\sqcup \mP^1$ & $\emptyset$ & $\T^2$      & $\mc{C}_3$ \\
 $D_4$ & $\emptyset$ & $\T^2$              & $\T^2$              & $\emptyset$ & $\emptyset$         & $\T^2$      & $\T^2$      & $\mc{C}_9$ \\
 $D_5$ & $\T^2$      & $\mP^1\sqcup \mP^1$ & $\mP^1\sqcup \mP^1$ & $\emptyset$ & $\mP^1\sqcup \mP^1$ & $\T^2$      & $\T^2$      & $\mc{C}_3$ \\
 $D_6$ & $\emptyset$ & $\T^2$              & $\emptyset$         & $\T^2$      & $\T^2$              & $\emptyset$ & $\T^2$      & $\mc{C}_9$ \\
 $D_7$ & $\emptyset$ & $\emptyset$         & $\T^2$              & $\T^2$      & $\T^2$              & $\T^2$      & $\emptyset$ & $\mc{C}_9$ \\
 $D_8$ & $\mc{C}_3$  & $\mc{C}_3$          & $\mc{C}_3$          & $\mc{C}_9$  & $\mc{C}_3$          & $\mc{C}_9$  & $\mc{C}_9$  & $\mc{C}_{81}$ \\
    \hline
  \end{tabular}
  \caption{Intersection curves of two coordinate divisors. Here $\mc{C}_g$ denotes a curve with Hodge numbers $h^{0,0} = 1$ and $h^{1,0} = g$.}
	\label{Tab2}
 \end{table}

\begin{table}[htb]
\centering
\resizebox{\textwidth}{!}{\begin{tabular}{|c||c|c|c|c|c|c|c|c|}
\hline 
{} & $D_1$  & $D_2$ & $D_3$ & $D_4$ & $D_5$ & $D_6$ & $D_7$ & $D_8$  \\
    \hline
 $D_1$ & $2\,t_1$  & $-2\,t_1$          & $-2\, t_1$       & $0$           & $-2\,t_1$        & $0$          & $0$           & $-4\,t_1$  \\
 $D_2$ & $-2\,t_1$ & $2\,t_1$           & $2(t_1+ t_4)$    & $2\,t_6$      & $2(t_1+t_6)$     & $2\,t_4$     & $0$           & $4(t_1+t_4+t_6)$  \\
 $D_3$ & $-2\,t_1$ & $2(t_1+ t_4)$      & $2\, t_1$        & $2\,t_7$      & $2(t_1+t_7)$     & $0$          & $2\,t_4$      & $4(t_1+t_4+t_7)$ \\
 $D_4$ & $0$       & $2\,t_6$           & $2\,t_7$         & $0$           & $0$              & $2\,t_7$     & $2\,t_6$      & $4(t_6+t_7)$ \\
 $D_5$ & $-2\,t_1$ & $2(t_1+ t_6)$      & $4(t_1+ t_7)$    & $0$           & $2\,t_1$         & $2\,t_7$     & $2\,t_6$      & $4(t_1+t_6+ t_7)$ \\
 $D_6$ & $0$       & $2\,t_4$           & $0$              & $2\,t_7 $     & $2\,t_7$         & $0$          & $2\,t_4$      & $4(t_4+ t_7)$  \\
 $D_7$ & $0$       & $0$                & $2\,t_4$         & $2\,t_6$      & $2\,t_6$         & $2\,t_4$     & $0$           & $4(t_4+ t_6)$ \\
 $D_8$ & $-4\,t_1$ & $4(t_1+ t_4+ t_6)$ & $4(t_1+t_4+t_7)$ & $4(t_6+ t_7)$ & $4(t_1+t_6+t_7)$ & $4(t_4+t_7)$ & $4(t_4+ t_6)$ & $8(t_1+2(t_4+ t_6+t_7))$  \\
    \hline
\end{tabular}}
\caption{Volumes of intersection curves between two coordinate divisors.}
\label{Tab3}
\end{table}

\subsection{Orientifold involution}

We focus on orientifold involutions of the form $\sigma: x_i \to - x_i$ with $i = 1, ...,8$  which feature an O7-plane on $D_i$ and O3-planes at the fixed points listed in Tab. \ref{FixedPointsA}. The effective non-trivial fixed point set in Tab. \ref{FixedPointsA} has been obtained after taking care of
the SR ideal symmetry. Moreover, the total number of O3-planes $N_{\rm O3}$ is obtained from the triple intersections restricted to the CY hypersurface, while the effective Euler number $\chi_{\rm eff}$ has been computed as:\footnote{The effective Euler number controls the strength of $N=1$ $\mc{O}(\alpha'^3)$ corrections due to O7-planes \cite{Minasian:2015bxa}.} 
\be
\chi_{\rm eff} = \chi(X) +2\int_X [{\rm O7}] \wedge [{\rm O7}] \wedge [{\rm O7}]\,.
\label{chieff}
\ee
In what follows we shall focus on the orientifold involution $\sigma: x_8\rightarrow-x_8$ which features just a single O7-plane located in $D_8$ and no O3-plane\,.
\begin{table}[H]
  \centering
 \begin{tabular}{|c|c|c|c|c|c|}
\hline
&  &  &  &  &  \\
$\sigma$ & O7  & O3  & $N_{{\rm O3}}$  & $\chi({\rm O7})$  & $\chi_{\rm eff}$       \\
&  &  &  &  &  \\
\hline
\hline
$x_1 \to -x_1$ &  $D_1$ & $\{D_2 D_3 D_4, D_2 D_4 D_6, D_2 D_5 D_6, $ & 14 & 10 & -184  \\
&  &  $D_3 D_4 D_7, D_3 D_5 D_7,$   &  &  &  \\
&  &  $D_4 D_6 D_7, D_5 D_6 D_7 \}$ &  &  &  \\
$x_2 \to -x_2$ &  $D_2\sqcup D_7$ & $D_1 D_3 D_5$ & 2 & 38 & -192  \\
$x_2 \to -x_3$ &  $D_3\sqcup D_6$ & $D_1 D_2 D_5$ & 2 & 38 & -192  \\
$x_4 \to -x_4$ &  $D_4\sqcup D_5$ & $D_1 D_2 D_3$ & 2 & 38 & -192  \\
$x_5 \to -x_5$ &  $D_4\sqcup D_5$ & $D_1 D_2 D_3$ & 2 & 38 & -192  \\
$x_6 \to -x_6$ &  $D_3\sqcup D_6$ & $D_1 D_2 D_5$ & 2 & 38 & -192  \\
$x_7 \to -x_7$ &  $D_2\sqcup D_7$ & $D_1 D_3 D_5$ & 2 & 38 & -192  \\
$x_8 \to -x_8$ &  $D_8$ & $\emptyset$ & 0 & 208 & -28  \\
\hline
\end{tabular}
\caption{Fixed point set for the involutions which are reflections of the eight coordinates $x_i$ with $i=1,...,8$.} 
\label{FixedPointsA} 
\end{table}

\subsection{Brane setup}

If the D7-tadpole cancellation condition is satisfied by placing four D7-branes on top of the O7-plane, the string loop corrections to the scalar potential can involve only KK effects between this D7-stack and O3-planes or D3-branes since winding contributions are absent due to the absence of any intersection between D7-branes and/or O7-planes. Thus loop effects are too simple to generate a viable inflationary plateau. They might even be completely absent in our case since there are no O3-planes and the D3-tadpole cancellation condition could be satisfied without the need to include D3-branes (i.e. just switching on appropriate background three-form fluxes). We shall therefore focus on a slightly more complicate D7-brane setup which gives rise to winding loop effects. This can be achieved by placing D7-branes not entirely on top of the O7-plane as follows:  
\be
8[{\rm O7}] \equiv 8([D_8]) = 16 \left([D_2]+[D_4]+[D_6] \right)\,.
\ee
This brane setup involves three stacks of D7-branes wrapped around the divisors $D_2$, $D_4$ and $D_6$. Moreover, the condition for D3-tadpole cancellation becomes:
\be
N_{\rm D3} + \frac{N_{\rm flux}}{2} + N_{\rm gauge} = \frac{N_{\rm O3}}{4} + \frac{\chi({\rm O7})}{12} + \sum_a\, \frac{N_a \left(\chi(D_a) + \chi(D_a^\prime) \right) }{48}  = 38 \,, \nn
\ee
showing that there is space for turning on both gauge and background three-form fluxes for complex structure and dilaton stabilisation.\footnote{We focus on flux vacua where the dilaton is fixed in a regime where our perturbative type IIB analysis is under control.} As shown in \cite{Camara:2004jj}, three-form fluxes stabilise also D7 position moduli and open string moduli living at the intersection between two different stacks of D7-branes since they generate soft supersymmetry breaking mass terms for each of these scalars. On the other hand, there are no Wilson line moduli in our model since $h^{1,0}(D_2)=h^{1,0}(D_4) = h^{1,0}(D_6)=0$. 

Let us point out that other orientifold involutions which could allow for D7-branes not entirely on top of the O7-plane are $x_4 \to - x_4$, $x_6 \to - x_6$ or $x_7 \to - x_7$. In each of these cases, the O7-plane is located on a K3 surface. However, given that $D_4 = D_1 + D_5$, $D_6 = D_1 + D_3$ and $D_7 = D_1 +D_2$, from Tab. \ref{Tab2} and \ref{Tab3} we see that the resulting D7-brane stacks are either non-intersecting (and so no winding corrections are generated) or the volumes of the intersection curves depend just on the `small' dP$_7$ divisor (and so winding loops are inflaton-independent). This is the reason why we chose the involution $x_8 \to - x_8$ where the O7-plane is located on the `special deformation' divisor $D_8$ which gives more freedom for D7-brane model building. 

\subsection{Gauge fluxes}

In order to obtain a chiral visible sector on the D7-brane stacks wrapping $D_2$, $D_4$ and $D_6$ we need to turn on worldvolume gauge fluxes of the form:
\be
\F_i = \sum_{j=1}^{h^{1,1}} f_{ij}\hat{D}_j - \frac12 \,c_1({D}_i) - \iota_{D_i}^*B \quad\text{with}\quad f_{ij}\in \mathbb{Z} \quad\text{and}\quad i=2,4,6\,,
\ee
where the half-integer contribution is due to Freed-Witten anomaly cancellation \cite{Minasian:1997mm, Freed:1999vc}. 

However we want to generate just one moduli-dependent Fayet-Iliopoulos term in order to fix only one K\"ahler modulus via D-term stabilisation. In fact, if the number of FI-terms is larger than one, there is no light K\"ahler modulus which can play the r\^ole of the inflaton. Moreover we wrap a D3-brane instanton on the rigid divisor $D_1$ in order to generate a non-perturbative contribution to the superpotential which is crucial for LVS moduli stabilisation. In order to cancel the Freed-Witten anomaly, the D3-instanton has to support a half-integer flux, and so the general expression of the total gauge flux on $D_1$ becomes (with $c_1(D_1) =-\hat{D}_1$):
\be
\F_1 = \sum_{j=1}^{h^{1,1}} f_{1j}\hat{D}_j + \frac12 \hat{D}_1 - \iota_{D_i}^*B \quad\text{with}\quad f_{1j}\in \mathbb{Z}\,.
\ee
However a non-vanishing $\F_1$ would not be gauge invariant, and so would prevent a non-perturbative contribution to the superpotential. We need therefore to check if it is possible to perform an appropriate choice of $B$-field which can simultaneously set $\F_4=\F_6=0$ (we choose to have a non-vanishing gauge flux only on $D_2$ to have just one moduli-dependent FI-term) and $\F_1=0$. Recalling that both $D_4$ and $D_6$ are K3 surfaces which are spin divisors with $c_1(D_4)=c_1(D_6)=0$ (since the K3 is a CY two-fold), if we set:
\be
B = \frac12 \,\hat{D}_1\,,
\label{BfieldA}
\ee
the condition $\F_1=\F_4=\F_6=0$ reduces to the requirement that the following forms are integer:
\be
\iota_{D_4}^* \left(\frac12 \hat{D}_1 \right) \qquad\text{and}\qquad
\iota_{D_6}^* \left(\frac12 \hat{D}_1 \right)\,,
\label{PullbacksA}
\ee
since in this case the integer flux quanta $f_{ij}$ can always be adjusted to yield vanishing gauge fluxes. 
Taking an arbitrary integer form $A\in H^2(\mathbb{Z},X)$ which can be expanded as $A=a_j\hat{D}_j$ with $a_j \in \mathbb{Z}$, the pullbacks in (\ref{PullbacksA}) give rise to integer forms if:
\bea
b_4 &\equiv& \int_X \left(\frac12 \hat{D}_1 \right) \wedge \hat{D}_4 \wedge A \in \mathbb{Z} \nn \\
b_6 &\equiv& \int_X \left(\frac12 \hat{D}_1 \right) \wedge \hat{D}_6 \wedge A \in \mathbb{Z} \nn
\eea
Using the intersection polynomial (\ref{I3A}) we find $b_4 = b_6 = 0$, showing how the choice of $B$-field in (\ref{BfieldA}) can indeed allow for $\F_1=\F_4=\F_6=0$. The only non-zero gauge flux is $\F_2$ whose half-integer contribution can be cancelled by adding an additional term to the $B$-field of the form $\frac12 \hat{D}_2$. Given that all the intersection numbers are even, this new term in $B$ does not modify our previous results on the pullbacks of the $B$-field on $D_1$, $D_4$ and $D_6$. Moreover the pullback of the $B$-field on $D_2$ will also generate an integer flux contribution. We shall therefore consider a non-vanishing gauge flux on the worldvolume of $D_2$ of the form:
\be
\F_2 = \sum_{j=1}^{h^{1,1}} f_{2j}\hat{D}_j \quad\text{with}\quad f_{2j}\in \mathbb{Z}\,.
\ee

\subsection{FI-term and chirality}

Given that the divisor $D_2$ is transversely invariant under the orientifold involution and it is wrapped by eight D7-branes, it supports an $Sp(16)$ gauge group which is broken down to $U(8)=SU(8)\times U(1)$ by a non-zero flux $\F_2$ along the diagonal $U(1)$. This non-trivial gauge flux $\F_2$ induces also a $U(1)$-charge $q_{i2}$ for the $i$-th K\"ahler modulus of the form:
\be
q_{i2} = \int_X \hat{D}_i \wedge \hat{D}_2 \wedge \F_2 \,.
\ee
Thus $\F_2\neq 0$ yields (using $D_2=D_7-D_1$):
\be
q_{12} =  - 2  f_{21} \qquad q_{42} =  2  f_{26} \qquad q_{62} =  2  f_{24} \qquad q_{72} = 0\,,
\ee
together with a flux-dependent correction to the gauge kinetic function which looks like:
\be
{\rm Re}(f_2)  = \alpha_2^{-1}=\frac{4\pi}{g_2^2}=\tau_2-h(\F_2) {\rm Re}(S)\,,
\ee
where: 
\be
h(\F_2) =\frac12 \int_X \hat{D}_2 \wedge \F_2 \wedge \F_2 =\frac12\left(f_{21} q_{12} + f_{24} q_{42} + f_{26} q_{62}\right)\,.
\ee
Moreover a non-vanishing gauge flux $\F_2$ induces a moduli-dependent FI-term of the form:
\be
\xi =\frac{1}{4\pi\vo}\int_X \hat{D}_2\wedge J\wedge\F_2=\frac{1}{4\pi\vo}\sum_{j=1}^{h^{1,1}} q_{j2}\,t_j
=\frac{1}{4\pi\vo} \left(q_{12}\,t_1+q_{42}\,t_4+q_{62}\,t_6\right)\,.
\ee
For vanishing open string VEVs (induced for example by non-tachyonic scalar masses), a leading-order supersymmetric stabilisation requires $\xi=0$ which implies:
\be
t_4 = - \frac{q_{12}}{q_{42}}\,t_1- \frac{q_{62}}{q_{42}}\,t_6\,.
\label{DfixA}
\ee
This $U(1)$ factor becomes massive via the St\"uckelberg mechanism and develops an $\mc{O}(M_s)$ mass by eating up a linear combination of an open and a closed string axion which is mostly given by the open string mode. 

Besides breaking the worldvolume gauge group and inducing moduli-dependent FI-terms, non-trivial gauge fluxes on D7-branes generate also 4D chiral modes. In fact, open strings stretching between the D7-branes on $D_2$ and the O7-planes or the image branes give rise to the following zero-modes in the symmetric and antisymmetric representations of $U(8)$:
\bea
I_2^{(S)} &=& - \frac12 \int_X \hat{D}_2 \wedge [{\rm O7}] \wedge \F_2 - \int_X \hat{D}_2 \wedge \hat{D}_2 \wedge \F_2 
= 2 q_{12} - q_{42} - q_{62}\,, \\
I_2^{(A)} &=& \frac12 \int_X \hat{D}_2 \wedge [{\rm O7}] \wedge \F_2 - \int_X \hat{D}_2 \wedge \hat{D}_2 \wedge \F_2 = q_{42} + q_{62}\,.
\eea
Due to the absence of worldvolume fluxes on the D7-branes wrapped around $D_4$ and $D_6$, both of these two D7-stacks support an $Sp(16)$ gauge group (since both $D_4$ and $D_6$ are transversely invariant) which are both unbroken. Thus open strings stretched between the D7-branes on $D_2$ and $D_4$ or $D_6$ (or their image branes) give rise to 4D chiral zero-modes in the bi-fundamental representation ($8$,$16$) of $U(8)$ and $Sp(16)$ whose number is:
\be
I_{24} = \int_X \hat{D}_2 \wedge \hat{D}_4 \wedge \F_2 = q_{42}\,, \qquad
I_{26} = \int_X \hat{D}_2 \wedge \hat{D}_6 \wedge \F_2 = q_{62}\,.
\ee
We need finally to check that there are no chiral intersections between the D7s on $D_2$ and the instanton on $D_1$ to make sure that the prefactor of the non-perturbative contribution to the superpotential is indeed non-zero. This is ensured if:
\be
I_{21}=\int_X \hat{D}_2 \wedge \hat{D}_1 \wedge \F_2 = q_{12} =  - 2  f_{21} = 0 \,.
\ee
This condition can be easily satisfied by choosing $f_{21}=0$. In turn, this choice simplifies the D-term constraint (\ref{DfixA}) to:
\be
t_4 = - \frac{q_{62}}{q_{42}}\,t_6 \equiv \alpha \,t_6 \,.
\label{DfixAA}
\ee

\subsection{Inflationary potential}

Using the D-term fixing relation (\ref{DfixAA}), the K\"ahler cone conditions (\ref{KCCA}) simplify to $t_7>-t_1>0$ together with $t_6>-t_1>0$ if $\alpha\geq 1$ or $\alpha t_6>-t_1>0$ if $\alpha\leq 1$. Moreover the CY volume (\ref{VolA}) reduces to:
\be
\vo = 2\alpha t_7 t_6^2 + \frac{t_1^3}{3} = t_7 \tau_7 -\frac13 \,\tau_1^{3/2} = \frac{1}{\sqrt{2\alpha}}\,\sqrt{\tau_7}\,\tau_6-\frac13\,\tau_1^{3/2}\,.
\ee
Given that this form is linear in $t_7$, the effective CY volume after D-term stabilisation looks like a single K3 fibre $\tau_7$ over a $\mP^1$ base $t_7$ and reduces to the typical form used in fibre inflation models. The blow-up mode $\tau_1$ and the overall volume $\vo$ are stabilised in the LVS fashion by means of a non-perturbative correction to $W$ generated by an Euclidean D3-brane instanton wrapping $D_1$. This leaves the fibre modulus $\tau_7$ as a flat direction which receives a potential at subleading order.

Let us now focus on the inflationary potential. The winding loop corrections can be written as (with $\kappa=g_s/(8\pi)$ for $e^{K_{\rm cs}}=1$):
\be
V^\W_{g_s}=-2\,\kappa\,\frac{W_0^2}{\vo^3}\,\sum_i\frac{C_i^\W}{t_i^\cap}\,,
\ee 
where $t_i^\cap$ are the volumes of the two-cycles where D7-branes/O7-planes intersect. Notice that if two coordinate divisors $D_i$ and $D_j$ are wrapped by D7-branes and/or O7-planes, the scalar potential receives $t^\cap$-dependent winding loop corrections only if their intersection curve contains
non-contractible 1-cycles, i.e. if $h^{1,0}(D_i \cap D_j)\neq 0$. In our case, we have an O7-plane located on $D_8$ and three stacks of D7-branes wrapping $D_2$, $D_4$ and $D_6$. Using Tab. \ref{Tab2} and \ref{Tab3}, we see all D7s intersect with each other and with the O7 and that winding corrections can arise from any of these intersections.  Thus we end up with: 
\be
V^\W_{g_s} =-\,\kappa\,\frac{W_0^2}{\vo^3}\,
\left[\frac{1}{\sqrt{\tau_7}} \left( C_\W - \tilde{C}_\W(\tau_7)\right) -\frac{\tau_7}{\vo} \left(|C_3^\W| - \hat{C}_\W(\tau_7)\right)\right],
\ee
where (setting $t_4=\alpha t_6$, $C_3^\W = - |C_3^\W|<0$ and $C_4^\W = - |C_4^\W|<0$):
\be
C_\W = \sqrt{2\alpha}\left(C_1^\W+\frac{C_2^\W}{\alpha}\right) \qquad 
\tilde{C}_\W (\tau_7)= \frac{|C_4^\W|}{(\alpha+1)}\sqrt{\frac{\alpha}{2}}\left(1-\frac{\sqrt{2\,\alpha}}{(\alpha+1)}\,\sqrt{\frac{\langle\tau_1\rangle}{\tau_7}}\right)^{-1}\,,
\ee
and:
\be
\hat{C}_\W (\tau_7) = \frac{C_5^\W}{2} \left(1+\frac{1}{\sqrt{2\,\alpha}}\frac{\tau_7^{3/2}}{\vo}\right)^{-1}
+ \frac{C_6^\W}{2} \left(1+\sqrt{\frac{\alpha}{2}}\,\frac{\tau_7^{3/2}}{\vo}\right)^{-1}\,.
\ee
Due to the absence of O3-planes (we also assume that the D3-tadpoles are cancelled without including any spacetime-filling D3-branes) and the fact that all D7s intersect with each other and with the O7-plane, there are no 1-loop corrections due to the exchange of closed strings carrying KK momentum.\footnote{Strictly speaking, there might be 1-loop corrections associated with the exchange of KK modes between the Euclidean D3-instanton on $D_1$ and the D7-branes which do not intersect $D_1$. However, we expect such corrections to be exponentially suppressed and, thus, not relevant for the analysis.}

On the other hand, higher derivative $\alpha'^3\,F^4$ corrections to the scalar potential can be written as \cite{Ciupke:2015msa}:\footnote{This expression displays merely the leading order $\mc{O}(\vo^{-4})$ terms which are corrected at subleading order in inverse volume by additional corrections as discussed in \cite{Cicoli:2016chb}. Furthermore, additional higher-derivative corrections mediated by the auxiliary fields sitting in the supergravity multiplet might emerge at order $\mc{O}(\vo^{-5})$ \cite{Cicoli:2016chb, Ciupke:2016agp}.}
\be
V_{F^4}=-\kappa^2\,\frac{\lambda\,W_0^4}{g_s^{3/2}\,\vo^4}\,\sum_{i=1}^{h^{1,1}}\Pi_i\,t_i\,,
\ee
where $\lambda$ is an unknown combinatorial factor which is expected to be of order $10^{-3}$ \cite{Ciupke:2015msa, Grimm:2017okk} and the topological quantities $\Pi_i$ are given in (\ref{Pi'sA}). After imposing the D-term condition (\ref{DfixAA}), the $F^4$ contributions can be rewritten as (ignoring the $t_1$-dependent term):
\be
V_{F^4} = -24\kappa^2\,\frac{\lambda\,W_0^4}{g_s^{3/2}\,\vo^3}\,\left[\frac{(\alpha+1)}{\sqrt{2\,\alpha}}\,\frac{\sqrt{\tau_7}}{\vo}+\frac{1}{\tau_7}\right].
\ee
Therefore the total inflationary potential becomes:
\be
V = V_{g_s}^\W+V_{F^4}= \kappa\,\frac{W_0^2}{\vo^3}\,
\left(\frac{A_1}{\tau_7}-\frac{A_2}{\sqrt{\tau_7}}  + \frac{B_1\sqrt{\tau_7}}{\vo}+\frac{B_2 \,\tau_7}{\vo}\right)\,,
\label{exA_tot_pot}
\ee
where (with $\lambda=-|\lambda|<0$):
\be
A_1 = \frac{3}{\pi}  \frac{|\lambda| W_0^2}{\sqrt{g_s}} \qquad A_2 = C_\W - \tilde{C}_\W(\tau_7)
\qquad B_1 = \frac{(\alpha+1)}{\sqrt{2\,\alpha}}\,A_1
\qquad B_2 = |C_3^\W| - \hat{C}_\W(\tau_7)\,. \nn
\ee

\section{Inflationary dynamics}
\label{InfAn}

In this section we shall analyse the inflationary dynamics by studying first the single-field approximation and then by focusing on the full multi-field evolution.

\subsection{Single-field evolution}
\label{exA_inf}

In order to realise single-field slow-roll inflation where the potential for the inflaton $\tau_7$ features a plateau-type region \cite{Cicoli:2008gp, Cicoli:2016chb}, the overall volume has to be approximately constant during the whole inflationary dynamics. Therefore, in order to get enough efoldings before reaching the dangerous limit where the base of the fibration $t_7$ becomes smaller than the string scale, we need to focus on the region in field space where the inflaton minimum is of order $\langle\tau_7\rangle\ll \vo^{2/3}$. For $g_s\lesssim\mc{O}(0.1)$, $|\lambda|\sim\mc{O}(10^{-3})$ and natural $\mc{O}(1)$ values of the coefficients of the string loop effects, in the vicinity of the minimum the terms in (\ref{exA_tot_pot}) proportional to $B_1$ and $B_2$ are therefore both negligible with respect to the terms proportional to $A_1$ and $A_2$. Numerical estimates show that we need values of order $\langle\tau_7\rangle \sim\mc{O}(1)$ and $\vo \sim\mc{O}(10^4)$ which, in turn, imply $W_0\sim\mc{O}(100)$ in order to match the observed amplitude of the density perturbations. 

The scalar potential (\ref{exA_tot_pot}) written in terms of the canonically normalised inflaton shifted from its minimum $\phi = \langle \phi \rangle + \hat\phi$, where $\tau_7 = \langle \tau_7 \rangle \,e^{k\hat\phi}$ with $k=2/\sqrt{3}$, becomes:
\be
V = \kappa\,\frac{A_2 W_0^2}{\vo^3 \sqrt{\langle\tau_7\rangle}}\, \left(C_\dS + c\,e^{-k\hat\phi}- e^{-\frac{k\hat\phi}{2}} 
 + \mc{R}_1\,e^{\frac{k\hat\phi}{2}} +\mc{R}_2\,e^{k\hat\phi}\right),
\label{VtotA}
\ee
where:
\be
c = \frac{3}{\pi\left(C_\W - \tilde{C}_\W(\tau_7)\right)}
\frac{|\lambda| W_0^2}{\sqrt{g_s \langle\tau_7\rangle}}\sim \mc{O}(1)\,, \nn
\ee
while for $\langle\tau_7\rangle\sim \mc{O}(1)\ll\vo^{2/3}$:
\be
\mc{R}_1 = \frac{(\alpha+1) c}{\sqrt{2\,\alpha}} \frac{\langle\tau_7\rangle^{3/2}}{\vo}\ll 1
\qquad \text{and}\qquad \mc{R}_2 = \frac{\left(|C_3^\W| - \hat{C}_\W(\tau_7)\right)}{\left(C_\W - \tilde{C}_\W(\tau_7)\right)} 
\frac{\langle\tau_7\rangle^{3/2}}{\vo}\ll 1 \,. \nn
\ee
Notice that in (\ref{VtotA}) we added a constant $C_\dS= 1-c - \mc{R}_1 - \mc{R}_2$ to obtain a Minkowski (or slightly dS) vacuum. Given that no O3-planes are present in our model, the usual uplift mechanism where an anti D3-brane is located in a resolved conifold region of the extra dimensions would require additional effort to implement. We leave the explicit embedding of the source of uplift to future research. 

The two negative exponentials in (\ref{VtotA}) compete to give a minimum at $\langle\tau_7\rangle \sim\mc{O}(1)$ while the two positive exponentials cause a steepening behaviour at large $\hat\phi$. Thus we need to make sure that both $\mc{R}_1\ll 1$ and $\mc{R}_2\ll 1$ to prevent the two positive exponentials from destroying the inflationary plateau before achieving enough efoldings of inflation.\footnote{If this is the case, these steepening terms could then be responsible for an interesting power loss at large angular scales \cite{Cicoli:2013oba}.} The condition $\mc{R}_1\ll 1$ could be satisfied for $c\ll 1$, for example for $W_0\sim\mc{O}(1)$ and $\langle\tau_7\rangle \gg 1$, in which case the minimum could be obtained by balancing the two terms in the coefficient $A_2$. However, as we shall see below, if $\langle\tau_7\rangle \gg 1$, the K\"ahler cone bounds restrict the allowed field space so much that it becomes impossible to realise enough efoldings of inflation. Hence we shall focus the region where $\mc{R}_1\ll 1$ and $\mc{R}_2\ll 1$ are satisfied by $\langle\tau_7\rangle\sim \mc{O}(1)\ll\vo^{2/3}$ (and possibly by allowing some tuning of the complex structure moduli-dependent coefficients of the loop corrections or by considering $|\lambda|\ll 1$).  

Turning now to the explicit numerical examples, let us formulate the necessary conditions that have to be satisfied in order to have a viable model:
\ben
\item Stringy effects can be neglected if each four-cycle in string frame has a volume larger than the string scale: ${\rm Vol}_s^{1/4}\gg \sqrt{\alpha'}$. Given that string and Einstein frame volumes are related as ${\rm Vol}_s = g_s {\rm Vol}_E = g_s \tau_E \ell_s$ with $\ell_s = 2\pi\sqrt{\alpha'}$, we end up with the condition:
\be
\epsilon_{\tau_i} \equiv \frac{1}{g_s (2\pi)^4 \, \tau_i} \ll 1\qquad\forall\,i\,.
\label{epsAppr}
\ee
\item The whole inflationary dynamics should take place inside the K\"ahler cone. This implies in particular that:
\bea
2\alpha \langle\tau_1 \rangle &<& \tau_7 < \frac{\vo}{\sqrt{\langle\tau_1 \rangle}}\qquad\text{if}\qquad \alpha\geq 1\,, \nn \\
\frac{2}{\alpha} \langle\tau_1 \rangle &<& \tau_7 < \frac{\vo}{\sqrt{\langle\tau_1 \rangle}}\qquad\text{if}\qquad  \alpha\leq 1\,. 
\label{KCA}
\eea
Notice that these conditions guarantee the absence of any singularity in the inflationary potential (\ref{VtotA}) which could originate from the shrinking of a two-cycle to zero size. Rewriting these conditions in terms of the canonically normalised inflaton field, we end up with:
\bea
\frac{\sqrt{3}}{2} \ln\left(\frac{2\alpha\langle\tau_1 \rangle}{\langle\tau_7\rangle}\right) &<& \hat{\phi} 
< \frac{\sqrt{3}}{2}\ln\left(\frac{\vo}{\langle\tau_7\rangle\sqrt{\langle\tau_1 \rangle}}\right) \qquad\text{if}\qquad \alpha\geq 1\,, \nn \\
\frac{\sqrt{3}}{2} \ln\left(\frac{2\langle\tau_1 \rangle}{\alpha\langle\tau_7\rangle}\right) &<& \hat{\phi} 
< \frac{\sqrt{3}}{2} \ln\left(\frac{\vo}{\langle\tau_7\rangle\sqrt{\langle\tau_1 \rangle}}\right)\qquad\text{if}\qquad  \alpha\leq 1\,. 
\label{KConeA}
\eea
In order to be able to describe within a consistent EFT, not just inflation but also the post-inflationary evolution of our model, $\hat\phi$ should reach its minimum before hitting the lower bounds in (\ref{KConeA}). Moreover the inflaton should drive enough efoldings of inflation before hitting the upper bounds in (\ref{KConeA}).

\item Horizon exit at $\hat\phi=\hat\phi_*$ should yield the required number of efoldings:
\be
N_e\simeq 57 + \frac14 \ln \left(r_*\, V_*\right)- \frac13 \ln\left(\frac{V_{\rm end}}{T_{\rm rh}}\right),
\label{Nenumber}
\ee
where the reheating temperature $T_{\rm rh}$ can be estimated in terms of the inflaton mass at the minimum $m_{\hat\phi}$ as:
\be
T_{\rm rh} \simeq \left(\frac{90}{\pi^2  g_*(T_{\rm rh})}\right)^{1/4} \sqrt{\Gamma_{\hat\phi} M_p}
\simeq 0.1 \,m_{\hat\phi}\, \sqrt{\frac{m_{\hat\phi}}{M_p}}\,.
\label{Trh}
\ee

\item Horizon exit at $\hat\phi=\hat\phi_*$ should reproduce the observed amplitude of the density perturbations:
\be
\frac{V_*^3}{V_*^{'2}} \simeq 2.6 \cdot 10^{-7}\,.
\label{COBE}
\ee

\item The $\alpha'$ expansion of the potential can be trusted only if:
\be
\epsilon_\xi = \frac{\xi}{2 g_s^{3/2} \vo} \ll 1\,.
\ee

\item The effective field theory is under control if throughout all the inflationary dynamics:
\be
m_{\rm inf} < H < m_{3/2} < M_\KK^{(i)} < M_s < M_p \qquad\forall i = {\rm bulk},2,4,6\,,
\label{Hierarchies}
\ee
where $m_{\rm inf}$ is the inflaton mass, $H\simeq \frac{V}{3 M_p^2}$ is the Hubble scale, $m_{3/2} = e^{K/2} W_0 = \sqrt{\kappa} \,\frac{W_0}{\vo}\,M_p$ is the gravitino mass which sets the mass scale of all complex structure moduli, the dilaton and the K\"ahler modulus $T_1 = \tau_1 + {\rm i}\int_{D_1} C_4$ and $M_\KK^{(i)} = \frac{\sqrt{\pi}}{\sqrt{\vo}\,\tau_i^{1/4}}\,M_p$ are the different KK scales in the model associated with bulk KK modes for $\tau_\b^{3/2} = \vo$ and KK replicas of open string modes living on D7-branes wrapped around $D_2$, $D_4$ and $D_6$. The bulk KK scale should be below the string scale $M_s = \frac{g_s^{1/4} \sqrt{\pi}}{\sqrt{\vo}}\,M_p$ while we do not need to impose $V^{1/4} < M_\KK^{(i)}$ since no energy can be extracted from the vacuum during an adiabatic inflationary expansion where $H\ll M_\KK^{(i)}$. 

\item Besides the two ultra-light axions associated with the base and the fibre which develop just negligible isocurvature fluctuations during inflation if they do not contribute significantly to dark matter, only the volume mode has a mass below $m_{3/2}$. In order to trust our single field approximation, we need therefore to check that the mass of the volume mode $m_\vo$ does not become smaller than the Hubble scale $H$. This condition boils down to:
\be
\delta = \frac{H}{m_{\vo}} \simeq \sqrt{\frac{V_*}{3 V_{\alpha'}}} \lesssim 1\,,
\label{singlefield}
\ee
where $V_{\alpha'}$ is the leading $\mc{O}(\alpha'^3)$ contribution to the scalar potential and reads \cite{Becker:2002nn}:
\be
V_{\alpha'} = \kappa \frac{3 \xi W_0^2}{4 g_s^{3/2} \vo^3}\qquad\text{with}\qquad \xi=-\frac{\zeta(3)\chi(X)}{2 (2\pi)^3} \,.
\label{a3F2}
\ee
If $\delta\simeq 1$, the inflationary energy density can either destabilise the volume direction or cause a significant shift of the volume minimum. Hence the inflationary dynamics can effectively become a multi-field evolution. However, as analysed in \cite{Cicoli:2008gp}, the motion might still remain mainly along the $\tau_7$ direction, and so the predictions for the inflationary observables could be basically unaltered apart from the fact that the number of allowed efoldings slightly increases. Notice also that in LVS models the CY Euler number together with the string coupling fixes the minimum of the blow-up mode $\tau_1$ as: $\langle\tau_1\rangle = (3\xi/2)^{2/3}\,g_s^{-1}$. This value is important to evaluate the K\"ahler cone conditions in (\ref{KConeA}).
\een
We shall now focus on single-field slow-roll inflation where:
\be
\epsilon (\hat\phi) = \frac12 \left(\frac{V'}{V}\right)^2 \ll 1\qquad\text{and}\qquad\eta (\hat\phi) = \frac{V''}{V}\ll 1\,. \nn
\label{eq:slow_roll}
\ee
Notice that the condition $\eta\ll 1$ guarantees that the inflaton is lighter than $H$ during inflation. In order to illustrate the main features of our inflationary model, we shall now consider two different choices of the underlying parameters characterised by different values of the coefficients $\xi$ and $\lambda$ which control the strength of the $\mc{O}(\alpha'^3)$ corrections to the effective action at $\mc{O}(F^2)$ and $\mc{O}(F^4)$. According to \cite{Minasian:2015bxa}, $N=1$ $\mc{O}(\alpha'^3)$ corrections due to O7-planes cause a shift of the CY Euler number $\chi(X)$ to $\chi_{\rm eff}(X)$ defined in (\ref{chieff}) and given in Tab. \ref{FixedPointsA}. From (\ref{a3F2}) this modification would give $\xi = 0.067$. Moreover the coefficient $\lambda$ of higher derivative $\mc{O}(\alpha'^3)$ effects has been estimated to be negative and of order $10^{-3}$ \cite{Ciupke:2015msa, Grimm:2017okk}. Hence the first set of parameters will be characterised by $\xi=0.067$ and $\lambda=-0.001$. However both of these corrections still lack a full supersymmetric analysis, and so in the second case we shall focus on a situation where the CY Euler number is not modified, and so $\xi = 0.456$, and the size of the coefficient $\lambda$ is much smaller: $|\lambda|\lesssim 10^{-6}$.

\subsubsection{Case 1: $\xi = 0.067$ and $|\lambda| = 0.001$}
\label{BadCase}

Let us now provide an explicit numerical example set to demonstrate the features of our inflationary model:
\bea
\alpha &=& 1\,, \quad C^\W_1=C^\W_2=15\,, \quad |C^\W_3|=0.013\,, \quad |C^\W_4| =18\,,\quad C^\W_5 = C^\W_6 =-5\,,  \nn \\ 
g_s &=& 0.114\,, \quad \vo=10^4 \,, \quad \langle\tau_1 \rangle = 1.91\,,\quad W_0=80\,, \quad |\lambda|=0.001\,,
\label{ex_set_exA}
\eea
with $\chi(X)=\chi_{\rm eff}(X)= -28$ in (\ref{a3F2}) which gives $\xi = 0.067$. Notice that the tuning of the steepening term here is mild since the difference between the largest and the smallest winding coefficient is between one and two orders of magnitude. The form of the inflationary potential is plotted in Fig. \ref{ExA_pot} and it is characterised by: 

\begin{figure}[h!]
\begin{center}
\includegraphics[width=0.55\textwidth, angle=0]{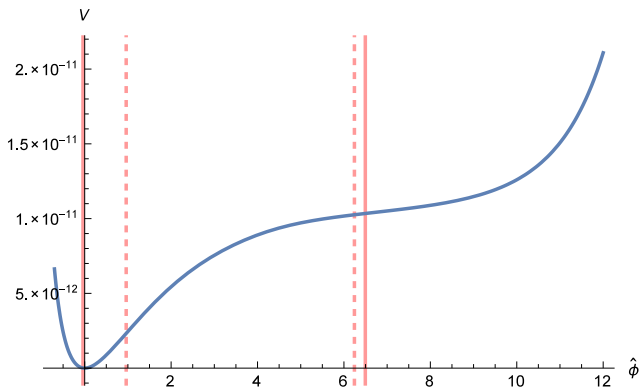}
\caption{Plot of the inflationary potential for the example set (\ref{ex_set_exA}). The red vertical lines correspond to the walls of the K\"ahler cone while the dashed vertical lines denote horizon exit and the end of inflation where $\epsilon=1$.} 
\label{ExA_pot}
\end{center}
\end{figure}

\bi
\item $\langle \tau_7 \rangle = 4.002$ leading to $\epsilon_{\langle \tau_7 \rangle} = 0.0014$. Moreover $2\langle \tau_1 \rangle \simeq 3.8$, and so the distance of the minimum from the lower bound of the K\"ahler cone is $\Delta \tau_7\simeq 0.178$ which is still larger than the string scale since, using (\ref{epsAppr}), we have that:
\be
\epsilon_{\Delta \tau_7} = \frac{1}{g_s (2\pi)^4 \Delta  \tau_7 } \simeq 0.03\,.
\ee

\item The K\"ahler cone bounds (\ref{KConeA}) in terms of the canonically normalised inflaton become $\hat\phi_{\rm min}\simeq -0.04 <\hat\phi < \hat\phi_{\rm max}\simeq 6.49$. Inflation ends at $\hat\phi=\hat\phi_{\rm end}\simeq 0.96$ where $\epsilon(\hat\phi_{\rm end})=1$ and $V_{\rm end}\simeq \left(7 \cdot 10^{15}\,{\rm GeV}\right)^4$. Horizon exit takes place at $\hat\phi=\hat\phi_*\simeq 6.24$ where $r=16\epsilon = 0.009$, $n_s = 1+ 2\eta_* - 6\epsilon_* =  0.983$, $V_* \simeq  \left(1\cdot 10^{16}\,{\rm GeV}\right)^4$ and the amplitude normalisation (\ref{COBE}) is satisfied. Notice that such a largish value of the scalar spectral index is in perfect agreement with Planck data in the presence of dark radiation since, using $\Delta N_{\rm eff}=0.39$ as a prior, \cite{Ade:2015xua} gives as best fit $n_s = 0.983 \pm 0.006$. This prior is fully justified in string models like ours where reheating is driven by the decay of the lightest modulus which naturally tends to produce extra axionic contributions to dark radiation \cite{Cicoli:2012aq, Higaki:2012ar, Hebecker:2014gka, Cicoli:2015bpq}. 

\item Horizon exit occurs well inside the K\"ahler cone since from (\ref{KCA}) we have:
\be
\tau_7^* = e^{\kappa (\langle\phi\rangle +\hat\phi_* )} \simeq 5404.82 < \tau_7^{\rm max} = \frac{\vo}{\sqrt{\langle\tau_1 \rangle}}\simeq 7231.87 \quad\Rightarrow\quad  \tau_7^{\rm max} - \tau_7^* \simeq 1827.06 \,. \nn
\ee

\item The mass of the inflaton around the minimum is $m_{\hat\phi} \simeq 4.25 \cdot 10^{13}$ GeV which from (\ref{Trh}) implies a reheating temperature $T_{\rm rh} \simeq 1.8 \cdot 10^{10}$ GeV.

\item The number of efoldings computed as:
\be
N_e = \int^{\hat\phi_*}_{\hat\phi_{\rm end}} \frac{V}{V'}\,d\hat\phi\,,
\ee
gives $N_e=52$ as required by the estimate (\ref{Nenumber}). The maximum number of efoldings between $\hat\phi_{\rm end}$ and $\hat\phi_{\rm max}$ is $N_e^{\rm max}\simeq 60$. 

\item The $\alpha'$ expansion is under control even if in our inflationary model the inflaton travels over a trans-Planckian distance of order $\Delta \hat\phi = \hat\phi_*-\hat\phi_{\rm end} = 5.28$ since we have $\epsilon_\xi\sim 10^{-4}$.

\item The mass of the volume mode is of order the Hubble scale during inflation since $\delta\simeq 1.6$. Hence the inflationary energy density could either cause a significant shift of the original LVS minimum or destabilise the volume direction. A definite answer to this question would require a more careful multi-field analysis. As mentioned above, a similar situation has been studied in \cite{Cicoli:2008gp}, where the authors found that for $\delta \sim 1$ the minimum for the volume mode gets a large shift but the inflationary evolution still remains mostly single-field since $m_{\rm inf}\ll m_\vo\sim H$. However if $\delta\sim 1$, the inflationary potential generated by string loops and $\alpha'^3$ $F^4$ terms is of the same order as the $\alpha'^3$ $F^2$ contribution, and so one also should carefully check if additional higher derivative corrections can be safely neglected.

\item The effective field theory approximation is valid during the whole inflationary evolution since $H \simeq 2 \cdot 10^{13} \,{\rm GeV} < m_{3/2}\simeq 1 \cdot 10^{15} \,{\rm GeV} < M_\KK^{\rm bulk} \simeq 9 \cdot 10^{15} \,{\rm GeV} < M_s \simeq 2.5 \cdot 10^{16} \,{\rm GeV}$. 
\ei

\begin{figure}[h!]
\begin{center}
\includegraphics[width=0.8\textwidth, angle=0]{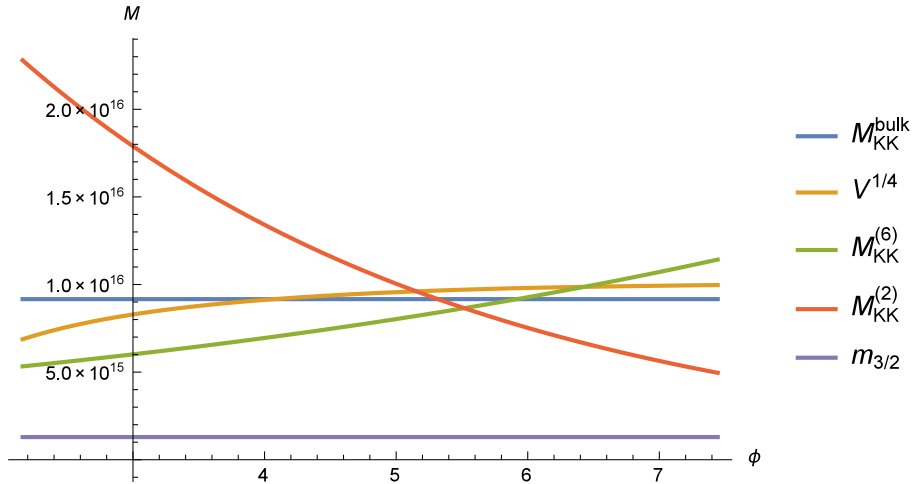}
\caption{Comparison between the different KK masses, $m_{3/2}$ and the inflationary energy density $V^{1/4}$ from horizon exit to the end of inflation. Note that $M_\KK^{(4)} = M_\KK^{(6)}$ which is why only one of them is displayed here.} 
\label{Fig:exA_KK_masses}
\end{center}
\end{figure}

We display the evolution of the different KK masses as compared to the gravitino mass and the inflationary scale $M_{\rm inf} = V^{1/4}$ in Fig.~\ref{Fig:exA_KK_masses}. Notice, in particular, that at the end of inflation the inflationary scale is of order $M_\KK^{\rm bulk}$ and, above all, mildly exceeds the KK scale $M_\KK^{(4)}$ by a factor of roughly $1.3$. As we stressed above, during an adiabatic expansion no energy can be extracted from the vacuum, and so our EFT is still valid even if some KK scales become smaller than $V^{1/4}$ since they are all always larger than $m_{3/2}$ which is, in turn, larger than $H$.
However, since all the inflationary energy density could instead be converted into particle production at reheating, one should make sure that there is enough Hubble friction between the end of inflation and reheating to bring the inflaton energy density below the relevant KK scale. This effect can be estimated by noticing that from: 
\be
\rho(\phi)=\frac12 \dot{\phi}^2 + V(\phi) = 3 H^2 M_p^2 \qquad\Leftrightarrow\qquad\partial_t \rho(\phi) = - 3 H\dot{\phi}^2\,,
\ee
we can obtain the following relation between the energy density at the end of inflation and at reheating:
\be
\rho_{\rm rh} = \rho_{\rm end} - 3 \langle \dot{\phi}^2 \rangle\, \int_{\rm end}^{\rm rh} \frac{da}{a} 
= \rho_{\rm end} - 3 N_{\rm rh} \langle \dot{\phi}^2 \rangle\,,
\ee
where $\langle \dot{\phi}^2 \rangle$ is the time average between the end of inflation and reheating and $N_{\rm rh}=\ln(a_{\rm rh}/a_{\rm end})$ is the number of efoldings of the reheating epoch. At the end of inflation when $\epsilon=1$ we have:
\be
\frac12 \dot{\phi}^2 = H^2 M_p^2 \qquad\Leftrightarrow\qquad \rho_{\rm end} = \frac32 V_{\rm end} \simeq 10 \left(M_\KK^{(4)}\right)^4\,.
\ee
On the other hand at reheating $V(\phi_{\rm rh})\simeq 0$, and so $\rho_{\rm rh}\simeq \dot{\phi}_{\rm rh}^2/2$. If we then write the time-average kinetic energy as $\langle \dot{\phi}^2 \rangle =  \dot{\phi}_{\rm rh}^2/x \simeq 2\rho_{\rm rh}/x$ with $x>0$, we end up with the following bound:
\be
\rho_{\rm rh} \simeq  \frac{10}{1+\frac6x N_{\rm rh}} \left(M_\KK^{(4)}\right)^4 < \left(M_\KK^{(4)}\right)^4\,.
\label{estimate}
\ee
Using the fact that:
\be\label{eq:N_rh}
N_{\rm rh} \simeq \frac13 \ln\left(\frac{H_{\rm end}^2 M_p^2}{T_{\rm rh}^4}\right)-\frac13\ln\left(\frac{\pi^2 g_*}{90}\right)\simeq 16\,,
\ee
the bound (\ref{estimate}) becomes $x < \tfrac{2}{3} N_{\rm rh} \simeq 10$. Our model should satisfy this bound since we expect $\dot{\phi}_{\rm end}$ to approach $\dot{\phi}_{\rm rh}$ relatively quickly due to the steepness of the potential near the end of inflation. However a definite answer would require a detailed study of the post-inflationary epoch which is beyond the scope of this paper.\footnote{Let us also point out that, even if $\rho_{\rm rh}\gtrsim \left(M_\KK^{(4)}\right)^4$, our model is not necessarily ruled out but we would just need to describe reheating within a 6D EFT where the base of the fibration is much larger than the characteristic size of the fibre. It would also be interesting to find brane setups where this problem is automatically absent since there is no D7-brane wrapped around the base.} 

Let us also mention that, due to the absence of KK corrections, this scenario represents a chiral global embedding of the $\alpha'$-inflation models discussed in \cite{Cicoli:2016chb}. Moreover, no KK scale becomes smaller than the gravitino mass even if $r\simeq 0.01$ and $\Delta \hat\phi \simeq 5$ in Planck units. In fact, if we focus for example on the KK scale $M_\KK^{(2)}$ associated with the K3 fibre (similar considerations apply to the KK scale $M_\KK^{(6)}$ associated with the base), we have:
\be
\frac{m_{3/2}}{M_\KK^{(2)}} = \alpha_1\,e^{\alpha_2 \phi}\simeq 0.03\,e^{\alpha_2 \phi}\,,
\label{KKratioo}
\ee
with:
\be
\alpha_1 = \sqrt{\frac{W_0}{2\pi}}\,\left(\frac{g_s}{2\pi}\right)^{1/4}\,\sqrt{\frac{m_{3/2}}{M_p}}\simeq 0.03 \qquad\text{and}\qquad \alpha_2 = \frac{1}{2\sqrt{3}}\,.
\ee
If we set $\phi=\phi_0 + \hat\phi_{\rm he} \simeq 7.44$, the ratio in (\ref{KKratioo}) becomes $m_{3/2}/M_\KK^{(2)} \simeq 0.26$, and so the KK scale $M_\KK^{(2)}$ is always larger than the gravitino mass throughout all the inflationary dynamics. Notice that this result seems to be in slight disagreement with the swampland conjecture of \cite{Klaewer:2016kiy, Blumenhagen:2017cxt} where the underlying parameters $\alpha_1$ and $\alpha_2$ were generically assumed to be of order unity. 

As explained above, given that in this case $\delta\simeq 1.6$, the inflationary dynamics can be fully trusted only after determining the proper multi-field evolution. Due to the difficulty to perform a full numerical analysis, in the next section we shall instead still focus on a single-field case where $\delta\sim 0.05$ since $\xi$ is larger, and so the volume mode mass is larger, while $|\lambda|$ is smaller, and so $F^4$ steepening terms can be easily neglected throughout the whole inflationary dynamics. The full three-field evolution for both of these cases will then be presented in Sec. \ref{Multi}.

\subsubsection{Case 2: $\xi = 0.456$ and $|\lambda| = 10^{-7}$}
\label{GoodCase}

According the discussion above, we shall now focus on the following different choice of the underlying parameters:
\bea
\alpha &=& 1\,, \quad C^\W_1=C^\W_2=0.034\,, \quad |C^\W_3|=10^{-5}\,, \quad |C^\W_4| =0.068\,,\quad C^\W_5 = C^\W_6 =-0.024\,,  \nn \\ 
g_s &=& 0.25\,, \quad \vo=4500 \,, \quad \langle\tau_1\rangle= 3.10\,, \quad W_0=150\,, \quad |\lambda|=10^{-7}\,,
\label{ex_set_exA2}
\eea
with $\chi(X)=\chi_{\rm eff}(X)= -188$ in (\ref{a3F2}) which gives $\xi = 0.456$. A larger value of the coefficient $\xi$ is helpful to increase the control on the single-field approximation since, as can be seen from (\ref{a3F2}), the leading $\mc{O}(\alpha'^3)$ contribution to the scalar potential is proportional to $\xi$. The form of the inflationary potential is plotted in Fig. \ref{ExA2_pot} and it is characterised by: 

\begin{figure}[h!]
\begin{center}
\includegraphics[width=0.55\textwidth, angle=0]{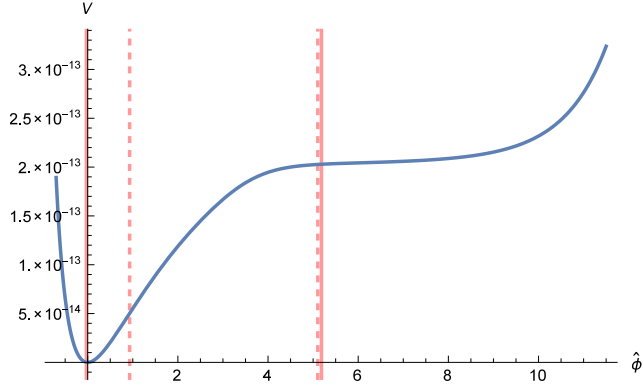}
\caption{Plot of the inflationary potential for the example set (\ref{ex_set_exA2}). The red vertical lines correspond to the walls of the K\"ahler cone while the dashed vertical lines denote horizon exit and the end of inflation where $\epsilon=1$.} 
\label{ExA2_pot}
\end{center}
\end{figure}

\bi
\item $\langle \tau_7 \rangle \simeq 6.41$ leading to $\epsilon_{\langle \tau_7 \rangle} \simeq 0.0004$ and $\langle\phi\rangle \simeq 1.61$. Moreover $2\langle \tau_1 \rangle \simeq 6.2$, and so the minimum is located close to the walls of the K\"ahler cone but at a distance $\Delta \tau_7\simeq 0.21$ which is still larger than the string scale since, using (\ref{epsAppr}), we have that:
\be
\epsilon_{\Delta \tau_7} = \frac{1}{g_s (2\pi)^4 \Delta  \tau_7 } \simeq 0.01\,.
\ee

\item The K\"ahler cone bounds (\ref{KConeA}) in terms of the canonically normalised inflaton become $\hat\phi_{\rm min}\simeq -0.028 <\hat\phi < \hat\phi_{\rm max}\simeq 5.19$. Inflation ends at $\hat\phi=\hat\phi_{\rm end}\simeq 0.93$ where $\epsilon(\hat\phi_{\rm end})=1$ and $V_{\rm end}=\left(4.4\cdot 10^{15}\,{\rm GeV}\right)^4$. Horizon exit takes place at $\hat\phi=\hat\phi_*\simeq 5.10$ where $r=16\epsilon = 0.0014$, $n_s = 1+ 2\eta_* - 6\epsilon_* =  0.963$, $V_* = \left(6.2\cdot 10^{15}\,{\rm GeV}\right)^4$ and the amplitude normalisation (\ref{COBE}) is satisfied. Notice that horizon exit occurs far away from the upper bound of the K\"ahler cone since from (\ref{KCA}) we have:
\be
\tau_7^* = e^{\kappa (\langle\phi\rangle +\hat\phi_* )} \simeq 2325.79 < \tau_7^{\rm max} = \frac{\vo}{\sqrt{\langle\tau_1 \rangle}}\simeq 2554.55 \quad\Rightarrow\quad  \tau_7^{\rm max} - \tau_7^* \simeq 228.76 \,. \nn
\ee

\item The mass of the inflaton around the minimum is $m_{\hat\phi} \simeq 1.85 \cdot 10^{13}$ GeV which from (\ref{Trh}) implies a reheating temperature $T_{\rm rh} \simeq 5.16 \cdot 10^9$ GeV.

\item The number of efoldings computed as:
\be
N_e = \int^{\hat\phi_*}_{\hat\phi_{\rm end}} \frac{V}{V'}\,d\hat\phi\,,
\ee
gives $N_e=51$ as required by the estimate (\ref{Nenumber}). The maximum number of efoldings between $\hat\phi_{\rm end}$ and $\hat\phi_{\rm max}$ is $N_e^{\rm max}\simeq 57.5$.  

\item The $\alpha'$ expansion is under control even if in our inflationary model the inflaton travels over a trans-Planckian distance of order $\Delta \hat\phi = \hat\phi_*-\hat\phi_{\rm end} = 4.17$ since we have $\epsilon_\xi\simeq 0.0004$.

\item The single-field approximation is under control since $\delta\simeq 0.05$.

\item The effective field theory approximation is valid during the whole inflationary evolution since $H\simeq 7\cdot 10^{12} \,{\rm GeV}< m_{3/2}\simeq 8\cdot 10^{15} \,{\rm GeV} < M_\KK^{\rm bulk} \simeq 1.6 \cdot 10^{16} \,{\rm GeV} < M_s \simeq 4.5 \cdot 10^{16} \,{\rm GeV}$. 
\ei

\begin{figure}[h!]
\begin{center}
\includegraphics[width=0.8\textwidth, angle=0]{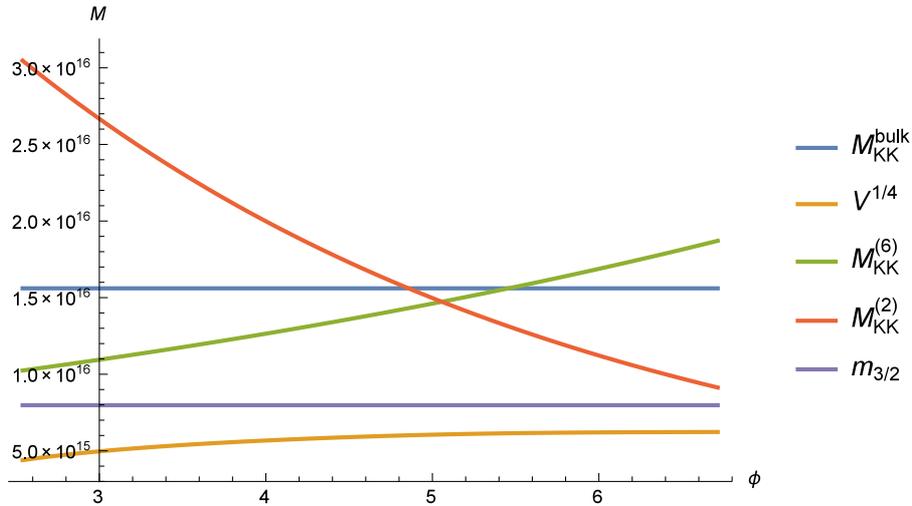}
\caption{Comparison between the different KK masses, the gravitino mass $m_{3/2}$ and the inflationary energy $V^{1/4}$ from horizon exit to the end of inflation. Note that $M_\KK^{(4)} = M_\KK^{(6)}$ which is why only one of them is displayed here.} 
\label{Fig:exA2_KK_masses}
\end{center}
\end{figure}

We display the evolution of the different KK masses as compared to the gravitino mass and the inflationary energy density $M_{\rm inf} = V^{1/4}$ in Fig.~\ref{Fig:exA2_KK_masses}. Notice that, contrary to case $1$ where $r=0.01$, all KK scales remain above $M_{\rm inf}$ throughout all the inflationary dynamics. The reason is that in this scale the tensor-to-scalar ratio, and so also the inflationary scale, is smaller since $r=0.001$. Moreover, as stressed above, no energy can be extracted from the vacuum during an adiabatic expansion, and so the consistency condition to be imposed during inflation is $H\ll M_\KK^{(i)}$ which is clearly satisfied since $H = \frac{M_{\rm inf}}{\sqrt{3}}\left(\frac{M_{\rm inf}}{M_p}\right) < M_{\rm inf}$. Moreover, no KK scale becomes smaller than the gravitino mass $m_{3/2} \simeq 8 \cdot 10^{15}$ GeV. If we focus for example on the KK scale $M_\KK^{(2)}$ associated with the K3 fibre (similar considerations apply to the KK scale $M_\KK^{(6)}$ associated with the base of the fibration), we have:
\be
\frac{m_{3/2}}{M_\KK^{(2)}} = \alpha_1\,e^{\alpha_2 \phi}\simeq 0.126\,e^{\alpha_2 \phi}\,,
\label{KKratio}
\ee
with:
\be
\alpha_1 = \sqrt{\frac{W_0}{2\pi}}\,\left(\frac{g_s}{2\pi}\right)^{1/4}\,\sqrt{\frac{m_{3/2}}{M_p}}\simeq 0.126 \qquad\text{and}\qquad \alpha_2 = \frac{1}{2\sqrt{3}}\,.
\ee
If we set $\phi=\phi_0 + \hat\phi_{\rm he} \simeq 6.71$, the ratio in (\ref{KKratio}) becomes $m_{3/2}/M_\KK^{(2)} \simeq 0.87$, and so the KK scale $M_\KK^{(2)}$ is always larger than the gravitino mass throughout all the inflationary dynamics. This result seems to be more in agreement with the swampland conjecture of \cite{Klaewer:2016kiy, Blumenhagen:2017cxt} than the one of case $1$ since $r$ is smaller, $r\simeq 0.001$, and the field range is slightly reduced, $\Delta \hat\phi\simeq 4$. Moreover larger values of $\phi$ would bring the effective field theory approach out of control. 

Even if this example satisfies all consistency and phenomenological constraints and the single-field inflationary analysis is under control, in Sec. \ref{Multi} we shall perform a more precise multifield analysis where the motion along the orthogonal directions enlarges the field space as well as the allowed number of efoldings.

\subsection{Multi-field evolution}
\label{Multi}

The following five consistency conditions require generically a multi-field study of the inflationary evolution (which might however still be mainly along a single direction in field space):
\ben
\item The whole inflationary dynamics takes place well inside the K\"ahler cone described by the conditions in (\ref{KCA});

\item The quantum fluctuations of the inflaton produce a correct amplitude of the density perturbations at horizon exit;

\item The directions orthogonal to the inflaton are not destabilised by the inflationary dynamics. This is guaranteed if inflation occurs in field space along a through which can however bend;

\item Throughout all the inflationary dynamics, no Kaluza-Klein scale becomes smaller than the gravitino mass;

\item The steepening of the inflationary potential due to $F^4$ corrections is negligible, so that enough efoldings can be obtained before destroying slow roll inflation.
\een
If $\vo\sim 10^3$ and $W_0\sim \mc{O}(1)$, the last four conditions can be easily satisfied but the K\"ahler cone conditions (\ref{KCA}) for such a small value of the volume would give an upper bound on the inflaton direction which would not allow to generate enough efoldings. In order to enlarge the inflaton field space, the value of the volume has therefore to be larger, of order $\vo \sim 10^4$. In the large volume regime where we can trust the 4D EFT, the inflationary potential then becomes more suppressed, and so the COBE normalisation condition (2) above can be satisfied only if $W_0\sim \mc{O}(100)$. However, given that the gravitino mass is proportional to $W_0$, for such a large value of the flux-generated superpotential, it is hard to satisfty the fourth condition above keeping $m_{3/2}$ below all KK scales during the whole inflationary evolution. Moreover, it becomes harder to suppress higher derivative corrections (condition (5) above) unless their numerical coefficient $\lambda$ turns out to be extremely small: $|\lambda|\lesssim 10^{-6}$. This is the example of case 2 above of Sec. \ref{GoodCase}.

Another option for $\vo \sim 10^4$ could be to keep $W_0\sim \mc{O}(1)$, so that the gravitino mass can remain small and the $F^4$ terms are still negligible, and to tune the background fluxes to increase the complex structure-dependent coefficients of the winding loop corrections. This would however 
make the inflaton-dependent potential of the same order of magnitude of the leading order $\alpha'$ correction. Hence the mass of the volume mode becomes of order the Hubble scale during inflation. This is the example of case 1 of Sec. \ref{BadCase} where $\delta \simeq 1.6$. This situation could either cause a considerable shift of the original LVS minimum or even a destabilisation, and so in this case one should perform a careful multi-field analysis to check that the condition (3) above is indeed satisfied.\footnote{A similar situation arises in K\"ahler moduli inflation where however a detailed multifield analysis shows that the minimum of the volume mode is shifted during inflation without developing a runaway direction \cite{Cicoli:2017shd, BlancoPillado:2009nw}.}

In what follows we shall therefore focus on the multifield case with $\vo\sim 10^4$, $W_0\sim \mc{O}(100)$ and $|\lambda|\lesssim 10^{-6}$. We shall also present an example with $W_0\sim \mc{O}(1)$ and $|\lambda| \sim 10^{-3}$ which satisfies all conditions above except for condition (2) since the amplitude of the density perturbations turns out to be too small. The correct value could be generated by the quantum fluctuations of the two light bulk axions which could play the r\^ole of curvaton fields \cite{curvaton}. This study is however beyond the scope of this paper, and so we leave it for future work. 

We analyse now the full three-field cosmological evolution involving the K\"ahler moduli $\tau_7$, $\vo$ and $\tau_1$. Their dynamics is governed by the 
following evolution equations for non-canonically normalised fields:
\be
 \left\{
 \begin{array}{c}
 \ddot{\phi}^{i}+3H\dot{\phi}^{i}
 +\Gamma_{jk}^{i}\dot{\phi}^{j}\dot{ \phi}^k+g^{ij}\frac{\partial V}{\partial \phi ^j}=0, \\
 H^2=\left( \frac{\dot{a}}{a}\right)^2=\frac13 \left(\frac12 g_{ij}\dot{\phi}^i\dot{\phi}^j+V \right),
 \end{array}
 \right.
\ee
where the $\phi_i$'s represent the scalar fields $\tau_7$, $\vo$ and $\tau_1$, $a$ is the scale factor and $\Gamma_{jk}^{i}$ are the target space Christoffel symbols using the metric $g_{ij}$ for the set of real scalars $\phi^i$ such that $\frac{\partial^2 K}{\partial \Phi^I
\partial \Phi^{*J}} \,\partial^{\mu}\Phi^I\partial_{\mu}\Phi^{*J} =\frac12\,g_{ij}\partial_{\mu}\phi^i\partial^{\mu}\phi^j$.

For numerical purposes it is more convenient to express the cosmological evolution of the fields as a function of the number of efoldings $N$ rather than time. In fact, by using $a(t)=e^N$ and $\frac{d}{dt}=H\frac{d}{dN}$, we can directly obtain $\tau_7 (N)$, $\vo(N)$ and $\tau_1(N)$ without having to solve for the scale factor. The equations of motion turn out to be (with $'$ denoting a derivative with respect to $N$):
\bea
\tau_7^{\prime \prime } &=&-\left( \mc{L}_{\rm kin}+3\right) \left(\tau_7^\prime +\tau_7\vo \frac{V_{,\vo}}{V}+2\tau_7^2 \frac{V_{,\tau_7}}{V}+2\tau_7\tau_1 \frac{V_{,\tau_1}}{V}\right) +\frac{\tau_7^{\prime 2}}{\tau_7}
+\frac{\tau_7 \tau_1^\prime}{\vo}\left(\frac{\tau_1^{\prime}}{\sqrt{\tau_1}} -\frac{\tau_7^\prime}{2\sqrt{\tau_7}}  \right)
\,, \nn \\
 \vo^{\prime \prime } &=&-\left( \mc{L}_{\rm kin}+3\right) \left( \vo^\prime +\frac{3\vo^2}{2}\frac{V_{,\vo}}{V}+\tau_7\vo\frac{V_{,\tau_7}}{V}+\tau_1\vo\frac{V_{,\tau_1}}{V}\right)
 +\frac{\vo^{\prime 2}}{\vo}\,, 
\label{NLEq} \\
\tau_1^{\prime \prime } &=& -\left( \mc{L}_{\rm kin}+3\right) \left(\tau_1^\prime +\tau_1\vo\frac{V_{,\vo}}{V}+ 2\tau_7\tau_1\,\frac{V_{,\tau_7}}{V}
+ 4\vo\sqrt{\tau_1}\,\frac{V_{,\tau_1}}{V}\right) \nn \\
&+& \frac{\tau_1^{\prime \,2}}{4\tau_1}
+ \frac{\tau_1 \vo^\prime}{\vo} \left( \frac{\tau_1^\prime}{\tau_1} - \frac{\tau_7^\prime}{\tau_7} \right) 
+\frac{\tau_1 \tau_7^\prime}{2\tau_7} \left(\frac{3 \tau_7^{\prime}}{2 \tau_7}
-\frac{\sqrt{\tau_1}}{\vo} \tau_1^\prime\right) \,,  \nn
\eea
where the kinetic Lagrangian reads:
\be
\mc{L}_{\rm kin} = \frac12 \left(-\frac{\vo^{\prime\,2}}{\vo^2} + \frac{\vo^\prime \tau_7^\prime}{\vo \tau_7 } - \frac{3 \tau_7^{\prime\,2}}{4 \tau_7^2} + \frac{\sqrt{\tau_1} \tau_7^\prime \tau_1^\prime}{ 2 \vo \tau_7} - \frac{\tau_1^{\prime\, 2}}{ 4 \vo \sqrt{\tau_1}}\right)\,,
\ee
and the full inflationary potential $V$ is given by the sum of the standard LVS potential, the $g_s$ loops and $F^4$ terms given in (\ref{exA_tot_pot}) and an uplifting contribution proportional to $\delta_{\rm up}$ which could come from an anti D3-brane at the tip of a warped throat:
\bea
V &=& \kappa \left[32 A_s^2 \pi^2\,\frac{\sqrt{\tau_1}}{\vo}\,e^{-4\pi\tau_1} - 8\pi A_s \frac{W_0 \tau_1}{\vo^2}\,e^{-2\pi\tau_1} + \frac{3\zeta}{4 g_s^{3/2}} \frac{W_0^2}{\vo^3} \right. \nn \\
&+& \left.  \frac{W_0^2}{\vo^3}\,\left(\frac{A_1}{\tau_7}-\frac{A_2}{\sqrt{\tau_7}}  + \frac{B_1\sqrt{\tau_7}}{\vo}+\frac{B_2 \,\tau_7}{\vo}\right)+\frac{\delta_{\rm up}}{\vo^{4/3}}\right]\,.
\label{Vinf}
\eea

\subsubsection{$|\lambda| = 10^{-6}$ and correct amplitude of the density perturbations}
\label{GreatCase}

Setting $\alpha=1$ and performing the following choice of the underlying parameters:
\bea
A_s &=& 6\cdot 10^5 \qquad \chi=-188 \quad\Rightarrow \quad \zeta = -\frac{\zeta(3) \chi(X)}{2 (2\pi)^3} = 0.456 \quad W_0 = 50 \quad g_s = 0.25 \nn \\
C_1^\W &=& C_2^\W = 0.05 \quad |C_3^\W| = 10^{-4} \quad |C_4^\W| = 0.1 \quad C_5^\W=C_6^\W = -0.05 \quad \lambda = -10^{-6}\,, \nn
\eea
the total potential (\ref{Vinf}) admits a Minkowski global minimum at:
\be
\langle \vo\rangle = 2690.625\,,\qquad \langle\tau_7\rangle = 6.503\,\qquad \langle \tau_1\rangle = 3.179\quad\text{for}\quad 
\delta_{\rm up} = 5.9598\cdot 10^{-4}\,. \nn
\ee
Notice that this minimum is inside the K\"ahler cone since $\langle\tau_7\rangle > 2\langle\tau_1\rangle = 6.358$, which respects the lower bound in (\ref{KCA}). At this level of approximation, the closed string axions associated to $\vo$ and $\tau_7$ are flat directions. They receive a tiny potential from highly suppressed non-perturbative effects, and so they remain very light. Being so light, they do not affect the inflationary dynamics but would acquire isocurvature fluctuations of order $H$ during inflation. If they do not play the r\^ole of dark matter, their final contribution to the amplitude of the isocurvature perturbations is negligible. On the other hand, if they are heavy enough to decay, their isocurvature fluctuations get converted into standard density perturbations, and so these bulk axions could behave as curvaton fields \cite{curvaton}.

Let us now shift $\tau_7$ away from its minimum at the initial condition $\tau_7 (N=0) = \langle\tau_7\rangle + 2030$ and recompute the new minimum for the other two directions $\langle \vo\rangle(\tau_7)$ and $\langle\tau_1\rangle(\tau_7)$. These values would set the initial conditions for these fields, ensuring that the inflationary dynamics takes place along a stable trough in field space:
\be
\vo(0) = \langle \vo\rangle (\tau_7(0)) = 3671.432\,,\quad \tau_7(0) = 2036.503\,,
\quad\tau_1(0) = \langle \tau_1\rangle(\tau_7(0)) = 3.227 \,. \nn
\ee
Notice that these initial conditions are again inside the K\"ahler cone since $\tau_7(0) < \frac{\vo(0)}{\sqrt{\tau_1(0)}} = 2043.7$, which satisfies the upper bound in (\ref{KCA}). We shall also focus on vanishing initial velocities for all scalar fields: $\vo^\prime (0) = \tau_7^\prime (0) = \tau_1^\prime(0) = 0$.

\begin{figure}[ht]
\begin{center}
\epsfig{file=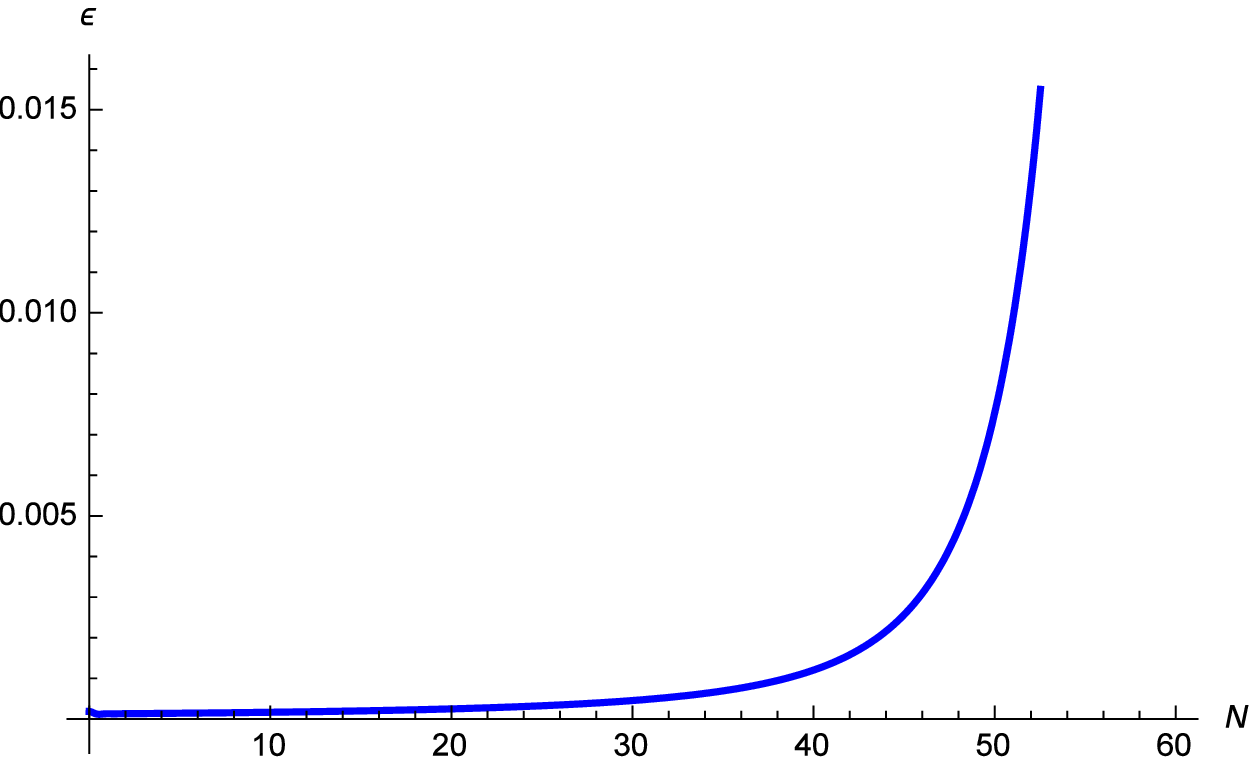, height=50mm,width=70mm}
\epsfig{file=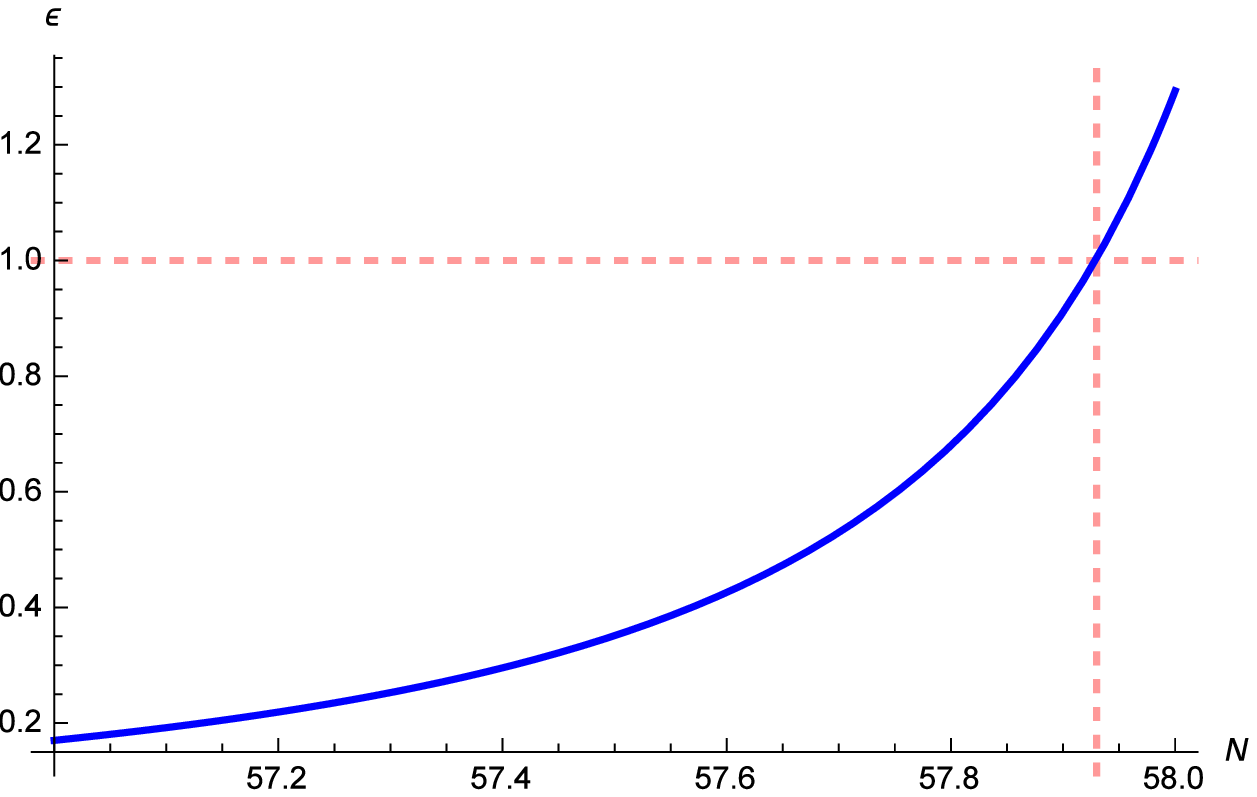, height=50mm,width=70mm}
\caption{Evolution of the $\epsilon$-parameter as a function of the number of efoldings $N$ for (left) the entire inflationary dynamics and (right) for the last efolding.}
\label{FigEps}
\end{center}
\end{figure}

Considering this set of initial conditions, we solved the system of equations of motion (\ref{NLEq}) finding the cosmological evolution of each scalar field as a function of the number of efoldings $N$. Inflation occurs in the region in field space where the generalised $\epsilon$-parameter:
\be
\epsilon (N) =-\frac{1}{4\mc{L}_{\rm kin} V^2}\left(V_{,\vo}\,\vo^\prime+V_{,\tau_7}\,\tau_7^\prime+V_{,\tau_1}\,\tau_1^\prime\right)^2\,,
\label{GenEps}
\ee
is much smaller than unity. As can be seen from Fig. \ref{FigEps}, $\epsilon\ll 1$ during the first $57$ efoldings and then quickly increases and reaches $\epsilon = 1$ at $N=57.93$ where inflation ends. 

Using the variable $N$ to parametrise the cosmological evolution of the scalar fields and denoting by $N_e$ the physical number of efoldings of inflation, $N_e = 52$, as estimated in Sec. \ref{exA_inf}, at $N_* = 5.93$. This is the point of horizon exit in field space where $\epsilon (N_*) = 1.456\cdot 10^{-4}$ which yields a tensor-to-scalar ratio $r = 16 \epsilon(N_*) = 0.0023$. The amplitude of the scalar power spectrum is:
\be
\sqrt{P(N_*)} = \frac{1}{10\pi}\sqrt{\frac{2\,V(N_*)}{3\,\epsilon(N_*)}} = 1.035 \cdot 10^{-5}\,,
\ee
reproducing the reference COBE value $\sqrt{P_{\scriptscriptstyle{\rm COBE}}}\simeq 2\cdot 10^{-5}$ with a good accuracy. Moreover the scalar spectral index is given by:
\be
n_s (N_*) = 1 + \left.\frac{d}{dN}\,\ln P(N)\right|_{N=N_*} = 0.9701\,,
\ee
in good agreement with Planck data \cite{Ade:2015lrj, Ade:2015xua}. 

\begin{figure}[ht]
\begin{center}
\epsfig{file=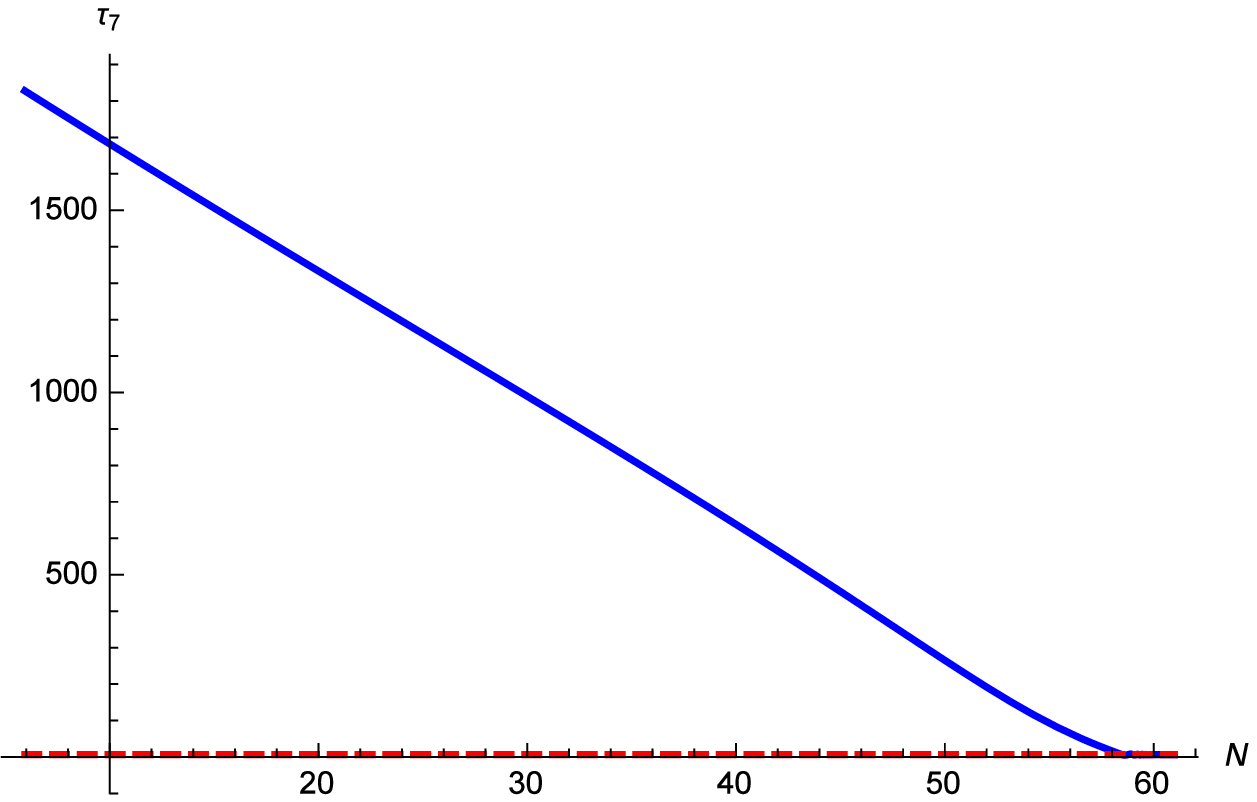, height=50mm,width=70mm}
\epsfig{file=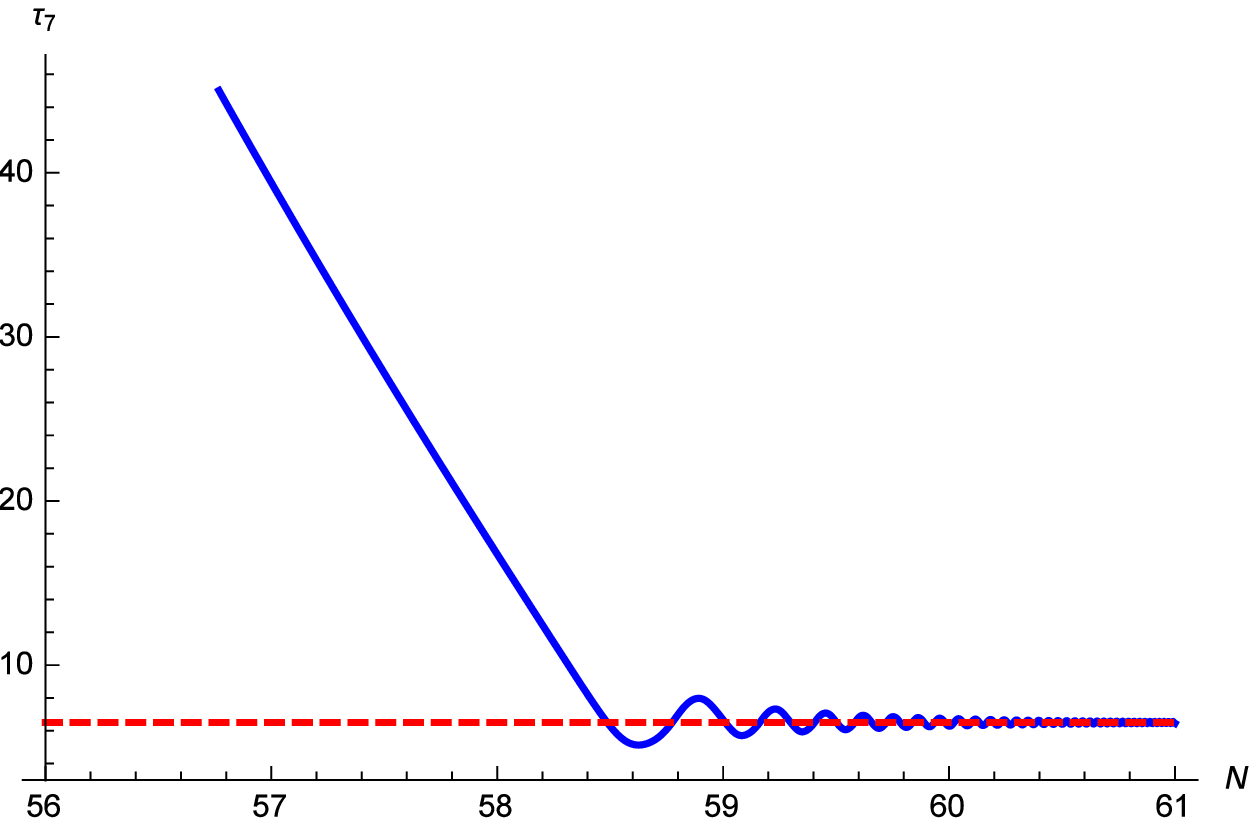, height=50mm,width=70mm}
\caption{Evolution of $\tau_7$ as a function of the number of efoldings $N$ for (left) the entire inflationary dynamics and (right) for the last $2$ efoldings. The dashed red line represents the position of the final global minimum.}
\label{Figtau7}
\end{center}
\end{figure}

\begin{figure}[ht]
\begin{center}
\epsfig{file=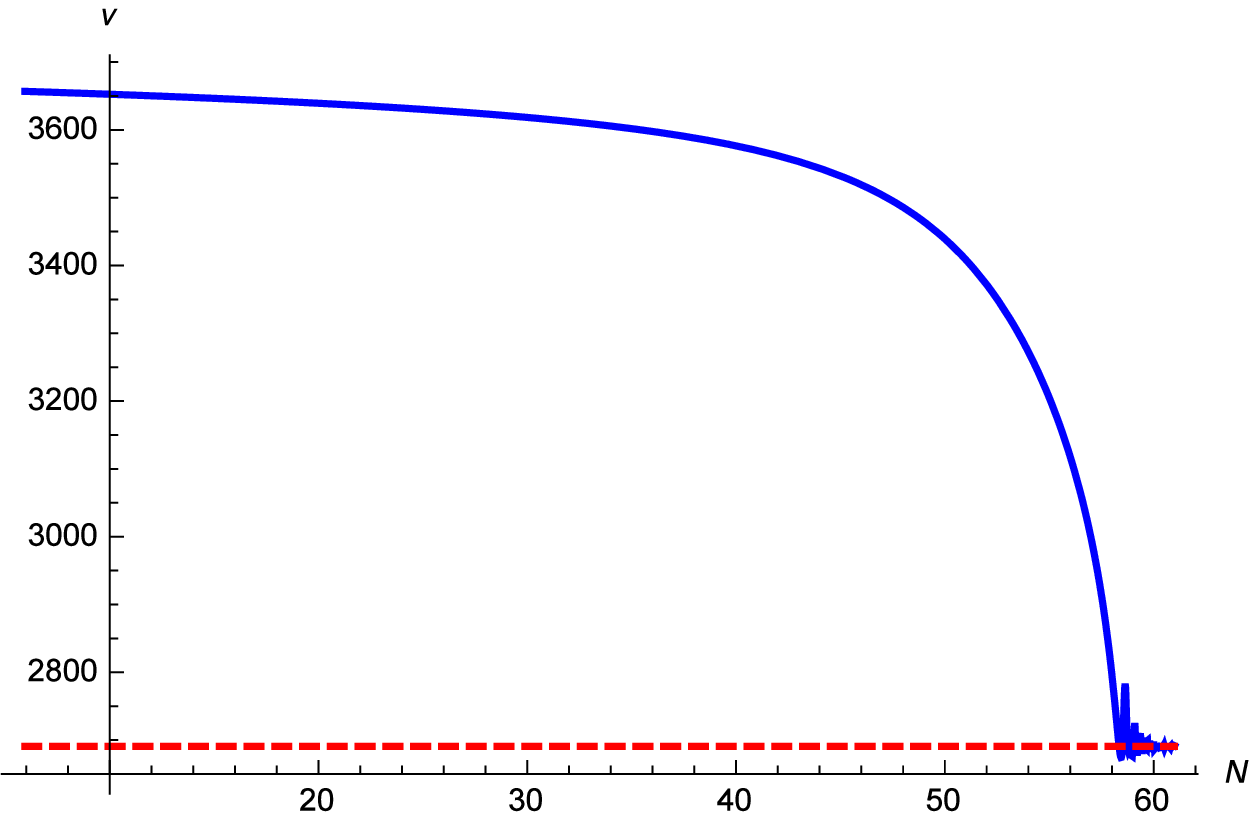, height=50mm,width=70mm}
\epsfig{file=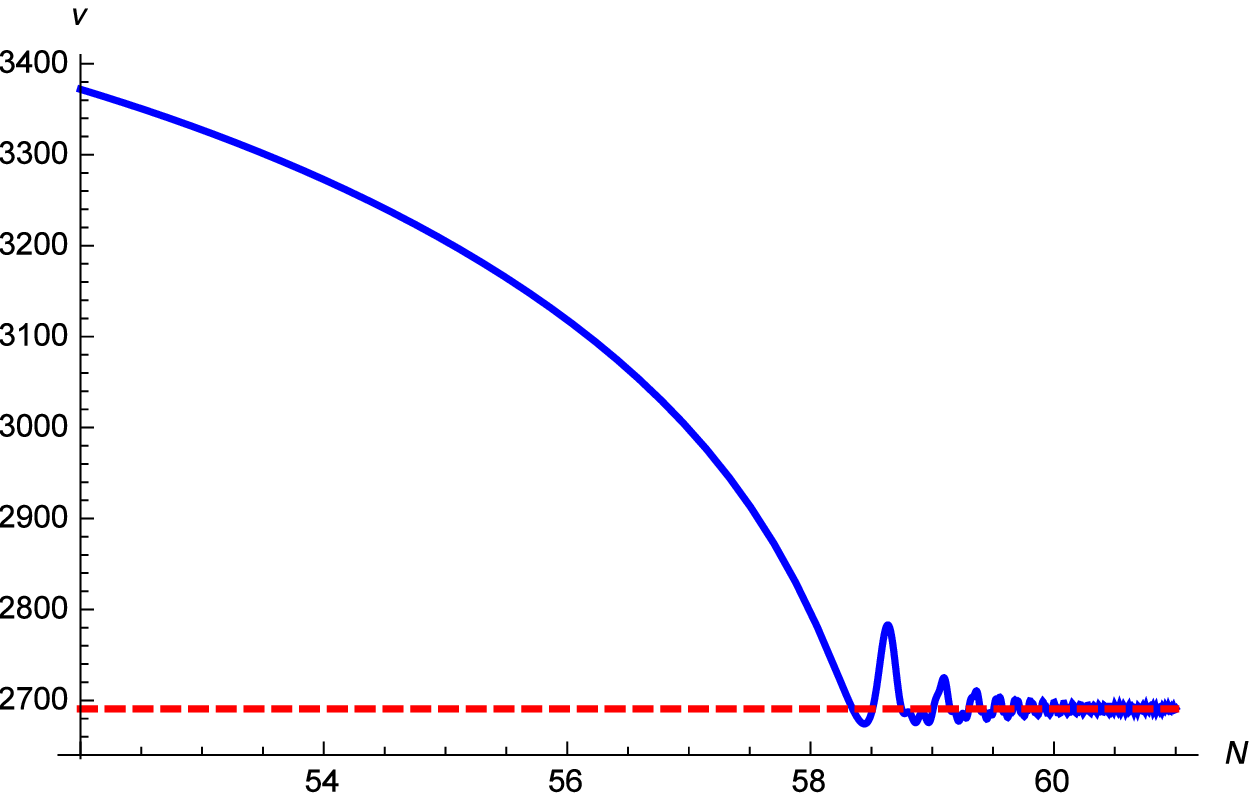, height=50mm,width=70mm}
\caption{Evolution of $\vo$ as a function of the number of efoldings $N$ for (left) the entire inflationary dynamics and (right) for the last $6$ efoldings. The dashed red line represents the position of the final global minimum.}
\label{FigVol}
\end{center}
\end{figure}

\begin{figure}[ht]
\begin{center}
\epsfig{file=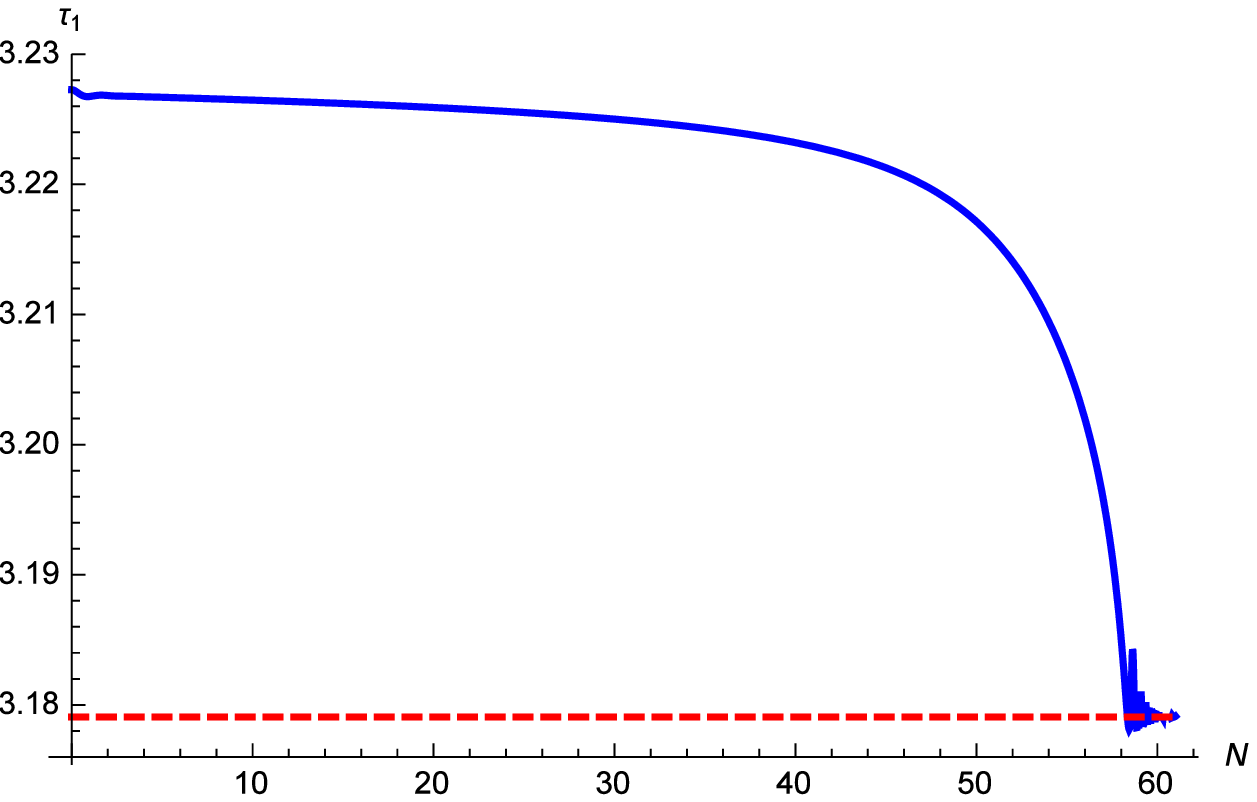, height=50mm,width=70mm}
\epsfig{file=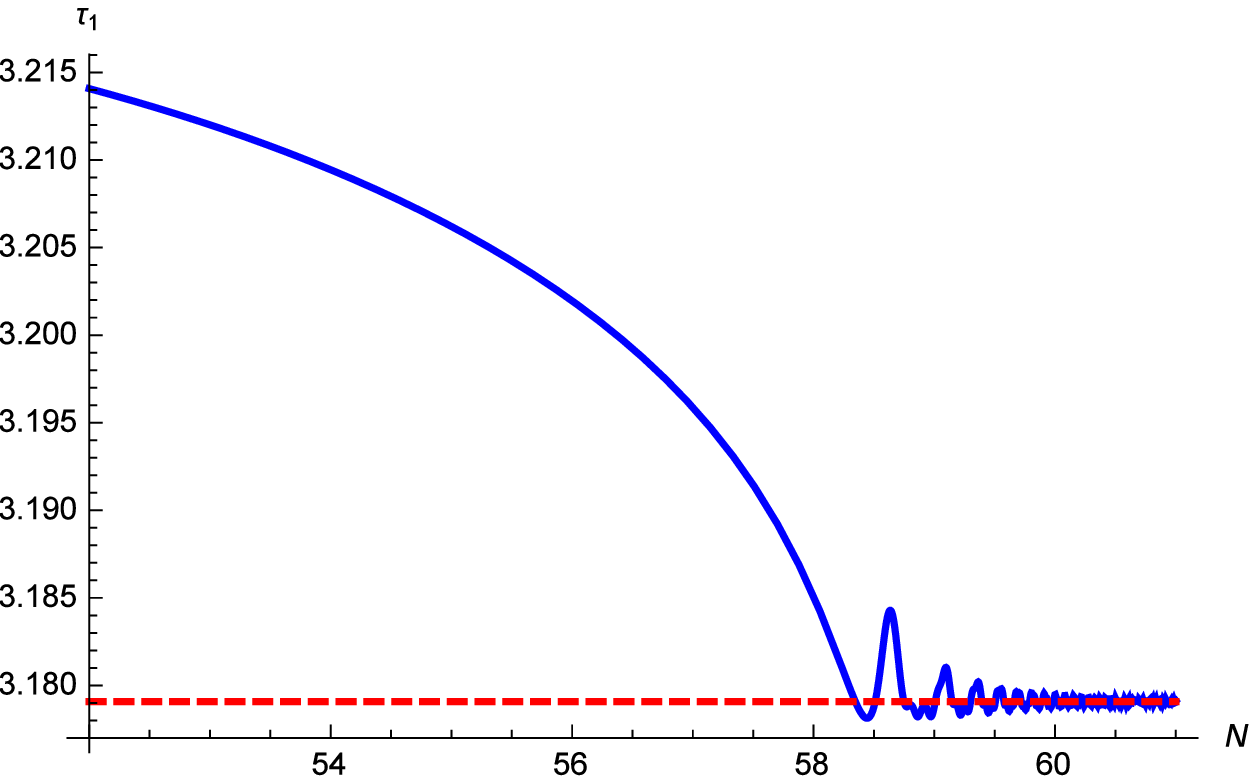, height=50mm,width=70mm}
\caption{Evolution of $\tau_1$ as a function of the number of efoldings $N$ for (left) the entire inflationary dynamics and (right) for the last $6$ efoldings. The dashed red line represents the position of the final global minimum.}
\label{Figtau1}
\end{center}
\end{figure}

\begin{figure}[ht]
\begin{center}
\epsfig{file=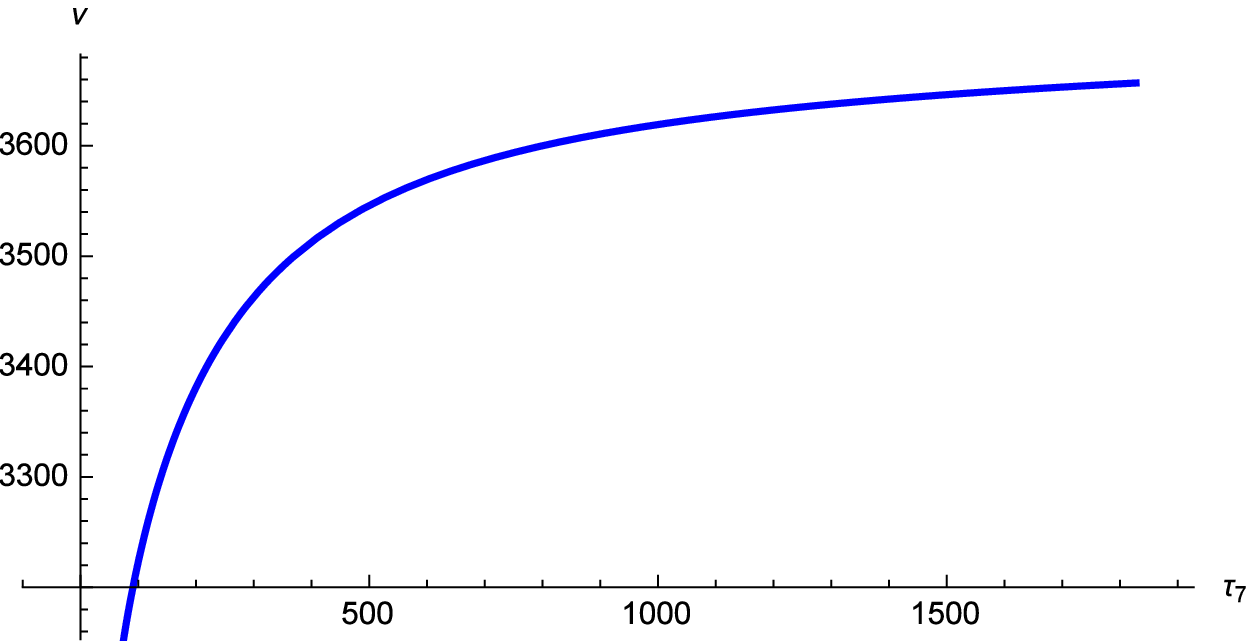, height=50mm,width=70mm}
\epsfig{file=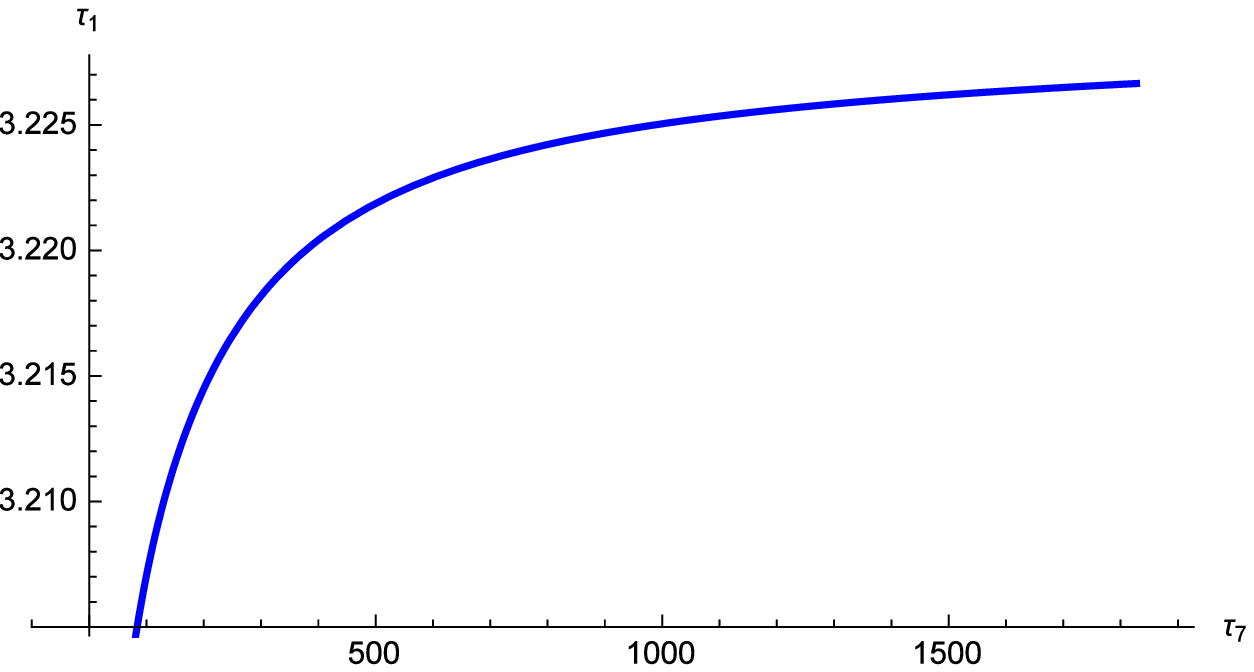, height=50mm,width=70mm}
\caption{Plot of the whole inflationary evolution in the $(\tau_7, \vo)$-plane (on the left) and in the $(\tau_7, \tau_1)$-plane (on the right). Notice that the scales are different on the two axes since the inflaton travels mainly along the $\tau_7$-direction.}
\label{FigParam}
\end{center}
\end{figure}

Fig. \ref{Figtau7}, \ref{FigVol} and \ref{Figtau1} show the cosmological evolution of the three scalar fields $\tau_7$, $\vo$ and $\tau_1$ during the whole inflationary dynamics and their final settling into the global minimum after a few oscillations. Fig. \ref{FigParam} shows instead the path of the inflationary trajectory in the $(\tau_7, \vo)$-plane (on the left) and in the $(\tau_7, \tau_1)$-plane (on the right). Clearly, as expected from the single-field analysis of Sec. \ref{exA_inf}, the inflaton travels mainly along the $\tau_7$-direction. 

\begin{figure}[h!]
\begin{center}
\includegraphics[width=0.8\textwidth, angle=0]{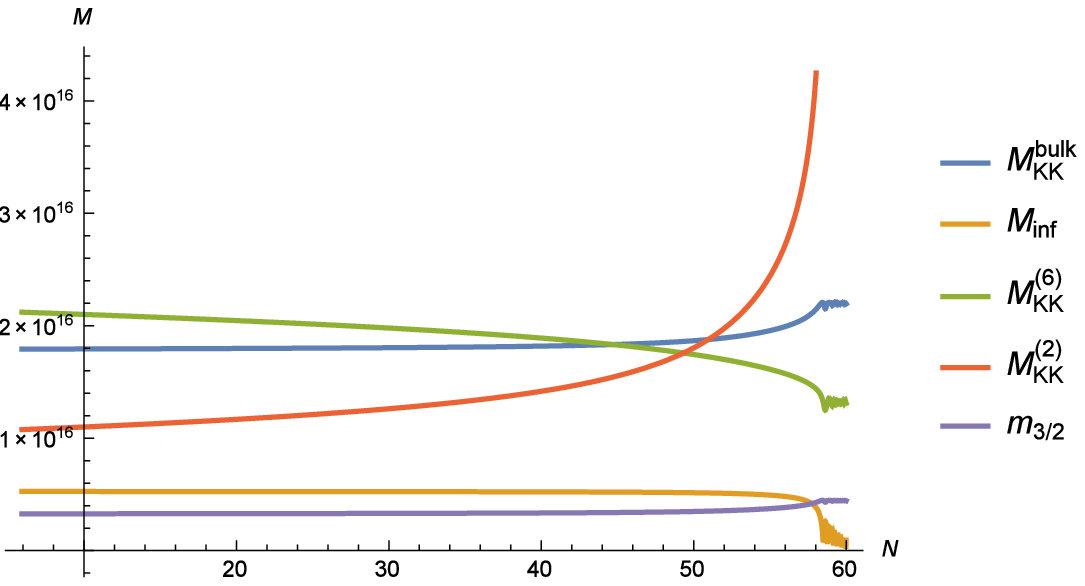}
\caption{Evolution of all KK masses (with $M_\KK^{(4)}=M_\KK^{(2)}$), the inflationary scale $M_{\rm inf} = V^{1/4}$ and the gravitino mass $m_{3/2}$ in GeV units from horizon exit to the final settling into the global minimum.}
\label{FigKK}
\end{center}
\end{figure}

Finally Fig. \ref{FigKK} presents a plot with the cosmological evolution of all KK mass scales, the inflationary scale $M_{\rm inf} = V^{1/4}$ and the gravitino mass $m_{3/2}$ from horizon exit to the final settling into the global minimum. The fact that $M_{\rm inf}$ remains always below all the KK scales, ensures that the Hubble scale during inflation $H = \frac{M_{\rm inf}}{\sqrt{3}}\left(\frac{M_{\rm inf}}{M_p}\right) < M_{\rm inf}$ is also always below each KK scale. The gravitino mass also remains always smaller than $M_\KK^{(i)}$ $\forall i$. This guarantees that the 4D effective field theory is under control. In particular, $M_\KK^{(2)}$, $M_\KK^{(6)}$ and the inflationary scale evolve from $M_\KK^{(2)}(N_*)\simeq 1.1\cdot 10^{16}$ GeV, $M_\KK^{(6)}(N_*)\simeq 2.1\cdot 10^{16}$ GeV and $M_{\rm inf}(N_*)\simeq 5.3\cdot 10^{15}$ GeV at horizon exit to $M_\KK^{(2)}(N=60)\simeq 6.2\cdot 10^{16}$ GeV, $M_\KK^{(6)}(N=60)\simeq 1.3\cdot 10^{16}$ GeV and $M_{\rm inf}(N=60)\simeq 9.3\cdot 10^{14}$ GeV around the final minimum. On the other hand the other scales remain approximately constant during the whole inflationary evolution around: $H\simeq 5\cdot 10^{12}\,{\rm GeV} < m_{3/2}\simeq 4\cdot 10^{15}\,{\rm GeV} < M_\KK^{\rm bulk}\simeq 2\cdot 10^{16}\,{\rm GeV}$.

\subsubsection{$|\lambda| = 10^{-3}$ and negligible amplitude of the density perturbations}
\label{NoCobeCase}

We shall now relax the condition of generating the correct amplitude of the density perturbations from the inflaton quantum fluctuations. As explained above, the right COBE value of the amplitude of the power spectrum could instead be reproduced in a non-standard way by a curvaton-like mechanism involving the quantum fluctuations of the two light bulk axions \cite{curvaton}. In this case we can focus on $\vo\sim 5\cdot 10^3$, $W_0\sim\mc{O}(1)$, $\lambda\sim 10^{-3}$ and relatively small values of the coefficients of the winding loop corrections which generate the plateau, so that all the remaining four conditions listed at the beginning of Sec. \ref{Multi} are fully satisfied. 

\begin{figure}[h!]
\begin{center}
\includegraphics[width=0.8\textwidth, angle=0]{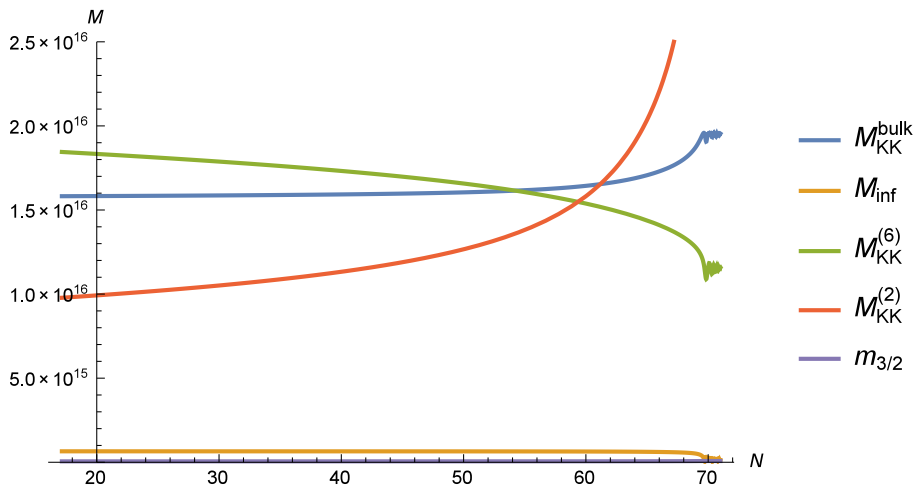}
\caption{Evolution of all KK masses (with $M_\KK^{(4)}=M_\KK^{(2)}$), the inflationary scale $M_{\rm inf} = V^{1/4}$ and the gravitino mass $m_{3/2}$ in GeV units from horizon exit to the final settling into the global minimum.}
\label{FigKK2}
\end{center}
\end{figure}

We shall set $\alpha=1$ and perform the following choice of the underlying parameters:
\bea
A_s &=& 1\cdot 10^4 \qquad \chi=-188 \quad\Rightarrow \quad \zeta = -\frac{\zeta(3) \chi(X)}{2 (2\pi)^3} = 0.455 \quad W_0 = 1 \quad g_s = 0.25 \nn \\
C_1^\W &=& C_2^\W = 0.05 \quad C_3^\W = -10^{-4} \quad C_4^\W = -0.1 \quad C_5^\W=C_6^\W = -0.05 \quad \lambda = -0.001\,, \nn
\eea
which yield a global Minkowski minimum inside the K\"ahler cone at:
\be
\langle \vo\rangle = 3220.899\,,\qquad \langle\tau_7\rangle = 6.403\,\qquad \langle \tau_1\rangle = 3.179\quad\text{for}\quad 
\delta_{\rm up} = 1.76588\cdot 10^{-7}\,. \nn
\ee
The initial conditions for the inflationary evolution are again derived in the same way: the fibre modulus $\tau_7$ is shifted away from its minimum at $\tau_7 (N=0) = \langle\tau_7\rangle + 2450$ and the other two directions $\langle \vo\rangle(\tau_7)$ and $\langle\tau_1\rangle(\tau_7)$ are set at the new minimum:
\be
\vo(0) = \langle \vo\rangle (\tau_7(0)) = 4436.094\,,\quad \tau_7(0) = 2456.403\,,
\quad\tau_1(0) = \langle \tau_1\rangle(\tau_7(0)) = 3.228 \,. \nn
\ee
Notice that these initial conditions are inside the K\"ahler cone since $\tau_7(0) < \frac{\vo(0)}{\sqrt{\tau_1(0)}} = 2468.95$, which satisfies the upper bound in (\ref{KCA}). Focusing again on vanishing initial velocities for all scalar fields, i.e. $\vo^\prime (0) = \tau_7^\prime (0) = \tau_1^\prime(0) = 0$, we worked out the cosmological evolution of each scalar field as a function of $N$ by solving the system of equations of motion (\ref{NLEq}). Looking for a slow-roll region in field space where the generalised $\epsilon$-parameter (\ref{GenEps}) is much smaller than unity, we found that $\epsilon\ll 1$ during the first $69$ efoldings and then quickly increases and reaches $\epsilon = 1$ at $N=69.15$ where inflation ends. The point of horizon exit corresponding to a physical number of efoldings of inflation $N_e = 52$ is localised at $N_* = 17.15$ where $\epsilon (N_*) = 1.36\cdot 10^{-4}$. The main cosmological observables at horizon exit take the following values:
\bea
n_s (N_*) &=& 1 + \left.\frac{d}{dN}\,\ln P(N)\right|_{N=N_*} = 0.9676\,, \qquad r = 16 \epsilon(N_*) = 0.0022\,, \nn \\
\sqrt{P(N_*)} &=& \frac{1}{10\pi}\sqrt{\frac{2\,V(N_*)}{3\,\epsilon(N_*)}} = 1.64 \cdot 10^{-7}\,. \nn
\eea
The scalar spectral index $n_s$ and the tensor-to-scalar ratio $r$ are in good agreement with Planck data \cite{Ade:2015lrj, Ade:2015xua} while the amplitude of the scalar power spectrum, as expected, is much smaller than the reference COBE value $\sqrt{P_{\scriptscriptstyle{\rm COBE}}}\simeq 2\cdot 10^{-5}$. As can be seen from Fig. \ref{FigKK2}, in this case the low-energy 4D effective field theory is fully under control since throughout all the inflationary evolution all KK scales are much higher than both the gravitino mass and the inflationary scale (and so also the Hubble scale).

In particular, $M_\KK^{(2)}$, $M_\KK^{(6)}$ and the inflationary scale evolve from $M_\KK^{(2)}(N_*)\simeq 9.8\cdot 10^{15}$ GeV, $M_\KK^{(6)}(N_*)\simeq 1.8\cdot 10^{16}$ GeV and $M_{\rm inf}(N_*)\simeq 6.5\cdot 10^{14}$ GeV at horizon exit to $M_\KK^{(2)}(N=70)\simeq 5.5\cdot 10^{16}$ GeV, $M_\KK^{(6)}(N=70)\simeq 1.2\cdot 10^{16}$ GeV and $M_{\rm inf}(N=70)\simeq 1.4\cdot 10^{14}$ GeV around the final minimum. On the other hand the other scales remain approximately constant during the whole inflationary evolution around: $H\simeq 8\cdot 10^{11}\,{\rm GeV} < m_{3/2}\simeq 6\cdot 10^{13}\,{\rm GeV} < M_\KK^{\rm bulk}\simeq 2\cdot 10^{16}\,{\rm GeV}$.

\section{Conclusions}
\label{Concl}

The study of large field inflationary models is particularly interesting from both a phenomenological and a theoretical point of view. In fact, from one side the next generation of CMB observations will be able to test values of the tensor-to-scalar ratio in the window $0.001\lesssim r\lesssim 0.01$, while on the other hand trans-Planckian inflaton excursions need a symmetry mechanism to trust the effective field theory approach. 

Natural inflaton candidates from type IIB string compactifications are K\"ahler moduli which enjoy non-compact shift-symmetries \cite{Burgess:2014tja}. In particular, fibre inflation models provide promising plateau-like potentials which seem to fit Planck data rather well and lead to the prediction of observable tensor modes \cite{Cicoli:2008gp, Broy:2015zba, Cicoli:2016chb, Burgess:2016owb}. These inflationary models are built within LVS moduli stabilisation scenarios and can be globally embedded in K3-fibred Calabi-Yau manifolds \cite{Cicoli:2016xae}. 

In this paper we extended previous work by constructing the first explicit realisations of fibre inflation models in concrete type IIB Calabi-Yau orientifolds with consistent brane setups, full closed string moduli fixing and chiral matter on D7-branes. The underlying compactification manifold features $h^{1,1}=4$ K\"ahler moduli which after D-term stabilisation get effectively reduced to the standard $3$ moduli of fibre inflation models. 

We found that the inflationary dynamics is strongly constrained by the K\"ahler cone conditions which never allow for enough efoldings of inflation if the internal volume is of order $\vo \sim 10^3$. For larger values of the Calabi-Yau volume of order $\vo\sim 10^4$, the K\"ahler cone becomes large enough for the inflaton to drive $N_e\simeq 52$ efoldings, as required by an estimate of the post-inflationary evolution. However such a large value of $\vo$ tends to suppress the amplitude of the density perturbations below the reference COBE value. This can be avoided by considering large values of either the coefficients of the winding loops which generate the plateau, or the flux superpotential $W_0$. Let us stress that in the string landscape this choice is guaranteed to be possible by the fact that both of these microscopic parameters are flux-dependent.

However, as shown in Sec. \ref{BadCase}, large values of the coefficients of the winding $g_s$ corrections make the Hubble scale during inflation of the same order of magnitude of the mass of the volume mode. This could either cause a large shift of the original LVS minimum or even a problem for the stability of the inflationary direction against orthogonal runaway directions. A definite answer to this issue hence requires a proper multi-field analysis even if the two-field study of \cite{Cicoli:2008gp} revealed that the inflationary motion is still mostly single-field. 

On the other hand, if the flux superpotential is of order $W_0\sim 100$, the gravitino mass can become too close to some KK scale in the model, destroying the 4D effective field theory. Moreover, $F^4$ terms are proportional to $|\lambda| W_0^4$. Thus if $W_0$ is large, these higher derivative effects can spoil the flatness of the inflationary potential before achieving enough efoldings of inflation if $|\lambda|$ is not small enough. Hence in Sec. \ref{GoodCase} we presented a model with $W_0\sim 100$ and a very small value of $|\lambda|$ of order $|\lambda|=10^{-7}$ which makes the $F^4$ terms harmless. The gravitino mass also turns out to be slightly smaller than any KK scale throughout the whole inflationary dynamics. 

Due to the fact that in the single-field case not all our approximations are fully under control, in Sec. \ref{Multi} we performed a complete numerical analysis of the $3$-field cosmological evolution. For $W_0\sim 100$ and $|\lambda|=10^{-6}$, the multi-field analysis of Sec. \ref{GreatCase} revealed that the accuracy of our approximations improves. In particular, the allowed number of efoldings of inflation increases due to the extra motion along the volume and blow-up directions. Hence inflation can successfully work also for smaller values of $\vo$ which cause a smaller K\"ahler cone for the fibre modulus. This, in turn, requires smaller values of $W_0$ to match the COBE normalisation of the density perturbations, which enlarges the hierarchy between $m_{3/2}$ and the KK scales in the model. 

We point out however that some of the underlying parameters are not flux-dependent, and so are not tunable in the string landscape. Two examples of this kind of parameters are the effective Euler number $\chi_{\rm eff}$ which controls the strength of $\mc{O}(\alpha'^3)$ corrections due to O7-planes \cite{Minasian:2015bxa} and the combinatorial factor $\lambda$ which is the coefficient of $\mc{O}(\alpha'^3)$ higher derivatives \cite{Ciupke:2015msa}. Both of these microscopic parameters have not been computed in full detail yet, even if $\lambda$ has been estimated to be of order $10^{-3}$ \cite{Grimm:2017okk}. Hence in Sec. \ref{NoCobeCase} we also presented a case with $|\lambda|=0.001$ where it is hard to obtain enough efoldings inside the K\"ahler cone and generate, at the same time, the correct amplitude of the density perturbations in a framework where all the approximations are fully under control. Hence we chose the flux superpotential so that the contribution of the inflaton quantum fluctuations to the scalar power spectrum is negligible. In this case a viable inflationary phenomenology can therefore be achieved only in the presence of a non-standard mechanism for the generation of the density perturbations. A promising case could be the curvaton scenario where the initial isocurvature fluctuations could be produced by the quantum oscillations of the two light bulk closed string axions \cite{curvaton}. 

Besides a complete computation of the exact value of both $\chi_{\rm eff}$ and $\lambda$, and the detailed derivation of a curvaton-like mechanism, there are several other important open issues for future work. A crucial one is a better determination of the actual Calabi-Yau K\"ahler cone since the one that we used is just an approximation inherited from the Mori cone of the ambient toric variety. It would also be interesting to develop a more systematic study of the constraints that the K\"ahler cone sets on the inflationary dynamics by performing a complete scan over all $h^{1,1}=3$ and $h^{1,1}=4$ K3 fibred CY threefolds with at least a del Pezzo divisor \cite{Cicoli:2018tcq}. Moreover our chiral global models still lack an explicit implementation of a mechanism responsible for the realisation of a dS vacuum. Finally, the study of the post-inflationary cosmological evolution of our universe is of primary importance in order to discriminate among different models that feature the same inflationary predictions of fibre inflation models. A first step forward towards understanding (p-)reheating has been taken in \cite{Cabella:2017zsa, Antusch:2017flz}. A full understanding of this mechanism requires further investigation of the underlying microscopic dynamics.

\acknowledgments

We would like to thank Ross Altman, Christoph Mayrhofer, Fernando Quevedo, Raffaele Savelli and Roberto Valandro for useful discussions. The work of M.C. is supported by the Programme Rita Levi Montalcini for young researcher of the Italian Ministry of Research. F. M. is funded in part by the European Research Council as Starting Grant 307605-SUSYBREAKING. P. S. is grateful to the Bologna INFN division for hospitality when most of this work was carried out.

\appendix

\section{Another chiral global example}
\label{ExB}

\subsection{Toric data}

Let us consider the following toric data for a CY threefold with $h^{1,1}=4$ which is a K3-fibration over a $\mP^1$ base along with a so-called `small' divisor:

\begin{table}[H]
  \centering
 \begin{tabular}{|c|cccccccc|}
\hline
     & $x_1$  & $x_2$  & $x_3$  & $x_4$  & $x_5$ & $x_6$  & $x_7$ & $x_8$   \\
    \hline
 8 & 0 & 0  & 0 & 1 & 1 & 1 & 1  & 4   \\
 6 & 0 & 0 &  1 & 0 & 1 & 0 & 1  & 3   \\
 6 & 0 & 1  & 0 & 0 & 0 & 1 & 1  & 3   \\   
 4 & 1 & 0  & 0 & 0 & 0 & 0 & 1  & 2   \\   
 \hline
    & dP$_5$ &NdP$_{11}$  &  NdP$_{11}$ & dP$_7$ & K3 & K3 & SD1 & SD2 \\
    \hline
  \end{tabular}
 \end{table}
 \noindent
with Hodge numbers $(h^{2,1}, h^{1,1}) = (106, 4)$ and Euler number $\chi=-204$.
The Stanley-Reisner ideal is:
\be
{\rm SR} = \{x_1 x_4, \, x_1 x_7, x_3 x_5, \, x_4 x_5, \, x_2 x_3 x_7, \, x_2 x_6 x_8,\, x_4 x_6 x_8\}\,. \nn
\ee
This corresponds to the CY threefold used in \cite{Cicoli:2011qg} to build global models with chiral matter on D7-branes and K\"ahler moduli stabilisation but without any inflationary dynamics. A detailed divisor analysis using \texttt{cohomCalg} \cite{Blumenhagen:2010pv, Blumenhagen:2011xn} shows that the divisor $D_4$ is a del Pezzo dP$_7$ which we find to be shrinkable after investigating the CY volume form. Further, each of the divisors $\{D_2, \, D_3\}$ are non-diagonal del Pezzo surfaces and $\{D_5, \, D_6\}$ are two K3 surfaces while the divisors $\{D_7, \, D_8\}$ are two `special deformation' divisors with Hodge diamond:
\bea
{\rm SD1} \equiv
\begin{tabular}{ccccc}
    & & 1 & & \\
   & 0 & & 0 & \\
  3 & & 38 & & 3 \\
   & 0 & & 0 & \\
    & & 1 & & \\
  \end{tabular} \qquad \qquad\text{and}\qquad\qquad
		{\rm SD2} \equiv
\begin{tabular}{ccccc}
    & & 1 & & \\
   & 0 & & 0 & \\
  25 & & 172 & & 25 \\
   & 0 & & 0 & \\
    & & 1 & & \\
  \end{tabular} \nn
	\eea
The intersection form in the basis of smooth divisors $\{D_1, D_4, D_5, D_6\}$ can be written as:
\be
I_3= 2 \, D_1 \, D_5 \, D_6 - 2\, D_1^2\,D_5 - 2\, D_1^2\, D_6+ 2\, D_4^3+  4 \, D_1^3\,.
\label{I3B}
\ee
Writing the K\"ahler form in the above basis of divisors as $J=t_1\, D_1 + t_4\, D_4 + t_5 \, D_5+ t_6\, D_6$  and using the intersection
polynomial (\ref{I3B}), the CY overall volume takes the form:
\be
\vo = 2\, t_1 \, t_5\, t_6  - t_1^2\, t_5 - t_1^2\, t_6 + \frac{t_4^3}{3} +\frac23\, t_1^3\,.
\ee
In order to express $\vo$ in terms of four-cycle moduli, we need to know the K\"ahler cone conditions which can be determined from the following K\"ahler cone generators:
\be
K_1 = D_1 + D_5 +D_6, \quad K_2 = D_1 - D_4 + D_5 +D_6, \quad K_3 = D_5, \quad K_4 = D_6\,.
\ee
Expanding the K\"ahler form $J$ in these K\"ahler cone generators as $J=\sum_{i=1}^4 r_i\, K_i$ results in the following conditions for the two-cycle moduli:
\be
r_1 = t_1 + t_4 >0\,, \qquad r_2 = - t_4 > 0\,, \qquad  r_3 = t_5- t_1 >0\,,  \qquad r_4 = t_6 - t_1>0\,.
\label{KCone}
\ee
Using the four-cycle moduli, $\tau_i = \partial_{t_i}\vo$, given by:
\be
\tau_1 = 2\,(t_5- t_1)(t_6-t_1), \quad \tau_4 =  t_4^2, \quad \tau_5 = t_1(2\,t_6- t_1), \quad \tau_6 = t_1(2\,t_5-t_1)\,,
\ee
the overall volume can be rewritten as:
\be
\vo = \frac13\left(t_1\tau_1 + t_5\tau_5+ t_6\tau_6- \tau_4^{3/2}\right)\,.
\label{VolB}
\ee
The second Chern class of the CY threefold $X$ is instead given by:
\be
c_2(X) = 2 \, D_6 \, D_8 + 8\, D_7\, D_8 - 2 \, D_6^2 - 4\, D_6\, D_7 - 12\, D_7^2\,,
\ee
which results in the following values of the topological quantities $\Pi_i$'s:
\be
\Pi_1 = 4, \quad \Pi_2 = \Pi_3 =16, \quad \Pi_4 =8, \quad \Pi_5 =\Pi_6 =24, \quad \Pi_7 =44, \quad\Pi_8 =136\,. \nn
\ee
The intersection curves between two coordinate divisors are given in Tab. \ref{TabIntB} while their volumes are listed in Tab. \ref{IntersectB}.

\begin{table}[H]
  \centering
 \begin{tabular}{|c||c|c|c|c|c|c|c|c|}
\hline
 & $D_1$ & $D_2$ & $D_3$ & $D_4$ & $D_5$ & $D_6$ & $D_7$ & $D_8$   \\
 \hline
 $D_1$ & $\mc{C}_5$  & $\mP^1$             & $\mP^1$             & $\emptyset$ & $\mP^1$     & $\mP^1$     & $\emptyset$   & $\T^2$ \\
 $D_2$ & $\mP^1$     & $\mP^1\sqcup \mP^1$ & $\mP^1\sqcup \mP^1$ & $\T^2$      & $\T^2$      & $\emptyset$ & $\mP^1$       & $\mc{C}_3$ \\
 $D_3$ & $\mP^1$     & $\mP^1\sqcup \mP^1$ & $\mP^1\sqcup \mP^1$ & $\T^2$      & $\emptyset$ & $\T^2$      & $\mP^1$       & $\mc{C}_3$ \\
 $D_4$ & $\emptyset$ & $\T^2$              & $\T^2$              & $\mc{C}_3$  & $\emptyset$ & $\emptyset$ & $\T^2$        & $\mc{C}_3$ \\
 $D_5$ & $\mP^1$     & $\T^2$              & $\emptyset$         & $\emptyset$ & $\emptyset$ & $\T^2$      & $\mc{C}_2$    & $\mc{C}_9$ \\
 $D_6$ & $\mP^1$     & $\emptyset$         & $\T^2$              & $\emptyset$ & $\T^2$      & $\emptyset$ & $\mc{C}_2$    & $\mc{C}_9$ \\
 $D_7$ & $\emptyset$ & $\mP^1$             & $\mP^1$             & $\T^2$      & $\mc{C}_2$  & $\mc{C}_2$  & $\mc{C}_3$    & $\mc{C}_{19}$ \\
 $D_8$ & $\T^2$      & $\mc{C}_3$          & $\mc{C}_3$          & $\mc{C}_3$  & $\mc{C}_9$  & $\mc{C}_9$  & $\mc{C}_{19}$ & $\mc{C}_{89}$ \\
 \hline
 \end{tabular}
 \caption{Intersection curves of two coordinate divisors. Here $\mc{C}_g$ denotes a curve with Hodge numbers $h^{0,0} = 1$ and $h^{1,0} = g$.}
 \label{TabIntB}
 \end{table}
 
\begin{table}[htb]
\centering
\resizebox{\textwidth}{!}{\begin{tabular}{|c||c|c|c|c|c|c|c|c|}
\hline 
{} & $D_1$  &$  D_2$   &$  D_3  $ & $ D_4 $  & $ D_5 $ & $ D_6 $  &$  D_7$    & $ D_8  $  \\
\hline
$ D_1 $ & $4t_1-2(t_5+t_6)$ & $2(t_5-t_1)$ & $2(t_6-t_1)$ & $0$     & $2(t_6-t_1)$ & $2(t_5-t_1)$ & $0$          & $2(t_5+t_6)-4t_1$  \\
$ D_2 $ & $2(t_5-t_1)$              & $2t_4$       & $2(t_1+t_4)$ & $-2t_4$ & $2t_1$       & $0$          & $2(t_5+t_4)$ & $2(t_1+2t_4+2t_5)$  \\
$ D_3 $ & $2(t_6-t_1)$              & $2(t_1+t_4)$ & $2t_4$       & $-2t_4$ & $0$          & $2t_1$       & $2(t_6+t_4)$ & $2(t_1+2t_4+2t_6)$ \\
$ D_4 $ & $0$                       & $-2t_4$      & $-2t_4$             & $2t_4$ & $0$                 & $0$                & $-2t_4$       & $-4t_4$ \\
$ D_5 $ & $2(t_6-t_1)$              & $2t_1$       & $0$                & $0$     & $0$            & $2t_1$   & $2t_6$ & $2(2t_6+t_1)$ \\
$ D_6 $ & $2(t_5-t_1)$              & $0$            & $2t_1$       & $0$     & $2t_1$        & $0$            & $2t_5$ & $2(2t_5+t_1)$ \\
$ D_7 $ & $0$                       & $2(t_5+t_4)$   & $2(t_6+t_4)$     & $-2t_4$  & $2t_6$    & $2t_5$    & $2(t_4+t_5+t_6)$ & $4t_4+6(t_5+t_6)$ \\
$ D_8 $ & $2(t_5+t_6)-4t_1$ & $2(t_1+2t_4+2t_5)$ & $2(t_1+2t_4+2t_6)$ & $-4t_4$ & $2(2t_6+t_1)$ & $2(2t_5+t_1)$ & $4t_4+6(t_5+t_6)$ &$4[t_1+2t_4+4(t_5+t_6)]$\\
\hline
\end{tabular}}
\caption{Volumes of intersection curves between two coordinate divisors.}
\label{IntersectB}
\end{table}

\subsection{Orientifold involution}

We focus on orientifold involutions of the form $\sigma: x_i \to - x_i$ with $i = 1, ...,8$  which feature an O7-plane on $D_i$ and O3-planes at the fixed points listed in Tab. \ref{FixedPointsB}. The effective non-trivial fixed point set in Tab. \ref{FixedPointsB} has been obtained after taking care of
the SR ideal symmetry. Moreover, the total number of O3-planes $N_{\rm O3}$ is obtained from the triple intersections restricted to the CY hypersurface, while the effective Euler number $\chi_{\rm eff}$ has been computed as:
\be
\chi_{\rm eff} = \chi(X) +2\int_X [{\rm O7}] \wedge [{\rm O7}] \wedge [{\rm O7}]\,.
\ee
In what follows we shall focus on the orientifold involution $\sigma: x_7\rightarrow-x_7$ which features two non-intersecting O7-planes located in $D_1$ and $D_7$ and two O3-planes at $\{D_2D_3D_4\}$\,.

\begin{table}[H]
 \hskip-0cm
\begin{tabular}{|c|c|c|c|c|c|}
\hline
&  &  &  &  &  \\
$\sigma$ & O7  & O3  & $N_{\rm O3}$  & $\chi({\rm O7})$  & $\chi_{\rm eff}$       \\
&  &  &  &  &  \\
\hline
\hline
$x_1 \to -x_1$ &  $D_1\sqcup D_7$ & $\{D_2 D_3 D_4 \}$ & 2 & 54 & -192  \\
$x_2 \to -x_2$ &  $D_2$ & $\{{D_1 D_6 D_8}, {D_3 D_4 D_7}, {D_6 D_7 D_8}\}$ & \{2, 2, 6\} & 14 & -208 \\
$x_3 \to -x_3$ &  $D_3$ & $\{{D_1 D_5 D_8}, {D_2 D_4 D_7}, {D_5 D_7 D_8}\}$ & \{2, 2, 6\} & 14 & -208 \\
$x_4 \to -x_4$ &  $D_4$ & $\{{D_1 D_2 D_3}, {D_1 D_5 D_6}, $ & \{2, 2, 4, 4, 2 \} & 10 & -200  \\
 & & ${D_2 D_5 D_8}, {D_3 D_6 D_8}, {D_5 D_6 D_7} \}$ & & & \\
$x_5 \to -x_5$ &  $D_5$ & $\{{D_1 D_3 D_8}, {D_3  D_7 D_8}, {D_2 D_4 D_8} \}$ & \{2, 2, 4\} & 24 & -204  \\
$x_6 \to -x_6$ &  $D_6$ & $\{{D_1 D_2 D_8}, {D_2  D_7 D_8}, {D_3 D_4 D_8} \}$ & \{2, 2, 4\} & 24 & -204  \\
$x_7 \to -x_7$ &  $D_1\sqcup D_7$ & $\{D_2 D_3 D_4 \}$ & 2 & 54 & -192  \\
$x_8 \to -x_8$ &  $D_8$ & $\emptyset$ & 0 & 224 & -28  \\
\hline
\end{tabular}
\caption{Fixed point set for the involutions which are reflections of the eight coordinates $x_i$ with $i=1,...,8$.} 
\label{FixedPointsB} 
\end{table}

\subsection{Brane setup}

If the D7-tadpole cancellation condition is satisfied by placing four D7-branes on top of the O7-plane, the string loop corrections to the scalar potential involve only KK effects since winding contributions are absent due to the absence of any intersection between D7-branes and/or O7-planes. Thus loop effects are too simple to generate a viable inflationary plateau. We shall therefore focus on a slightly more complicate D7-brane setup which gives rise also to winding loop effects. This can be achieved by placing D7-branes not entirely on top of the O7-plane as follows:  
\be
8[{\rm O7}] \equiv 8([D_1]+[D_7]) =8 \left(2[D_1] + [D_2]+[D_5] \right)\,.
\ee
This brane setup involves three stacks of D7-branes wrapped around the divisors $D_1$, $D_2$ and $D_5$. Moreover, the condition for D3-tadpole cancellation becomes:
\be
N_{\rm D3} + \frac{N_{\rm flux}}{2} + N_{\rm gauge} = \frac{N_{\rm O3}}{4} + \frac{\chi({\rm O7})}{12} + \sum_a\, \frac{N_a \left(\chi(D_a) + \chi(D_a^\prime) \right) }{48}  = 14 \,, \nn
\ee
showing that there is space for turning on both gauge and background three-form fluxes for complex structure and dilaton stabilisation.

\subsection{Gauge fluxes}

In order to obtain a chiral visible sector on the D7-brane stacks wrapping $D_1$, $D_2$ and $D_5$ we need to turn on worldvolume gauge fluxes of the form:
\be
\F_i = \sum_{j=1}^{h^{1,1}} f_{ij}\hat{D}_j + \frac12 \hat{D}_i - \iota_{D_i}^*B \quad\text{with}\quad f_{ij}\in \mathbb{Z} \quad\text{and}\quad i=1,2,5\,,
\ee
where the half-integer contribution is due to Freed-Witten anomaly cancellation \cite{Minasian:1997mm, Freed:1999vc}. 

However we want to generate just one moduli-dependent Fayet-Iliopoulos term in order to fix only one K\"ahler modulus via D-term stabilisation. In fact, if the number of FI-terms is larger than one, there is no light K\"ahler modulus which can play the r\^ole of the inflaton. Moreover we wrap a D3-brane instanton on the rigid divisor $D_4$ in order to generate a non-perturbative contribution to the superpotential which is crucial for LVS moduli stabilisation. In order to cancel the Freed-Witten anomaly, the D3-instanton has to support a half-integer flux, and so the general expression of the total gauge flux on $D_4$ becomes:
\be
\F_4 = \sum_{j=1}^{h^{1,1}} f_{4j}\hat{D}_j + \frac12 \hat{D}_4 - \iota_{D_i}^*B \quad\text{with}\quad f_{4j}\in \mathbb{Z}\,.
\ee
However a non-vanishing $\F_4$ would not be gauge invariant, and so would prevent a non-perturbative contribution to the superpotential. We need therefore to check if it is possible to perform an appropriate choice of $B$-field which can simultaneously set $\F_1=\F_2=0$ (we choose to have a non-vanishing gauge flux only on $D_5$ to have just one moduli-dependent FI-term) and $\F_4=0$. If we set:
\be
B = \frac12 \hat{D}_1 + \frac12 \hat{D}_2 + \frac12 \hat{D}_4\,,
\label{Bfield}
\ee
the condition $\F_1=\F_2=\F_4=0$ reduces to the requirement that the following forms are integer:
\be
\iota_{D_1}^* \left(\frac12 \hat{D}_2 + \frac12 \hat{D}_4 \right) \qquad 
\iota_{D_2}^* \left(\frac12 \hat{D}_1 + \frac12 \hat{D}_4 \right) \qquad
\iota_{D_4}^* \left(\frac12 \hat{D}_1 + \frac12 \hat{D}_2 \right)\,,
\label{Pullbacks}
\ee
since in this case the integer flux quanta $f_{ij}$ can always be adjusted to yield vanishing gauge fluxes. 
Taking an arbitrary integer form $A\in H^2(\mathbb{Z},X)$ which can be expanded as $A=a_j\hat{D}_j$ with $a_j \in \mathbb{Z}$, the pullbacks in (\ref{Pullbacks}) give rise to integer forms if:
\bea
b_1 &\equiv& \int_X \left(\frac12 \hat{D}_2 + \frac12 \hat{D}_4 \right) \wedge \hat{D}_1 \wedge A \in \mathbb{Z} \nn \\
b_2 &\equiv& \int_X \left(\frac12 \hat{D}_1 + \frac12 \hat{D}_4 \right) \wedge \hat{D}_2 \wedge A \in \mathbb{Z} \nn \\
b_4 &\equiv& \int_X \left(\frac12 \hat{D}_1 + \frac12 \hat{D}_2 \right) \wedge \hat{D}_4 \wedge A \in \mathbb{Z} \nn
\eea
Using the intersection polynomial (\ref{I3B}) we find $b_1= a_5-a_1 \in \mathbb{Z}$, $b_2 = b_1 -a_4 \in \mathbb{Z}$ and $b_4 = -a_4 \in \mathbb{Z}$, 
showing how the choice of $B$-field in (\ref{Bfield}) can indeed allow for $\F_1=\F_2=\F_4=0$. The only non-zero gauge flux is $\F_5$ which does not feature any half-integer contribution since $c_1(D_5)=0$ given that $D_5$ is a K3 surface. Given that all the intersection numbers are even, the pullback of the $B$-field on $D_5$ does also not generate an half-integer flux. We shall therefore consider a non-vanishing gauge flux on the worldvolume of $D_5$ of the form:
\be
\F_5 = \sum_{j=1}^{h^{1,1}} f_{5j}\hat{D}_j \quad\text{with}\quad f_{5j}\in \mathbb{Z}\,.
\ee

\subsection{FI-term and chirality}

Given that the divisor $D_5$ is transversely invariant under the orientifold involution and it is wrapped by four D7-branes, it supports an $Sp(8)$ gauge group which is broken down to $U(4)=SU(4)\times U(1)$ by a non-zero flux $\F_5$ along the diagonal $U(1)$. This non-trivial gauge flux $\F_5$ induces also a $U(1)$-charge $q_{i5}$ for the $i$-th K\"ahler modulus of the form:
\be
q_{i5} = \int_X \hat{D}_i \wedge \hat{D}_5 \wedge \F_5 \,.
\ee
Thus $\F_5\neq 0$ yields:
\be
q_{15} = 2 (f_{56} - f_{51}) \qquad q_{45} = q_{55} = 0 \qquad q_{65} = 2 f_{51}\,,
\ee
together with a flux-dependent correction to the gauge kinetic function which looks like:
\be
{\rm Re}(f_5)  = \alpha_5^{-1}=\frac{4\pi}{g_5^2}=\tau_5-h(\F_5) {\rm Re}(S)\,,
\ee
where: 
\be
h(\F_5) =\frac12 \int_X \hat{D}_5 \wedge \F_5 \wedge \F_5 =\frac12\left(f_{51} q_{15} + f_{56} q_{65}\right)\,.
\ee
Moreover a non-vanishing gauge flux $\F_5$ induces a moduli-dependent FI-term of the form:
\be
\xi =\frac{1}{4\pi\vo}\int_X \hat{D}_5\wedge J\wedge\F_5=\frac{1}{4\pi\vo}\sum_{j=1}^{h^{1,1}} q_{j5}\,t_j
=\frac{1}{4\pi\vo} \left(q_{15}\,t_1+q_{65}\,t_6\right)\,.
\ee
For vanishing open string VEVs (induced for example by non-tachyonic scalar masses), a leading-order supersymmetric stabilisation requires $\xi=0$ which implies:
\be
t_6 = - \frac{q_{15}}{q_{65}}\,t_1= \left(1-\frac{f_{56}}{f_{51}}\right) \,t_1 \equiv \alpha\,t_1\,.
\label{Dfix}
\ee
This $U(1)$ factor becomes massive via the St\"uckelberg mechanism and develops an $\mc{O}(M_s)$ mass by eating up a linear combination of an open and a closed string axion which is mostly given by the open string mode. 

Besides breaking the worldvolume gauge group and inducing moduli-dependent FI-terms, non-trivial gauge fluxes on D7-branes generate also 4D chiral modes. In fact, open strings stretching between the D7-branes on $D_5$ and the O7-planes or the image branes give rise to the following zero-modes in the symmetric and antisymmetric representations of $U(4)$:
\bea
I_5^{(S)} &=& - \frac12 \int_X \hat{D}_5 \wedge [{\rm O7}] \wedge \F_5 - \int_X \hat{D}_5 \wedge \hat{D}_5 \wedge \F_5 
= -\left(q_{15}+\frac{q_{65}}{2}\right), \\
I_5^{(A)} &=& \frac12 \int_X \hat{D}_5 \wedge [{\rm O7}] \wedge \F_5 - \int_X \hat{D}_5 \wedge \hat{D}_5 \wedge \F_5 = - I_5^{(S)}\,.
\eea
Due to the absence of worldvolume fluxes on the D7-branes wrapped around $D_1$ and $D_2$, the gauge groups supported by these two D7-stacks are respectively $SO(16)$ (since $D_1$ is an O7-locus) and $Sp(8)$ (since $D_2$ is transversely invariant) which are both unbroken. Thus open strings stretched between the D7-branes on $D_5$ and $D_1$ (or its image brane) give rise to chiral zero-modes in the bi-fundamental representation ($4$,$16$) of $U(4)$ and $SO(16)$ whose number is:
\be
I_{51}=\int_X \hat{D}_5 \wedge \hat{D}_1 \wedge \F_5 = q_{15}\,.
\ee
On the other hand, the number of 4D chiral zero-modes in the bi-fundamental representation ($4$,$8$) of $U(4)$ and $Sp(8)$ (corresponding to open strings stretching between the D7s on $D_5$ and $D_2$) is:
\be
I_{52}=\int_X \hat{D}_5 \wedge \hat{D}_2 \wedge \F_5 = q_{65}\,.
\ee
We need finally to check that there are no chiral intersections between the D7s on $D_5$ and the instanton on $D_4$ to make sure that the prefactor of the non-perturbative contribution to the superpotential is indeed non-zero. This is ensured by the fact that:
\be
I_{54}=\int_X \hat{D}_5 \wedge \hat{D}_4 \wedge \F_5 = 0 \,.
\ee

\subsection{Inflationary potential}

Using the D-term fixing relation (\ref{Dfix}), the K\"ahler cone conditions (\ref{KCone}) simplify to $t_5>t_1>-t_4>0$ and $\alpha>1$. 
Moreover the CY volume (\ref{VolB}) reduces to:
\be
\vo = \left(2\alpha-1\right) t_5 t_1^2 - \left(\alpha-\frac23 \right) t_1^3 + \frac{t_4^3}{3}=t_b \tau_f -\frac13 \,\tau_4^{3/2}\,.
\ee
Given that this form is linear in $t_5$, the effective CY volume after D-term stabilisation looks like a K3 fibre $\tau_f$ over a $\mP^1$ base $t_b$ whose volumes are given by:
\be
\tau_f = \tau_5 = \left(2\alpha-1\right)  t_1^2\qquad\text{and}\qquad t_b = t_5 - \frac{\left(\alpha-\frac23 \right)}{\left(2\alpha-1\right)}\, t_1\,.
\ee
Notice that the K\"ahler cone condition $t_5>t_1$ can be rewritten as:
\be
\tau_f < \sigma(\alpha) \,\vo^{2/3}\,,
\label{KCc}
\ee
where:
\be
\sigma(\alpha) \equiv \left(2\alpha-1\right)\left(\frac{3}{3\alpha-1}\right)^{2/3}\quad\text{with}\quad \alpha>1\,.
\ee
In terms of the canonically normalised inflaton shifted from its minimum, the condition (\ref{KCc}) reads:
\be
\tau_f = \langle\tau_f\rangle\,e^{2\hat\phi/\sqrt{3}} < \sigma\,\vo^{2/3}\qquad\Leftrightarrow\qquad
\hat\phi<\frac{\sqrt{3}}{2}\ln\left(\frac{\sigma}{\langle\tau_f\rangle}\,\vo^{2/3}\right).
\label{KCbound}
\ee
Let us now focus on the inflationary potential. The winding loop corrections look like (with $\kappa=g_s/(8\pi)$ for $e^{K_{\rm cs}}=1$):
\be
V_{g_s}^\W=-\kappa\frac{W_0^2}{\vo^3}\frac{C_\W}{\sqrt{\tau_f}}\,,
\ee
where:
\be
C_\W= \sqrt{2\alpha-1}\left(C_1^\W+ \frac{C_2^\W}{\alpha}\right)\,.
\ee
On the other hand, the KK loop corrections read (neglecting $\tau_4$-dependent terms which yield subdominant contributions):
\be
V_{g_s}^\KK = \kappa g_s^2\frac{W_0^2}{\vo^2}\sum_{i,j=1,5,6} C_i^\KK C_j^\KK K_{ij}\,.
\ee
After substituting $t_6=\alpha t_1$, we obtain:
\be
Z\vo^2\sum_{i,j} C_i^\KK C_j^\KK K_{ij} = a t_1^2 + C_5^2 t_5\left(t_5 - t_1\right) -(1-Z)\left(b t_1^2 + c t_1 t_5 + \frac{C_5^2}{2} t_5^2 \right),  \nn
\ee
where:
\bea
a &=& C_1 \left(C_1+ C_5 + C_6\right) + C_5 \left(C_6+\frac{C_5}{2}\right) +C_6^2 \left(\alpha^2-\alpha+\frac12\right) \nn \\
b &=& \alpha C_1 C_6 + \frac{\alpha^2}{2} C_6^2  + \frac{C_1^2}{2}
\qquad \qquad c = C_5\left( C_1 + \alpha C_6\right), \nn
\eea
and:
\be
Z= 1-\frac{2}{3 \alpha-1}\left(\frac{\tau_f}{\sigma\,\vo^{2/3}}\right)^{3/2}. \nn
\ee
Notice that the K\"ahler cone conditions $\tau_f<\sigma\,\vo^{2/3}$ and $\alpha>1$ imply $0<Z<1$. 
This guarantees the absence of any singularity in the K\"ahler metric. Expressing the scalar potential in terms of the 4-cycle moduli, we end up with:
\be
V_{g_s}^\KK = \kappa g_s^2\frac{W_0^2}{Z\vo^2} \left[\frac{C_5^2}{\tau_f^2}  - \frac{2}{3\left(2\alpha-1\right)^{3/2}}\frac{C_5^2}{\vo\sqrt{\tau_f}}
+ d\,\frac{\tau_f}{\vo^2}\left(1 -h\, \frac{\tau_f^{3/2}}{\vo}\right)\right], 
\label{Vgskk}
\ee
where $h=u/d$ with:
\bea
d &=&  \frac{a}{\left(2\alpha-1\right)}
- \frac23 \frac{c}{\left(2\alpha-1\right)^2}
- \frac{C_5^2}{\left(2\alpha-1\right)^3}\left(\alpha^2-\frac{\alpha}{3}-\frac29\right) \nn \\
u &=& \frac{2\,b}{3\left(2\alpha-1\right)^{5/2}} 
 +  \frac{2\,c}{3}\,\frac{\left(\alpha-\frac23 \right)}{\left(2\alpha-1\right)^{7/2}} 
+\frac{C_5^2}{3}\,\frac{\left(\alpha-\frac23 \right)^2}{\left(2\alpha-1\right)^{9/2}}\,. \nn
\eea
If all the coefficients of the KK corrections take natural $\mc{O}(1)$ values, the term in (\ref{Vgskk}) proportional to $h$ is suppressed by $h\ll1$, 
and so it can be safely neglected. 

On the other hand, higher derivative $\alpha'^3$ $F^4$ corrections take the form (neglecting the $t_4$-dependent term and setting $t_6=\alpha t_1$):
\be
V_{F^4}=-4\kappa^2 \frac{\lambda W_0^4}{g_s^{3/2}\vo^4}\left[(6\alpha+1) t_1+6t_5 \right]\,,
\ee
which in terms of four-cycle moduli looks like:
\be
V_{F^4}=-4\kappa^2 \frac{\lambda W_0^4}{g_s^{3/2}\vo^4}\left[ \frac{12\alpha^2+2\alpha-5}{(2\alpha-1)^{3/2}}
\sqrt{\tau_f}+ 6\frac{\vo}{\tau_f} \right].
\ee
Therefore the total inflationary potential becomes:
\be
V = V_{g_s}^\W+V_{g_s}^\KK + V_{F^4}  
= \kappa \frac{W_0^2}{\vo^2}\left(\frac{A_1}{\tau_f^2}+\frac{A_2}{\vo\tau_f}-\frac{A_3}{\vo\sqrt{\tau_f}}+\frac{B_1\,\sqrt{\tau_f}}{\vo^2}+\frac{B_2\,\tau_f}{\vo^2}\right),
\label{Vtot}
\ee
where (with $\lambda=-|\lambda|<0$):
\be
A_1=\frac{g_s^2}{Z}\,C_5^2\qquad 
A_2 = \frac{3}{\pi} \frac{|\lambda| W_0^2}{\sqrt{g_s}}
\qquad A_3= C_\W+ \frac{g_s^2}{Z}\frac{2\,C_5^2}{3\left(2\alpha-1\right)^{3/2}}\simeq C_\W
\ee
and:
\be
B_1 = \frac{12\alpha^2+2\alpha-5}{6(2\alpha-1)^{3/2}}\,A_2 \qquad B_2= \frac{g_s^2\,d}{Z} \,.
\ee
The potential (\ref{Vtot}) could support single-field slow-roll inflation driven by $\tau_f$ \cite{Cicoli:2008gp, Cicoli:2016chb}. In order to get enough efoldings before hitting the walls of the K\"ahler cone given in (\ref{KCbound}), we need to focus on the region in field space where the inflaton minimum is of order $\langle\tau_f\rangle\ll \vo^{2/3}$. Numerical estimates show that we need values of order $\langle\tau_f\rangle \sim\mc{O}(1)$ and $\vo \sim\mc{O}(10^4)$ which, in turn, imply $W_0\sim\mc{O}(100)$ in order to match the observed amplitude of the density perturbations. For $g_s\lesssim\mc{O}(0.1)$, $|\lambda|\sim\mc{O}(10^{-3})$ and natural $\mc{O}(1)$ values of the coefficients of the string loop effects, the terms in (\ref{Vtot}) proportional to $B_1$ and $B_2$ are both negligible with respect to the first three terms in the vicinity of the minimum where $\tau_f\sim \mc{O}(1)\ll\vo^{2/3}$. 

The scalar potential (\ref{Vtot}) written in terms of the canonically normalised inflaton $\phi = \langle \phi \rangle + \hat\phi$ looks like (with $k=2/\sqrt{3}$):
\be
V = \kappa\,\frac{A_1 W_0^2}{\langle \tau_f \rangle^2 \vo^2} \left(C_\dS+ e^{-2 k \hat\phi}+\lambda_1 Z\,e^{-k\hat\phi}- \lambda_2 Z\,e^{-\frac{k \hat\phi}{2}} + \mc{R}_1 Z\, e^{\frac{k \hat\phi}{2}}+ \mc{R}_2\, e^{k \hat\phi}\right)\,,
\label{Vref}
\ee
where we added a constant $C_\dS=\lambda_2 Z-\lambda_1 Z -1 - \mc{R}_1 Z -\mc{R}_2$ to obtain a Minkowski (or slightly dS) vacuum and:
\be
\lambda_1 = \frac{3\langle\tau_f\rangle}{\pi C_5^2} \frac{|\lambda| W_0^2}{g_s^{5/2}\,\vo}\sim\mc{O}(1-10)
\qquad \qquad\lambda_2 \simeq \frac{\langle\tau_f\rangle^{3/2}}{C_5^2} \frac{C_\W }{g_s^2\,\vo}\sim\mc{O}(1-10)\,, \nn
\ee
while:
\be
\mc{R}_1 = \frac{12\alpha^2+2\alpha-5}{6(2\alpha-1)^{3/2}} \frac{\lambda_1\langle\tau_f\rangle^{3/2}}{\vo} \ll 1
\qquad\qquad
\mc{R}_2 = \frac{\langle\tau_f\rangle^3}{C_5^2} \frac{d}{\vo^2} \ll 1\,. \nn
\ee
The three negative exponentials in (\ref{Vref}) compete to give a minimum at $\langle\tau_f\rangle \sim\mc{O}(1)$ while the two positive exponentials cause a steepening behaviour at large $\hat\phi$. 

In this section we shall not present a detailed quantitative analysis of inflation. We however point out that, if the approximated expression (\ref{KCbound}) is correct, in this case the K\"ahler cone bounds seem to be more constraining than in the case discussed in the main text since the inflaton direction $\tau_f$ is bounded by $\vo^{2/3}$ instead of $\vo/\sqrt{\tau_s}$. Thus a viable inflationary dynamics in this case would require a more severe tuning of the underlying parameters and a better understanding of the validity of our effective field theory approach.


\begin{thebibliography}{99}

\bibitem{Lyth:1996im}
  D.~H.~Lyth,
  ``What would we learn by detecting a gravitational wave signal in the cosmic microwave background anisotropy?,''
  Phys.\ Rev.\ Lett.\  {\bf 78} (1997) 1861
  doi:10.1103/PhysRevLett.78.1861
  [hep-ph/9606387].
  
\bibitem{McAllister:2007bg}
  L.~McAllister and E.~Silverstein,
  ``String Cosmology: A Review,''
  Gen.\ Rel.\ Grav.\  {\bf 40} (2008) 565
  doi:10.1007/s10714-007-0556-6
  [arXiv:0710.2951 [hep-th]].

\bibitem{Baumann:2009ni}
  D.~Baumann and L.~McAllister,
  ``Advances in Inflation in String Theory,''
  Ann.\ Rev.\ Nucl.\ Part.\ Sci.\  {\bf 59} (2009) 67
  doi:10.1146/annurev.nucl.010909.083524
  [arXiv:0901.0265 [hep-th]].

\bibitem{Cicoli:2011zz}
  M.~Cicoli and F.~Quevedo,
  ``String moduli inflation: An overview,''
  Class.\ Quant.\ Grav.\  {\bf 28} (2011) 204001
  doi:10.1088/0264-9381/28/20/204001
  [arXiv:1108.2659 [hep-th]].
  
\bibitem{Burgess:2013sla}
  C.~P.~Burgess, M.~Cicoli and F.~Quevedo,
  ``String Inflation After Planck 2013,''
  JCAP {\bf 1311} (2013) 003
  doi:10.1088/1475-7516/2013/11/003
  [arXiv:1306.3512 [hep-th]].
	
\bibitem{Burgess:2014tja}
  C.~P.~Burgess, M.~Cicoli, F.~Quevedo and M.~Williams,
  ``Inflating with Large Effective Fields,''
  JCAP {\bf 1411} (2014) 045
  doi:10.1088/1475-7516/2014/11/045
  [arXiv:1404.6236 [hep-th]].

\bibitem{Blumenhagen:2007sm}
  R.~Blumenhagen, S.~Moster and E.~Plauschinn,
  ``Moduli Stabilisation versus Chirality for MSSM like Type IIB Orientifolds,''
  JHEP {\bf 0801} (2008) 058
  doi:10.1088/1126-6708/2008/01/058
  [arXiv:0711.3389 [hep-th]].

  \bibitem{Green:2007gs}
  D.~R.~Green,
  ``Reheating Closed String Inflation,''
  Phys.\ Rev.\ D {\bf 76} (2007) 103504
  doi:10.1103/PhysRevD.76.103504
  [arXiv:0707.3832 [hep-th]].

\bibitem{Brandenberger:2008kn}
  R.~H.~Brandenberger, A.~Knauf and L.~C.~Lorenz,
  ``Reheating in a Brane Monodromy Inflation Model,''
  JHEP {\bf 0810} (2008) 110
  doi:10.1088/1126-6708/2008/10/110
  [arXiv:0808.3936 [hep-th]].

\bibitem{Barnaby:2009wr}
  N.~Barnaby, J.~R.~Bond, Z.~Huang and L.~Kofman,
  ``Preheating After Modular Inflation,''
  JCAP {\bf 0912} (2009) 021
  doi:10.1088/1475-7516/2009/12/021
  [arXiv:0909.0503 [hep-th]].

\bibitem{Cicoli:2010ha}
  M.~Cicoli and A.~Mazumdar,
  ``Reheating for Closed String Inflation,''
  JCAP {\bf 1009} (2010) no.09,  025
  doi:10.1088/1475-7516/2010/09/025
  [arXiv:1005.5076 [hep-th]];
	M.~Cicoli and A.~Mazumdar,
  ``Inflation in string theory: A Graceful exit to the real world,''
  Phys.\ Rev.\ D {\bf 83} (2011) 063527
  doi:10.1103/PhysRevD.83.063527
  [arXiv:1010.0941 [hep-th]].

\bibitem{Dutta:2014tya}
  K.~Dutta and A.~Maharana,
  ``Inflationary constraints on modulus dominated cosmology,''
  Phys.\ Rev.\ D {\bf 91} (2015) no.4,  043503
  doi:10.1103/PhysRevD.91.043503
  [arXiv:1409.7037 [hep-ph]].
  
\bibitem{Cicoli:2016olq}
  M.~Cicoli, K.~Dutta, A.~Maharana and F.~Quevedo,
  ``Moduli Vacuum Misalignment and Precise Predictions in String Inflation,''
  JCAP {\bf 1608} (2016) no.08,  006
  doi:10.1088/1475-7516/2016/08/006
  [arXiv:1604.08512 [hep-th]].	
	
\bibitem{Bhattacharya:2017ysa}
  S.~Bhattacharya, K.~Dutta and A.~Maharana,
  ``Constrains on K\"ahler Moduli Inflation from Reheating,''
  arXiv:1707.07924 [hep-ph].	
	
\bibitem{Cicoli:2012aq}
  M.~Cicoli, J.~P.~Conlon and F.~Quevedo,
  ``Dark radiation in LARGE volume models,''
  Phys.\ Rev.\ D {\bf 87} (2013) no.4,  043520
  doi:10.1103/PhysRevD.87.043520
  [arXiv:1208.3562 [hep-ph]].
  
\bibitem{Higaki:2012ar}
  T.~Higaki and F.~Takahashi,
  ``Dark Radiation and Dark Matter in Large Volume Compactifications,''
  JHEP {\bf 1211} (2012) 125
  doi:10.1007/JHEP11(2012)125
  [arXiv:1208.3563 [hep-ph]].

\bibitem{Hebecker:2014gka}
  A.~Hebecker, P.~Mangat, F.~Rompineve and L.~T.~Witkowski,
  ``Dark Radiation predictions from general Large Volume Scenarios,''
  JHEP {\bf 1409} (2014) 140
  doi:10.1007/JHEP09(2014)140
  [arXiv:1403.6810 [hep-ph]].
	
\bibitem{Cicoli:2015bpq}
  M.~Cicoli and F.~Muia,
  ``General Analysis of Dark Radiation in Sequestered String Models,''
  JHEP {\bf 1512} (2015) 152
  doi:10.1007/JHEP12(2015)152
  [arXiv:1511.05447 [hep-th]].

\bibitem{Acharya:2008bk}
  B.~S.~Acharya, P.~Kumar, K.~Bobkov, G.~Kane, J.~Shao and S.~Watson,
  ``Non-thermal Dark Matter and the Moduli Problem in String Frameworks,''
  JHEP {\bf 0806} (2008) 064
  doi:10.1088/1126-6708/2008/06/064
  [arXiv:0804.0863 [hep-ph]].

\bibitem{Allahverdi:2013noa}
  R.~Allahverdi, M.~Cicoli, B.~Dutta and K.~Sinha,
  ``Nonthermal dark matter in string compactifications,''
  Phys.\ Rev.\ D {\bf 88} (2013) no.9,  095015
  doi:10.1103/PhysRevD.88.095015
  [arXiv:1307.5086 [hep-ph]];
R.~Allahverdi, M.~Cicoli, B.~Dutta and K.~Sinha,
  ``Correlation between Dark Matter and Dark Radiation in String Compactifications,''
  JCAP {\bf 1410} (2014) 002
  doi:10.1088/1475-7516/2014/10/002
  [arXiv:1401.4364 [hep-ph]].

\bibitem{Aparicio:2015sda}
  L.~Aparicio, M.~Cicoli, B.~Dutta, S.~Krippendorf, A.~Maharana, F.~Muia and F.~Quevedo,
  ``Non-thermal CMSSM with a 125 GeV Higgs,''
  JHEP {\bf 1505} (2015) 098
  doi:10.1007/JHEP05(2015)098
  [arXiv:1502.05672 [hep-ph]];
  L.~Aparicio, B.~Dutta, M.~Cicoli, F.~Muia and F.~Quevedo,
  ``Light Higgsino Dark Matter from Non-thermal Cosmology,''
  JHEP {\bf 1611} (2016) 038
  [arXiv:1607.00004 [hep-ph]].

\bibitem{Kane:2011ih}
  G.~Kane, J.~Shao, S.~Watson and H.~B.~Yu,
  ``The Baryon-Dark Matter Ratio Via Moduli Decay After Affleck-Dine Baryogenesis,''
  JCAP {\bf 1111} (2011) 012
  doi:10.1088/1475-7516/2011/11/012
  [arXiv:1108.5178 [hep-ph]].
	
\bibitem{Allahverdi:2016yws}
  R.~Allahverdi, M.~Cicoli and F.~Muia,
  ``Affleck-Dine Baryogenesis in Type IIB String Models,''
  JHEP {\bf 1606} (2016) 153
  doi:10.1007/JHEP06(2016)153
  [arXiv:1604.03120 [hep-th]].

\bibitem{Conlon:2008cj}
  J.~P.~Conlon, R.~Kallosh, A.~D.~Linde and F.~Quevedo,
  ``Volume Modulus Inflation and the Gravitino Mass Problem,''
  JCAP {\bf 0809} (2008) 011
  doi:10.1088/1475-7516/2008/09/011
  [arXiv:0806.0809 [hep-th]];
M.~Cicoli, F.~Muia and F.~G.~Pedro,
  ``Microscopic Origin of Volume Modulus Inflation,''
  JCAP {\bf 1512} (2015) no.12,  040
  doi:10.1088/1475-7516/2015/12/040
  [arXiv:1509.07748 [hep-th]].

\bibitem{He:2010uk}
  T.~He, S.~Kachru and A.~Westphal,
  ``Gravity waves and the LHC: Towards high-scale inflation with low-energy SUSY,''
  JHEP {\bf 1006} (2010) 065
  doi:10.1007/JHEP06(2010)065
  [arXiv:1003.4265 [hep-th]].

\bibitem{Antusch:2011wu}
  S.~Antusch, K.~Dutta and S.~Halter,
  ``Combining High-scale Inflation with Low-energy SUSY,''
  JHEP {\bf 1203} (2012) 105
  doi:10.1007/JHEP03(2012)105
  [arXiv:1112.4488 [hep-th]].

\bibitem{Buchmuller:2014pla}
  W.~Buchmuller, E.~Dudas, L.~Heurtier and C.~Wieck,
  ``Large-Field Inflation and Supersymmetry Breaking,''
  JHEP {\bf 1409} (2014) 053
  doi:10.1007/JHEP09(2014)053
  [arXiv:1407.0253 [hep-th]].

\bibitem{Conlon:2005jm}
  J.~P.~Conlon and F.~Quevedo,
  ``Kahler moduli inflation,''
  JHEP {\bf 0601} (2006) 146
  doi:10.1088/1126-6708/2006/01/146
  [hep-th/0509012].

\bibitem{Cicoli:2017shd}
  M.~Cicoli, I.~Garcìa-Etxebarria, C.~Mayrhofer, F.~Quevedo, P.~Shukla and R.~Valandro,
  ``Global Orientifolded Quivers with Inflation,''
  arXiv:1706.06128 [hep-th].

\bibitem{Balasubramanian:2005zx}
  V.~Balasubramanian, P.~Berglund, J.~P.~Conlon and F.~Quevedo,
  ``Systematics of moduli stabilisation in Calabi-Yau flux compactifications,''
  JHEP {\bf 0503} (2005) 007
  doi:10.1088/1126-6708/2005/03/007
  [hep-th/0502058].
	
\bibitem{Cicoli:2008va}
  M.~Cicoli, J.~P.~Conlon and F.~Quevedo,
  ``General Analysis of LARGE Volume Scenarios with String Loop Moduli Stabilisation,''
  JHEP {\bf 0810} (2008) 105
  doi:10.1088/1126-6708/2008/10/105
  [arXiv:0805.1029 [hep-th]].

\bibitem{Cicoli:2008gp}
  M.~Cicoli, C.~P.~Burgess and F.~Quevedo,
  ``Fibre Inflation: Observable Gravity Waves from IIB String Compactifications,''
  JCAP {\bf 0903} (2009) 013
  doi:10.1088/1475-7516/2009/03/013
  [arXiv:0808.0691 [hep-th]].
  
\bibitem{Berg:2005ja}
  M.~Berg, M.~Haack and B.~Kors,
  ``String loop corrections to Kahler potentials in orientifolds,''
  JHEP {\bf 0511} (2005) 030
  doi:10.1088/1126-6708/2005/11/030
  [hep-th/0508043].	
	
\bibitem{Berg:2007wt}
  M.~Berg, M.~Haack and E.~Pajer,
  ``Jumping Through Loops: On Soft Terms from Large Volume Compactifications,''
  JHEP {\bf 0709} (2007) 031
  doi:10.1088/1126-6708/2007/09/031
  [arXiv:0704.0737 [hep-th]].

\bibitem{Berg:2014ama}
  M.~Berg, M.~Haack, J.~U.~Kang and S.~Sjörs,
  ``Towards the one-loop Kähler metric of Calabi-Yau orientifolds,''
  JHEP {\bf 1412} (2014) 077
  doi:10.1007/JHEP12(2014)077
  [arXiv:1407.0027 [hep-th], arXiv:1407.0027].

\bibitem{Haack:2015pbv}
  M.~Haack and J.~U.~Kang,
  ``One-loop Einstein-Hilbert term in minimally supersymmetric type IIB orientifolds,''
  JHEP {\bf 1602} (2016) 160
  doi:10.1007/JHEP02(2016)160
  [arXiv:1511.03957 [hep-th]].

\bibitem{Ciupke:2015msa}
  D.~Ciupke, J.~Louis and A.~Westphal,
  ``Higher-Derivative Supergravity and Moduli Stabilization,''
  JHEP {\bf 1510} (2015) 094
  doi:10.1007/JHEP10(2015)094
  [arXiv:1505.03092 [hep-th]].
  
\bibitem{Grimm:2017okk}
  T.~W.~Grimm, K.~Mayer and M.~Weissenbacher,
  ``Higher derivatives in Type II and M-theory on Calabi-Yau threefolds,''
  arXiv:1702.08404 [hep-th].

\bibitem{Cicoli:2007xp}
  M.~Cicoli, J.~P.~Conlon and F.~Quevedo,
  ``Systematics of String Loop Corrections in Type IIB Calabi-Yau Flux Compactifications,''
  JHEP {\bf 0801} (2008) 052
  doi:10.1088/1126-6708/2008/01/052
  [arXiv:0708.1873 [hep-th]].
	
\bibitem{Broy:2015zba}
  B.~J.~Broy, D.~Ciupke, F.~G.~Pedro and A.~Westphal,
  ``Starobinsky-Type Inflation from $\alpha'$-Corrections,''
  JCAP {\bf 1601} (2016) 001
  doi:10.1088/1475-7516/2016/01/001
  [arXiv:1509.00024 [hep-th]].	
	
\bibitem{Cicoli:2016chb}
  M.~Cicoli, D.~Ciupke, S.~de Alwis and F.~Muia,
  ``$\alpha'$ Inflation: moduli stabilisation and observable tensors from higher derivatives,''
  JHEP {\bf 1609} (2016) 026
  doi:10.1007/JHEP09(2016)026
  [arXiv:1607.01395 [hep-th]].

\bibitem{Burgess:2016owb}
  C.~P.~Burgess, M.~Cicoli, S.~de Alwis and F.~Quevedo,
  ``Robust Inflation from Fibrous Strings,''
  JCAP {\bf 1605} (2016) no.05,  032
  doi:10.1088/1475-7516/2016/05/032
  [arXiv:1603.06789 [hep-th]].

\bibitem{Ade:2015lrj}
  P.~A.~R.~Ade {\it et al.} [Planck Collaboration],
  ``Planck 2015 results. XX. Constraints on inflation,''
  Astron.\ Astrophys.\  {\bf 594} (2016) A20
  doi:10.1051/0004-6361/201525898
  [arXiv:1502.02114 [astro-ph.CO]].

\bibitem{Ade:2015xua}
  P.~A.~R.~Ade {\it et al.} [Planck Collaboration],
  ``Planck 2015 results. XIII. Cosmological parameters,''
  Astron.\ Astrophys.\  {\bf 594} (2016) A13
  doi:10.1051/0004-6361/201525830
  [arXiv:1502.01589 [astro-ph.CO]].

\bibitem{Kallosh:2017wku}
  R.~Kallosh, A.~Linde, D.~Roest, A.~Westphal and Y.~Yamada,
  ``Fibre Inflation and $\alpha$-attractors,''
  arXiv:1707.05830 [hep-th].

\bibitem{Cicoli:2016xae}
  M.~Cicoli, F.~Muia and P.~Shukla,
  ``Global Embedding of Fibre Inflation Models,''
  JHEP {\bf 1611} (2016) 182
  doi:10.1007/JHEP11(2016)182
  [arXiv:1611.04612 [hep-th]].

\bibitem{Cicoli:2011it}
  M.~Cicoli, M.~Kreuzer and C.~Mayrhofer,
  ``Toric K3-Fibred Calabi-Yau Manifolds with del Pezzo Divisors for String Compactifications,''
  JHEP {\bf 1202} (2012) 002
  doi:10.1007/JHEP02(2012)002
  [arXiv:1107.0383 [hep-th]].

\bibitem{Kreuzer:2000xy}
  M.~Kreuzer and H.~Skarke,
  ``Complete classification of reflexive polyhedra in four-dimensions,''
  Adv.\ Theor.\ Math.\ Phys.\  {\bf 4} (2002) 1209
  [hep-th/0002240].

\bibitem{Angus:2014bia}
  S.~Angus,
  ``Dark Radiation in Anisotropic LARGE Volume Compactifications,''
  JHEP {\bf 1410} (2014) 184
  doi:10.1007/JHEP10(2014)184
  [arXiv:1403.6473 [hep-ph]].

\bibitem{Cicoli:2013swa}
  M.~Cicoli, J.~P.~Conlon, A.~Maharana and F.~Quevedo,
  ``A Note on the Magnitude of the Flux Superpotential,''
  JHEP {\bf 1401} (2014) 027
  doi:10.1007/JHEP01(2014)027
  [arXiv:1310.6694 [hep-th]].

\bibitem{Becker:2002nn}
  K.~Becker, M.~Becker, M.~Haack and J.~Louis,
  ``Supersymmetry breaking and alpha-prime corrections to flux induced potentials,''
  JHEP {\bf 0206} (2002) 060
  doi:10.1088/1126-6708/2002/06/060
  [hep-th/0204254].

\bibitem{Minasian:2015bxa}
  R.~Minasian, T.~G.~Pugh and R.~Savelli,
  ``F-theory at order $\alpha'^3$,''
  JHEP {\bf 1510} (2015) 050
  doi:10.1007/JHEP10(2015)050
  [arXiv:1506.06756 [hep-th]].

\bibitem{Bonetti:2016dqh}
  F.~Bonetti and M.~Weissenbacher,
  ``The Euler characteristic correction to the Kähler potential — revisited,''
  JHEP {\bf 1701} (2017) 003
  doi:10.1007/JHEP01(2017)003
  [arXiv:1608.01300 [hep-th]].

\bibitem{Kachru:2003aw}
  S.~Kachru, R.~Kallosh, A.~D.~Linde and S.~P.~Trivedi,
  ``De Sitter vacua in string theory,''
  Phys.\ Rev.\ D {\bf 68} (2003) 046005
  doi:10.1103/PhysRevD.68.046005
  [hep-th/0301240].

\bibitem{antiDdS}
  R.~Kallosh and T.~Wrase,
  ``Emergence of Spontaneously Broken Supersymmetry on an Anti-D3-Brane in KKLT dS Vacua,''
  JHEP {\bf 1412} (2014) 117
  doi:10.1007/JHEP12(2014)117
  [arXiv:1411.1121 [hep-th]]; 
   E.~A.~Bergshoeff, K.~Dasgupta, R.~Kallosh, A.~Van Proeyen and T.~Wrase,
  ``$ \overline{\mathrm{D}3} $ and dS,''
  JHEP {\bf 1505} (2015) 058
  doi:10.1007/JHEP05(2015)058
  [arXiv:1502.07627 [hep-th]];
  R.~Kallosh, F.~Quevedo and A.~M.~Uranga,
  ``String Theory Realizations of the Nilpotent Goldstino,''
  JHEP {\bf 1512} (2015) 039
  doi:10.1007/JHEP12(2015)039
  [arXiv:1507.07556 [hep-th]];
  L.~Aparicio, F.~Quevedo and R.~Valandro,
  ``Moduli Stabilisation with Nilpotent Goldstino: Vacuum Structure and SUSY Breaking,''
  JHEP {\bf 1603} (2016) 036
  doi:10.1007/JHEP03(2016)036
  [arXiv:1511.08105 [hep-th]];
   I.~García-Etxebarria, F.~Quevedo and R.~Valandro,
  ``Global String Embeddings for the Nilpotent Goldstino,''
  JHEP {\bf 1602} (2016) 148
  doi:10.1007/JHEP02(2016)148
  [arXiv:1512.06926 [hep-th]];
	M.~P.~Garcia del Moral, S.~Parameswaran, N.~Quiroz and I.~Zavala,
  ``Anti-D3 branes and moduli in non-linear supergravity,''
  arXiv:1707.07059 [hep-th];
 J.~Moritz, A.~Retolaza and A.~Westphal,
  ``Towards de Sitter from 10D,''
  arXiv:1707.08678 [hep-th].

\bibitem{Cicoli:2015ylx}
  M.~Cicoli, F.~Quevedo and R.~Valandro,
  ``De Sitter from T-branes,''
  JHEP {\bf 1603} (2016) 141
  doi:10.1007/JHEP03(2016)141
  [arXiv:1512.04558 [hep-th]].

\bibitem{Cicoli:2012fh}
  M.~Cicoli, A.~Maharana, F.~Quevedo and C.~P.~Burgess,
  ``De Sitter String Vacua from Dilaton-dependent Non-perturbative Effects,''
  JHEP {\bf 1206} (2012) 011
  doi:10.1007/JHEP06(2012)011
  [arXiv:1203.1750 [hep-th]].

\bibitem{Westphal:2015eva}
  A.~Westphal,
  ``String Cosmology — Large-Field Inflation in String Theory,''
  Adv.\ Ser.\ Direct.\ High Energy Phys.\  {\bf 22} (2015) 351.

\bibitem{Remazeilles:2015hpa}
  M.~Remazeilles, C.~Dickinson, H.~K.~K.~Eriksen and I.~K.~Wehus,
  ``Sensitivity and foreground modelling for large-scale cosmic microwave background B-mode polarization satellite missions,''
  Mon.\ Not.\ Roy.\ Astron.\ Soc.\  {\bf 458} (2016) no.2,  2032
  doi:10.1093/mnras/stw441
  [arXiv:1509.04714 [astro-ph.CO]].

\bibitem{Remazeilles:2017zlb}
  M.~Remazeilles, C.~Dickinson, H.~K.~Eriksen and I.~K.~Wehus,
  ``Joint Bayesian estimation of tensor and lensing B-modes in the power spectrum of CMB polarization data,''
  arXiv:1707.02981 [astro-ph.CO].
	
\bibitem{Minasian:1997mm}
  R.~Minasian and G.~W.~Moore,
  ``K theory and Ramond-Ramond charge,''
  JHEP {\bf 9711} (1997) 002
  doi:10.1088/1126-6708/1997/11/002
  [hep-th/9710230].

\bibitem{Freed:1999vc}
  D.~S.~Freed and E.~Witten,
  ``Anomalies in string theory with D-branes,''
  Asian J.\ Math.\  {\bf 3} (1999) 819
  [hep-th/9907189].

\bibitem{Dine:1987xk}
  M.~Dine, N.~Seiberg and E.~Witten,
  ``Fayet-Iliopoulos Terms in String Theory,''
  Nucl.\ Phys.\ B {\bf 289} (1987) 589.
  doi:10.1016/0550-3213(87)90395-6
	
\bibitem{Dine:1987gj}
  M.~Dine, I.~Ichinose and N.~Seiberg,
  ``F Terms and d Terms in String Theory,''
  Nucl.\ Phys.\ B {\bf 293} (1987) 253.
  doi:10.1016/0550-3213(87)90072-1

\bibitem{Altman:2014bfa}
  R.~Altman, J.~Gray, Y.~H.~He, V.~Jejjala and B.~D.~Nelson,
  ``A Calabi-Yau Database: Threefolds Constructed from the Kreuzer-Skarke List,''
  JHEP {\bf 1502} (2015) 158
  doi:10.1007/JHEP02(2015)158
  [arXiv:1411.1418 [hep-th]].

\bibitem{Blumenhagen:2010pv}
  R.~Blumenhagen, B.~Jurke, T.~Rahn and H.~Roschy,
  ``Cohomology of Line Bundles: A Computational Algorithm,''
  J.\ Math.\ Phys.\  {\bf 51} (2010) 103525
  doi:10.1063/1.3501132, 10.1063/1.3523343
  [arXiv:1003.5217 [hep-th]].

\bibitem{Blumenhagen:2011xn}
  R.~Blumenhagen, B.~Jurke and T.~Rahn,
  ``Computational Tools for Cohomology of Toric Varieties,''
  Adv.\ High Energy Phys.\  {\bf 2011} (2011) 152749
  doi:10.1155/2011/152749
  [arXiv:1104.1187 [hep-th]].

\bibitem{Cicoli:2012vw} 
  M.~Cicoli, S.~Krippendorf, C.~Mayrhofer, F.~Quevedo and R.~Valandro,
  ``D-Branes at del Pezzo Singularities: Global Embedding and Moduli Stabilisation,''
  JHEP {\bf 1209}, 019 (2012)
  doi:10.1007/JHEP09(2012)019
  [arXiv:1206.5237 [hep-th]].

\bibitem{Camara:2004jj}
  P.~G.~Camara, L.~E.~Ibanez and A.~M.~Uranga,
  ``Flux-induced SUSY-breaking soft terms on D7-D3 brane systems,''
  Nucl.\ Phys.\ B {\bf 708} (2005) 268
  doi:10.1016/j.nuclphysb.2004.11.035
  [hep-th/0408036].

\bibitem{Ciupke:2016agp}
  D.~Ciupke,
  ``Scalar Potential from Higher Derivative $\mathcal{N} = 1$ Superspace,''
  arXiv:1605.00651 [hep-th].
  
\bibitem{Cicoli:2013oba}
  M.~Cicoli, S.~Downes and B.~Dutta,
  ``Power Suppression at Large Scales in String Inflation,''
  JCAP {\bf 1312} (2013) 007
  doi:10.1088/1475-7516/2013/12/007
  [arXiv:1309.3412 [hep-th]];
	M.~Cicoli, S.~Downes, B.~Dutta, F.~G.~Pedro and A.~Westphal,
  ``Just enough inflation: power spectrum modifications at large scales,''
  JCAP {\bf 1412} (2014) no.12,  030
  doi:10.1088/1475-7516/2014/12/030
  [arXiv:1407.1048 [hep-th]].
	
\bibitem{Klaewer:2016kiy}€
  D.~Klaewer and E.~Palti,
  ``Super-Planckian Spatial Field Variations and Quantum Gravity,''
  JHEP {\bf 1701} (2017) 088
  doi:10.1007/JHEP01(2017)088
  [arXiv:1610.00010 [hep-th]].
	
\bibitem{Blumenhagen:2017cxt}
  R.~Blumenhagen, I.~Valenzuela and F.~Wolf,
  ``The Swampland Conjecture and F-term Axion Monodromy Inflation,''
  arXiv:1703.05776 [hep-th].	

\bibitem{BlancoPillado:2009nw}
  J.~J.~Blanco-Pillado, D.~Buck, E.~J.~Copeland, M.~Gomez-Reino and N.~J.~Nunes,
  ``Kahler Moduli Inflation Revisited,''
  JHEP {\bf 1001} (2010) 081
  doi:10.1007/JHEP01(2010)081
  [arXiv:0906.3711 [hep-th]].

\bibitem{curvaton}	
D.~H.~Lyth and D.~Wands,
  ``Generating the curvature perturbation without an inflaton,''
  Phys.\ Lett.\ B {\bf 524} (2002) 5
  doi:10.1016/S0370-2693(01)01366-1
  [hep-ph/0110002];
T.~Moroi and T.~Takahashi,
  ``Effects of cosmological moduli fields on cosmic microwave background,''
  Phys.\ Lett.\ B {\bf 522} (2001) 215
   Erratum: [Phys.\ Lett.\ B {\bf 539} (2002) 303]
  doi:10.1016/S0370-2693(02)02070-1, 10.1016/S0370-2693(01)01295-3
  [hep-ph/0110096];
 C.~P.~Burgess, M.~Cicoli, M.~Gomez-Reino, F.~Quevedo, G.~Tasinato and I.~Zavala,
  ``Non-standard primordial fluctuations and nongaussianity in string inflation,''
  JHEP {\bf 1008} (2010) 045
  doi:10.1007/JHEP08(2010)045
  [arXiv:1005.4840 [hep-th]].

\bibitem{Cicoli:2018tcq}
  M.~Cicoli, D.~Ciupke, C.~Mayrhofer and P.~Shukla,
  ``A Geometrical Upper Bound on the Inflaton Range,''
  arXiv:1801.05434 [hep-th].
	
\bibitem{Cabella:2017zsa}
  P.~Cabella, A.~Di Marco and G.~Pradisi,
  ``Fiber inflation and reheating,''
  Phys.\ Rev.\ D {\bf 95} (2017) no.12,  123528
  doi:10.1103/PhysRevD.95.123528
  [arXiv:1704.03209 [astro-ph.CO]].	
	
\bibitem{Antusch:2017flz}
  S.~Antusch, F.~Cefala, S.~Krippendorf, F.~Muia, S.~Orani and F.~Quevedo,
  ``Oscillons from String Moduli,''
  arXiv:1708.08922 [hep-th].

\bibitem{Cicoli:2011qg}
  M.~Cicoli, C.~Mayrhofer and R.~Valandro,
  ``Moduli Stabilisation for Chiral Global Models,''
  JHEP {\bf 1202} (2012) 062
  [arXiv:1110.3333 [hep-th]].

\end{thebibliography}
\end{document}